\def\spose#1{\hbox to 0pt{#1\hss}}
\def\simlt{\mathrel{\spose{\lower 3pt\hbox{$\mathchar"218$}}
     \raise 2.0pt\hbox{$\mathchar"13C$}}}
\def\simgt{\mathrel{\spose{\lower 3pt\hbox{$\mathchar"218$}}
     \raise 2.0pt\hbox{$\mathchar"13E$}}}
\def\simpropto{\mathrel{\spose{\lower 3pt\hbox{$\mathchar"218$}}
     \raise 2.0pt\hbox{$\propto$}}}

\documentclass[twocolumn,aps,prd,nofootinbib,showpacs]{revtex4-1}
\usepackage{amsmath,graphicx,bm,color}
\begin{document}
\title{Global $21\,\textrm{cm}$ signal experiments: A designer's guide}

\author{Adrian Liu}
\email{acliu@berkeley.edu}
\affiliation{Dept. of Astronomy and Berkeley Center for Cosmological Physics, UC Berkeley, Berkeley, CA 94720, USA}
\affiliation{Dept. of Physics and MIT Kavli Institute, Massachusetts Institute of Technology, Cambridge, MA 02139, USA}
\author{Jonathan R. Pritchard}
\affiliation{Imperial Center for Inference and Cosmology, Imperial College London, Blackett Laboratory, Prince Consort Road, London SW7 2AZ, United Kingdom}
\author{Max Tegmark}
\affiliation{Dept. of Physics and MIT Kavli Institute, Massachusetts Institute of Technology, Cambridge, MA 02139, USA}
\author{Abraham Loeb}
\affiliation{Dept. of Astronomy, Harvard University, Cambridge, MA 02138, USA}
\date{\today}


\begin{abstract}
The global (\emph{i.e.} spatially averaged) spectrum of the redshifted $21\,\textrm{cm}$ line has generated much experimental interest lately, thanks to its potential to be a direct probe of the Epoch of Reionization and the Dark Ages, during which the first luminous objects formed.  Since the cosmological signal in question has a purely spectral signature, most experiments that have been built, designed, or proposed have essentially no angular sensitivity.  This can be problematic because with only spectral information, the expected global $21\,\textrm{cm}$ signal can be difficult to distinguish from foreground contaminants such as Galactic synchrotron radiation, since both are spectrally smooth and the latter is many orders of magnitude brighter.  In this paper, we establish a systematic mathematical framework for global signal data analysis.  The framework removes foregrounds in an optimal manner, complementing spectra with angular information.  We use our formalism to explore various experimental design trade-offs, and find that 1) with spectral-only methods, it is mathematically impossible to mitigate errors that arise from uncertainties in one's foreground model; 2) foreground contamination can be significantly reduced for experiments with fine angular resolution; 3) most of the statistical significance in a positive detection during the Dark Ages comes from a characteristic high-redshift trough in the $21\,\textrm{cm}$ brightness temperature; 4) Measurement errors decrease more rapidly with integration time for instruments with fine angular resolution; and 5) Better foreground models can help reduce errors, but once a modeling accuracy of a few percent is reached, significant improvements in accuracy will be required to further improve the measurements.  We show that if observations and data analysis algorithms are optimized based on these findings, an instrument with a $5^\circ$ wide beam can achieve highly significant detections (greater than $5\sigma$) of even extended (high $\Delta z$) reionization scenarios after integrating for $500\,\textrm{hrs}$.  This is in strong contrast to instruments without angular resolution, which cannot detect gradual reionization.  Ionization histories that are more abrupt can be detected with our fiducial instrument at the level of 10's to 100's of $\sigma$.  The expected errors are similarly low during the Dark Ages, and can yield a $25\sigma$ detection of the expected cosmological signal after only $100\,\textrm{hrs}$ of integration.
\end{abstract}
\maketitle

\section{Introduction}
\label{intro}
Measurements of the highly redshifted $21\,\textrm{cm}$ line are thought to be the primary way to make direct observations of the epoch of reionization and the preceding dark ages, when the first luminous objects were formed from primordial fluctuations \cite{aviBook}.  Theoretical studies have shown that observations of the $21\,\textrm{cm}$ line will not only provide crucial constraints on a relatively poorly understood period of structure formation, during which a complex interplay of dark matter physics and baryon astrophysics produced large scale changes in the intergalactic medium \cite{fob,morales2010,pritchard2012}; eventually, the enormous volume probed by the $21\,\textrm{cm}$ line will also allow one to make exquisite measurements of fundamental physics parameters \cite{barkanaLoebSep,mattParameterEst,yiParameterEst,Clesse2012}.  It is thus no surprise that numerous experimental groups are making concerted efforts to arrive at the first positive detection of the cosmological $21\,\textrm{cm}$ signal.

To date, most observational efforts have focused on understanding the \emph{fluctuations} in the $21\,\textrm{cm}$ line by measuring the $21\,\textrm{cm}$ brightness temperature power spectrum (although there have also been theoretical proposals to capture the non-Gaussianity of the signal \cite{wyitheMoralesSkewness,barkanaLoebDiff,gluscevicBarkanaDiff,panBarkanaDiff}).  These include the Murchison Widefield Array (MWA) \cite{tingayMWA}, the Precision Array for Probing the Epoch of Reionization (PAPER) \cite{aaronPAPER}, the Low Frequency Array (LOFAR) \cite{LOFAR}, and the Giant Metrewave Radio Telescope Epoch of Reionization (GMRT-EoR) \cite{gregGMRT} projects.  These experiments are difficult: high sensitivity requirements dictate long integration times to reach the expected amplitude of the faint cosmological signal, which is dominated by strong sources of foreground emission (such as Galactic synchrotron radiation) by several orders of magnitude in all parts of the sky \cite{gianniFG1,gianniFG2,GSM}.  While lately these experiments have made much progress towards a detection of the cosmological signal, the challenges remain daunting.

As an alternative way to detect the $21\,\textrm{cm}$ cosmological signal, there have recently been attempts to measure the \emph{global} $21\,\textrm{cm}$ signal, where one averages the signal over all directions in the sky and focuses on extracting a globally averaged spectrum, probing the evolution of the mean signal through cosmic history \cite{furlanettoFirstTheory}.  These observations are complementary to the power spectrum measurements, and have been shown to be an incisive probe of many physical parameters during reionization \cite{jonathanAvi}.  Examples of global signal experiments include the Experiment to Detect the Global EoR Signature (EDGES) \cite{bowmanRogersMeasurement}, the Large aperture Experiment to detect the Dark Ages (LEDA) \cite{LEDA}, the Long Wavelength Array (LWA) \cite{JuddPrivComm}, and if funded, the Dark Ages Radio Explorer (DARE) \cite{DARE}.

Compared to power spectrum measurements, global signal experiments are in some ways easier, and in other ways more difficult \cite{shaver}.  A rough calculation of the thermal noise reveals that global signal experiments need far less integration time to reach the required sensitivities for a detection of the cosmological signal.  However, the problem of foreground mitigation is much more challenging.  In a power spectrum measurement, one probes detailed maps of brightness temperature fluctuations in all three dimensions.  The maps are particularly detailed in the line-of-sight direction, since for a spectral line like the $21\,\textrm{cm}$ line this translates to the frequency spectrum of one's measurement, and typical instruments have extremely fine spectral resolution.  The result is that the cosmological signal fluctuates extremely rapidly with frequency, since one is probing local structure.  In contrast, the foregrounds are spectrally smooth.  This difference has formed the basis of most foreground subtraction schemes that have been proposed for power spectrum measurements, and theoretical calculations have been encouraging \cite{xiaomin,nusserforegrounds,Judd08,paper1,paper2,LOFARfg,Harker1,Harker2,PetrovicOh,paper4}.  Global signal experiments cannot easily take advantage of this difference in fluctuation scale, for the globally averaged spectrum does not trace local structures, but instead probes the average evolution with redshift.  The resulting signals are thus expected to be rather smooth functions of frequency, which makes them difficult to separate from the smooth foregrounds.  Traditionally, experiments have attempted to perform spectral separations anyway, and have thus been limited to ruling out sharp spectral signatures such as those that might arise from rapid reionization \cite{bowmanRogersMeasurement}.

In this paper, we confront the general problem of extracting the global $21\,\textrm{cm}$ signal from measurements that are contaminated by instrumental noise and foregrounds, placing the problem in a systematic mathematical framework.  We adopt the philosophy that one should take advantage of every possible difference between the cosmological signal and contaminants, not just the spectral differences.  To do so, we first develop optimal data analysis methods for cosmological signal estimation, using angular information to complement spectral information.  Traditional spectral-only methods are a special case in our formalism, and we prove that in such methods, foreground subtraction simply amounts to subtracting a ``best-guess" foreground model from the data and doing nothing to mitigate possible errors in the model itself.

Additionally, we build on existing foreground models in the literature, and use our improved model along with our data analysis formalism to predict expected measurement errors given various experimental parameters.  Since our formalism allows these measurement errors to be calculated analytically (given a foreground model), we are able to explore parameter space efficiently to identify the best experimental designs and the best data analysis methods.  We find that the most important such ``lesson learned" is that angular information is necessary to suppress foregrounds to a low enough level to differentiate between different cosmological models.

Our paper therefore has two logically separate (but closely related) goals.  The first is to develop a robust signal extraction formalism, and we fulfill this goal without assuming a precise model for the cosmological signal, thus immunizing ourselves to uncertainties in theoretical modeling.  The second is to build intuition for instrumental design and to forecast the performance of different types of instruments.  This part does assume a theoretical form for the cosmological signal.

In this paper our focus is on foreground removal, so for simplicity we do not consider calibration errors.  However, we do note that an accurate bandpass calibration (or at least a good estimate of the calibration error) is a prerequisite for accurate foreground removal.  It is for this reason that much effort has been spent on better instrument characterization recently \cite{rogersCalib}.  We also do not include frequency-dependent beam effects, although the formalism of this paper is general enough to include them.  Both calibration errors and frequency-dependent beams are inevitably instrument-specific issues, so we defer detailed investigations of them to future work.

The rest of this paper is organized as follows.  In Section \ref{models} we detail the properties of the cosmological signal and our foreground and instrumental noise model.  Readers that are more interested in data analysis methods may wish to skim this section (perhaps reading in detail only the summary in Section \ref{genNoiseModel}), and skip ahead to Section \ref{measurementformalism}, where we establish our mathematical framework for signal extraction and error computation.  Those that are interested in experimental design may wish to focus on Sections \ref{darkDesigner} and \ref{reionizationDesigner}, where we explore various experimental trade-offs for observations targeting the dark ages and reionization respectively.  Forecasts for the performances of some fiducial experiments are presented in Section \ref{fidExperiments}, and we summarize our conclusions in Section \ref{conc}.

\section{The cosmological global spectrum and sources of measurement contamination}
\label{models}
\subsection{Model of the signal}
\label{signalmodel}

We begin by sketching the ingredients of our $21\,\textrm{cm}$ global signal model. For more details there exist several good reviews on the physics of the $21\,\textrm{cm}$ signal \cite{fob,morales2010,pritchard2012}. The $21\,\textrm{cm}$ signal arises from the physical properties of neutral hydrogen in the intergalactic medium (IGM), specifically from the density of neutral hydrogen and the $21\,\textrm{cm}$ spin temperature, which describes the relative number of hydrogen atoms with proton and electron spin aligned or anti-aligned. These quantities respond to radiation from luminous sources and so are expected to vary from place to place. 

\begin{figure*}
\centering
\includegraphics[scale=0.8]{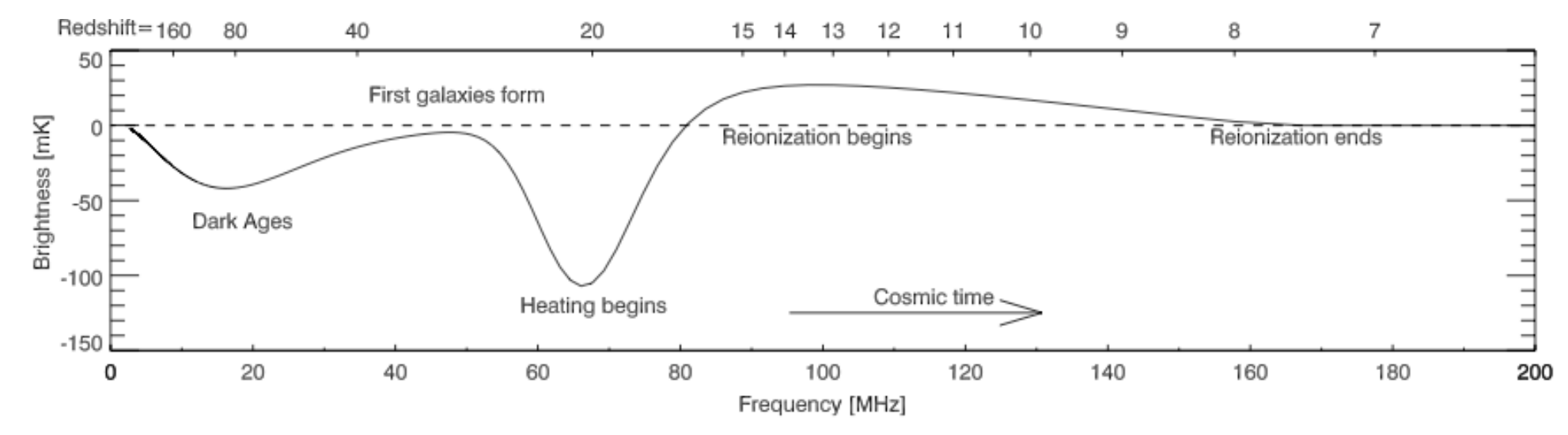}
\caption{Target $21\,\textrm{cm}$ global signal as predicted by the model of \cite{jonathanAvi}. The exact details of this signal are uncertain and depend upon the nature of the first galaxies.}
\label{fig:target_signal}
\end{figure*}

Fluctuations in the $21\,\textrm{cm}$ signal are being targeted by radio interferometers such as LOFAR, MWA, PAPER, and GMRT, as mentioned in Section \ref{intro}. These fluctuation are scale dependent with most power on the characteristic scale of ionized or heated regions, which are believed to be tens of arc-minutes across. When viewed with a beam larger than this characteristic size these fluctuations will average out, giving a measure of the mean or ``global" $21\,\textrm{cm}$ signal. In this paper, we concentrate on measuring this isotropic part of the $21\,\textrm{cm}$ signal and consider the anisotropic fluctuations as a source of noise. Studies by \cite{BittnerLoeb} showed that during reionization a beam size of a few degrees is sufficient to smooth out most of the fluctuations. In this paper, any contribution of the fluctuations left over after convolving with the large beam will be considered as irreducible noise.

The basic dependence of the differential $21\,\textrm{cm}$ brightness temperature $T_b$ on the average ionized fraction $\overline{x}_i$ and spin temperature $T_S$ is
\begin{equation}\label{modelTb}
T_b\approx27 (1-\overline{x}_i)\left(\frac{T_S-T_{\rm CMB}}{T_S}\right)\left(\frac{1+z}{10}\right)^{1/2} {\rm mK},
\end{equation}
where we have used the WMAP7 cosmological parameters to fix the baryon and mass abundances as $\Omega_bh^2=0.023$ and $\Omega_mh^2=0.15$ \cite{WMAP7}. The key redshift dependence comes via $x_i(z)$ and $T_S(z)$, which we model following the approach of \citep{pritchard2008}, where the reader will find technical details.  This model incorporates (1) the ionizing radiation from galaxies, (2) X-ray heating from galaxies \citep{pritchard2007}, (3) Lyman-alpha emission from galaxies, and assumes a simple prescription linking the star formation rate to the fraction of mass in collapsed structure above a critical mass threshold required for atomic hydrogen cooling.


This model predicts a $21\,\textrm{cm}$ signal that divides into three qualitatively different regimes (see Figure \ref{fig:target_signal}). The first, a shallow absorption feature at $30\lesssim z\lesssim200$ begins as the gas thermally decouples from the CMB and ends as our Universe become too rarified for collisions to couple $T_S$ to $T_{\rm gas}$. Next, a second and possibly deeper absorption feature occurs as the first galaxies form at $z\gtrsim30$. This is initiated as Lyman alpha photons illuminate the Universe, coupling spin and gas temperatures strongly, and ends as increasing X-ray emission heats the IGM above the CMB, leading to a $21\,\textrm{cm}$ emission signal. This emission signal is the third key feature, which slowly dies away in a ``reionization step" as ionizing UV photons ionize the IGM. As described in \cite{jonathanAvi, barkanaMorandi} there is considerable uncertainty in the exact positions and details of these features, but the basic picture seems robust.

The last two features---an absorption trough driven by the onset of galaxy formation\footnote{For linguistic convenience, we will include the absorption trough at $z\sim20$ as part of the Dark Ages, even though it really marks the very \emph{end} of the Dark Ages.} and an emission step accompanying reionization---form the focus of this paper, since these seem most likely to be detectable (a fact that we will explain and rigorously justify in Section \ref{spectral}). The earliest absorption trough seems unlikely to be detected in the near future, since it occurs at frequencies $\nu<50$ MHz that are more strongly affected by the Earth's ionosphere and where galactic foregrounds are considerably brighter.

It is important to note that the data analysis methods presented later in this paper do not depend on the precise form of the cosmological signal.  This differs from other studies in the literature (such as \cite{DAREMCMC}), which assume a parametric form for the signal and fit for parameters.  In contrast, we use the theoretical signal shown in Figure \ref{fig:target_signal} \emph{only} to assess detection significance and to identify trade-offs in instrumental design; our foreground subtraction and signal extraction algorithms will be applicable even if the true signal differs from our theoretical expectations, making our approach quite a robust one.  
\subsection{Generalized noise model}
\label{gennoise}
We now construct our noise model.  We define the generalized noise (or henceforth just ``noise") to be any contribution to the measurement that is not the global $21\,\textrm{cm}$ signal as described in the previous section.  As mentioned above, by this definition our noise contains more than just instrumental noise and foregrounds.  It also includes the \emph{anisotropic} portion of the cosmological signal.  In other words, the ``signal" in a tomographic measurement (where one measures angular anisotropies on various scales) is an unwanted contaminant in our case, since we seek to measure the \emph{global} signal (\emph{i.e.} the monopole).

If we imagine performing an experiment that images $N_{\textrm{pix}}$ pixels on the sky over $N_{\textrm{freq}}$ frequency channels, the noise contribution in various pixels at various frequencies can be grouped into a vector $\mathbf{n}$ of length equal to the number of voxels $N_{\textrm{vox}} \equiv N_{\textrm{pix}} N_{\textrm{freq}}$ in our survey.  It is comprised of three contaminants:
\begin{equation}
\label{noisedecomp}
\mathbf{n}_{\alpha i} \equiv \mathbf{n}^{\textrm{fg}}_{\alpha i} +  \mathbf{n}^{\textrm{inst}}_{\alpha i} + \mathbf{n}^{\textrm{s}}_{\alpha i},
\end{equation}
where $\mathbf{n}^{\textrm{fg}}$, $\mathbf{n}^{\textrm{inst}}$, and $\mathbf{n}^{\textrm{s}}$ signify the foregrounds, instrumental noise, and anisotropic cosmological signal, respectively.  Throughout this paper, we use Greek indices to signify the radial/frequency direction, and Latin indices to signify the spatial directions.  Note that $\mathbf{n}$ is formally a 
vector even though we assign separate spatial and spectral indices to it for clarity.  In the following subsections we discuss each of these three contributions to the noise, with an eye towards how each can be mitigated or removed in a real measurement.  We will construct detailed models containing parameters that are mostly constrained empirically.  However, since these constraints are often somewhat uncertain, we will vary many of them as we explore parameter space in Sections \ref{darkDesigner} and \ref{reionizationDesigner}.  Our conclusions should therefore be robust to reasonable changes in our assumptions.

Finally, we stress that in what follows, our models are comprised of two conceptually separate---but closely related---pieces.  To understand this, note that Equation \eqref{noisedecomp} is a random vector, both because the instrumental noise is sourced by random thermal fluctuations and because the foregrounds and the cosmological signal have modeling uncertainties associated with them.  Thus, to fully describe the behavior of $\mathbf{n}$, we need to specify two pieces of information: a mean (our ``best guess" of what the foregrounds and other noise sources look like as a function of frequency and angle) and a covariance (which quantifies the uncertainty and correlations in our best guess).  We will return to this point in Section \ref{genNoiseModel} when we summarize the essential features of our model.  Readers may wish to skip directly to that section if they are more interested in the ``designer's guide" portion of the paper than the mathematical details of our generalized noise model.

\subsubsection{Foreground Model}
\label{fgModel}
Given that foregrounds are likely to be the largest contaminant in a measurement of the global signal, it is important to have a foreground model that is an accurate reflection of the actual contamination faced by an experiment, as a function of both angle and frequency.  Having such a model that describes the particular realization of foregrounds contaminating a certain measurement is crucial for optimizing the foreground removal process, as we shall see in Section \ref{measurementformalism}.  However, constructing such a model is difficult to do from first principles, and is much more difficult than what is typically done, which is to capture only the \emph{statistical} behavior of the foregrounds (\emph{e.g.} by measuring quantities such as the spatial average of a spectral index).  It is thus likely that a full foreground model will have to be based at least partially on empirical data.

Unfortunately, the community currently lacks full-sky, low noise, high angular resolution survey data in the low frequency regime relevant to global signal experiments.  Foreground models must therefore be constructed via interpolations and extrapolations from measurements that are incomplete both spatially and spectrally.  One such effort is the Global Sky Model (GSM) of \cite{GSM}.  In that study, the authors obtained foreground survey data at $11$ different frequencies, and formed a series of foreground maps, stored in the vector $\mathbf{g}$.  The maps were then used to define a spectral covariance matrix $\mathbf{G}$:
\begin{equation}
\mathbf{G}^{\textrm{GSM}}_{\alpha \beta} \equiv \frac{1}{N} \sum_{i=1}^{N} \mathbf{g}_{\alpha i} \mathbf{g}_{\beta i},
\end{equation}
where $N$ is the number of pixels in a spectrally well-sampled region of the sky, and in accordance with our previous notation, $\mathbf{g}_{\alpha i}$ denotes the measured foregrounds in the $i^{th}$ pixel at the $\alpha^{th}$ frequency channel.  From this covariance, a dimensionless frequency correlation matrix was formed:
\begin{equation}
\label{whitenedcovar1}
\widetilde{\mathbf{G}}_{\alpha \beta} \equiv \frac{\mathbf{G}^{\textrm{GSM}}_{\alpha \beta} }{\sqrt{\mathbf{G}^{\textrm{GSM}}_{\alpha \alpha} \mathbf{G}^{\textrm{GSM}}_{\beta \beta}}}.
\end{equation}
By performing an eigenvalue decomposition of $\widetilde{\mathbf{G}}$ into its principal components, the authors found that the spectral features of the foregrounds were dominated by the first three principal components, which could be used as spectral templates for constructing empirically-based foreground models.  The GSM approach was found to be accurate to $\sim10\%$.

Being relatively quick and accurate, the GSM has been used as a fiducial foreground model in many studies of the global $21\,\textrm{cm}$ signal to date \cite{jonathanAvi,DAREMCMC}.  However, this may be insufficient for two reasons.  First, as mentioned above, the GSM approach predicts the magnitude of foregrounds by forming various linear combinations of three principal component (\emph{i.e.} spectral eigenmode) templates.  Thus, if the GSM is considered the ``true" foreground contribution in our models, it becomes formally possible to achieve \emph{perfect} foreground removal simply by projecting out just three spectral modes from the data.  This is too optimistic an assumption\footnote{The reader should thus be cautious when comparing our predictions to forecasts in the literature that use the GSM as their only foreground model.  For identical experimental parameters, one should expect our work to give larger error bars.  However, our new signal extraction algorithms (particularly those that make use of angular information) more than overcome this handicap, and our conclusions will in fact be quite optimistic.}.  The other weakness of the GSM is that it does not include bright point sources, which are expected to be quite numerous at the sensitivities of most $21\,\textrm{cm}$ experiments.

In this paper, we use the GSM as a starting point, but add our own spectral extensions to deal with the aforementioned shortcomings.  The extensions come partly from the phenomenological model of \cite{Paper5}, which in our notation can be summarized by writing down a matrix $\mathbf{G}^{\textrm{ext}}$ that is analogous to $\mathbf{G}^{\textrm{GSM}}$ defined above:
\begin{equation}
\mathbf{G}^{\textrm{ext}} \equiv \mathbf{G}^{\textrm{ps}} + \mathbf{G}^{\textrm{sync}} + \mathbf{G}^{\textrm{ff}},
\end{equation}
where $\mathbf{G}^{\textrm{ps}}$, $\mathbf{G}^{\textrm{sync}}$, and $\mathbf{G}^{\textrm{ff}}$ refer to foreground contributions from unresolved extragalactic point sources, Galactic synchrotron radiation, and Galactic free-free emission, respectively.  Each contribution takes the generic form
\begin{equation}
\mathbf{G}_{\eta \beta} = A^2 \left( \frac{\nu_\eta \nu_\beta}{\nu_*^2} \right)^{-\alpha + \frac{\Delta \alpha^2}{2} \ln \left( \frac{\nu_\eta \nu_\beta}{\nu_*^2} \right)},
\end{equation}
where $A = 335.4\,\textrm{K}$, $\alpha = 2.8$, and $\Delta \alpha = 0.1$ for the synchrotron contribution; $A =70.8 \,\textrm{K}$, $\alpha = 2.5$, and $\Delta \alpha = 0.5$ for the unresolved point source contribution; $A = 33.5\,\textrm{K}$, $\alpha = 2.15$, and $\Delta \alpha = 0.01$ for the free-free contribution, and $\nu_*$ is a reference frequency, which we take to be $150\,\textrm{MHz}$.  Our strategy is to perform a principal component decomposition on this model, and to use its higher order principal components to complement the three components provided by the GSM, completing the basis.

\begin{figure}
\centering
\includegraphics[width=0.45\textwidth]{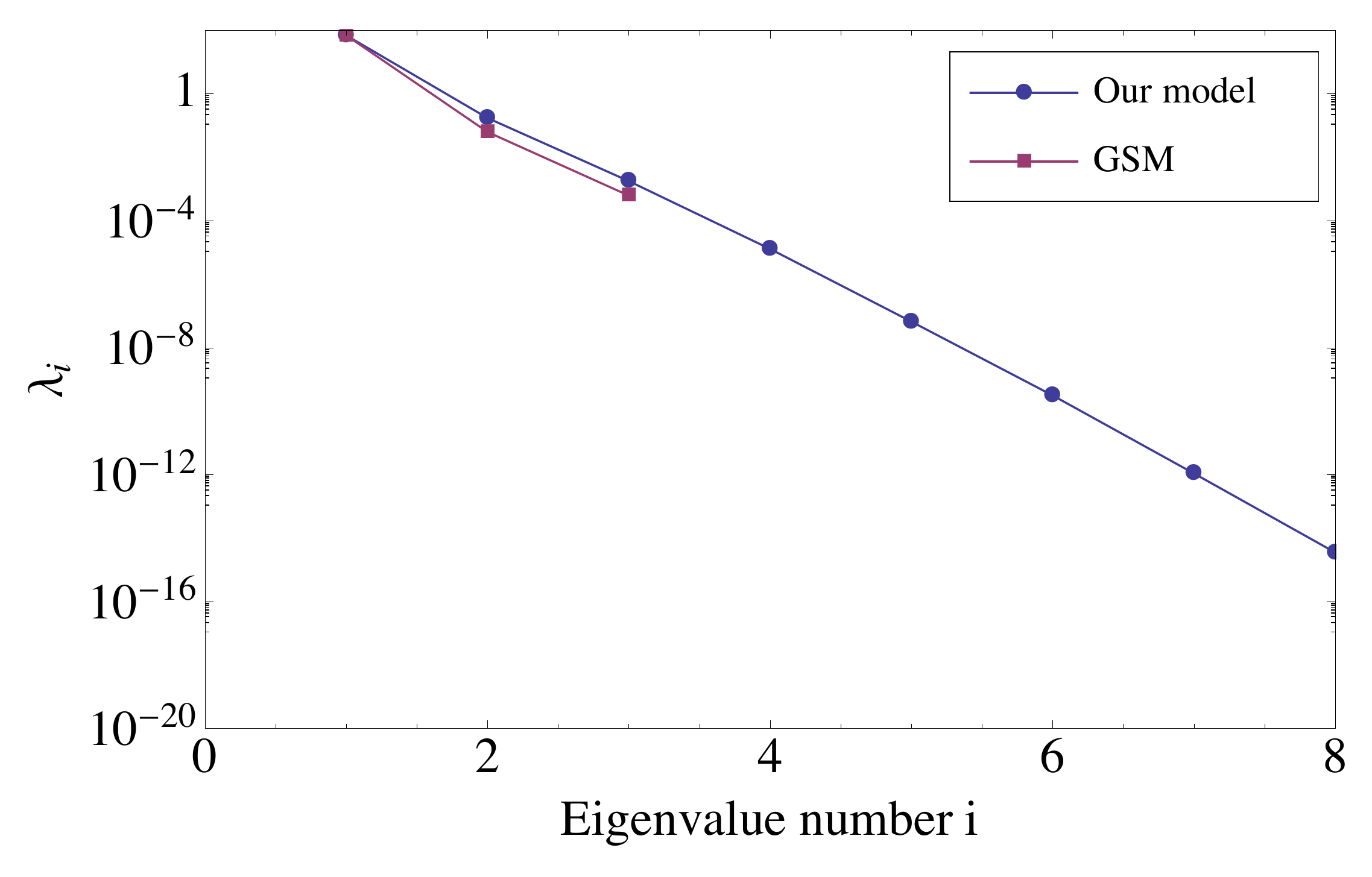}
\caption{Comparison with the global sky model \cite{GSM}.  Our eigenvalues are represented by the blue circles, while those of the GSM are denoted by the purple squares.  The GSM contains only three eigenvalues, which are in reasonable agreement with our model.}
\label{GSMComp}
\end{figure}

We can check that our strategy is a sensible one by forming $\widetilde{\mathbf{G}}$ [Equation \eqref{whitenedcovar1}] for both the GSM and our phenomenological model.  For this test, we form the matrices for a set of observations over 70 frequency channels, each with bandwidth $1\,\textrm{MHz}$ spanning the $30\,\textrm{MHz}$ to $100\,\textrm{MHz}$ frequency range (relevant to observations probing the first luminous sources).  We then compute the eigenvalue spectrum of both models, and the result is shown in Figure \ref{GSMComp}.  The GSM, built from three principal components, contains just three eigenvalues.  The first three eigenvalues of the phenomenological model agree with these values quite well, with the phenomenological model slightly more conservative.  It is thus reasonable to use the higher eigenmodes of the phenomenological model to ``complete" the foreground model spectrally.  We expect this completion to be necessary because there are more than three eigenvalues above the thermal noise limit for a typical experiment \cite{Paper5}.  Spatially, we form an angular model by averaging the GSM maps over all frequencies, and give all of the higher spectral eigenmodes this averaged angular dependence.  While in general we expect the spatial structure to be different from eigenmode to eigenmode, our simple assumption is a conservative one, since any additional variations in spatial structure provide extra information which can be used to aid foreground subtraction.

Next, we add bright point sources to our model.  The brightnesses of these sources are distributed according to the source count function
\begin{equation}
\frac{dn}{dS} = (4\,\textrm{sources mJy}^{-1}\,\textrm{sr}^{-1})\left( \frac{S}{880\,\textrm{mJy}} \right)^{-1.75},
\end{equation}
and we give each source a spectrum of
\begin{equation}
S(\nu) = S_0 \left( \frac{\nu}{150\,\textrm{MHz}}\right)^{-\alpha},
\end{equation}
where the $S_0$ is the brightness drawn from the source count function, and $\alpha$ is the spectral index, drawn from a Gaussian distribution with mean $0.5$ and a standard deviation of $0.25$ \cite{paper1}.  Spatially, we distribute the bright point sources randomly across the sky.  This is strictly speaking an unrealistic assumption, since point sources are known to be clustered.  However, a uniform distribution suffices for our purposes because it is a conservative assumption---any spatial clustering would constitute an additional foreground signature to aid foreground removal.

\begin{figure*}
\centering
\includegraphics[width=0.75\textwidth]{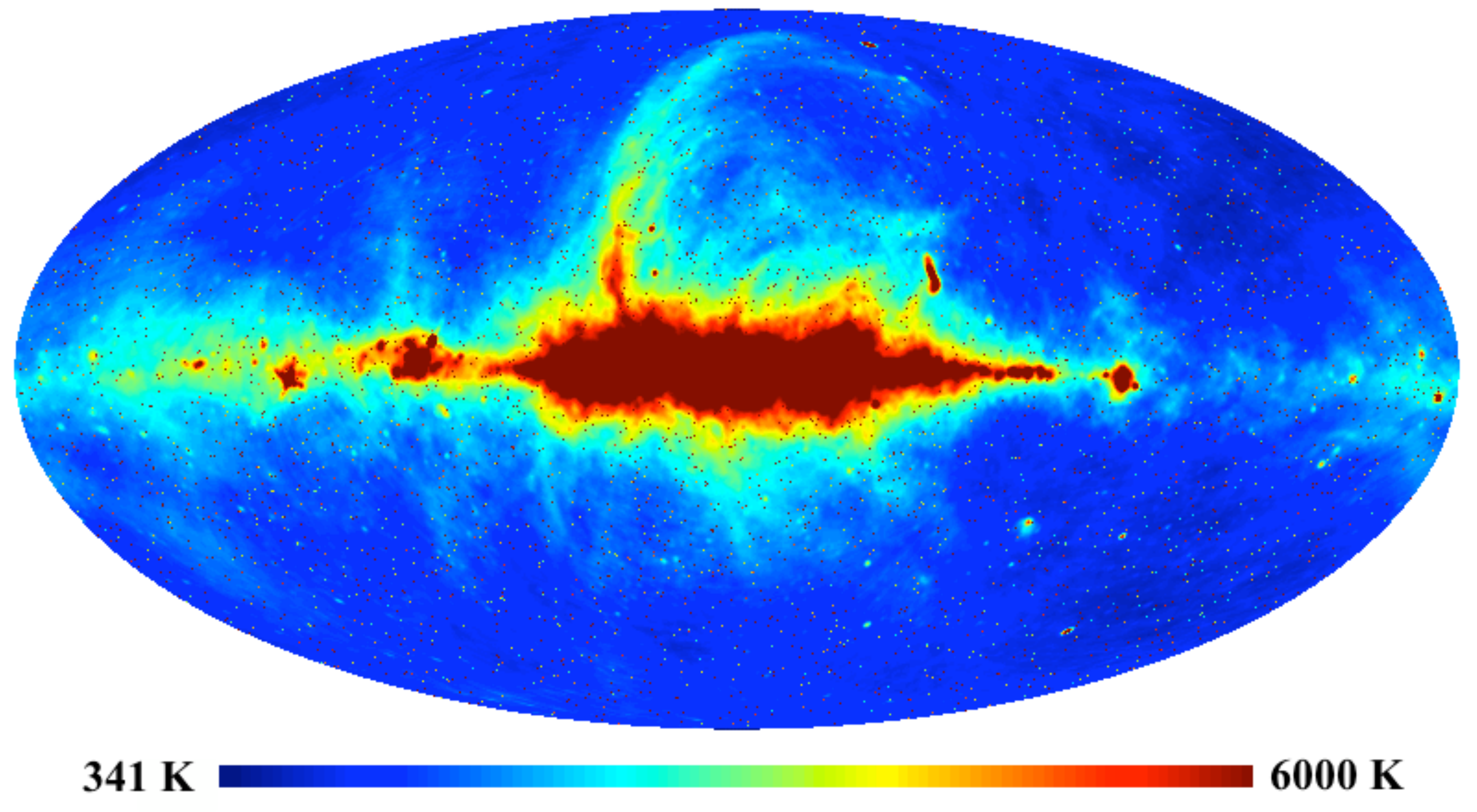}
\caption{Foreground template at $79\,\textrm{MHz}$.  The color scale is linear and has been capped at $6000\,\textrm{K}$ to bring out visual contrast even though much of the sky is far brighter.}
\label{30MHzFGTemplate}
\end{figure*}

Putting all these ingredients together, the result is a series of foreground maps like that shown in Figure \ref{30MHzFGTemplate}, one at every frequency.  This model constitutes a set of ``best guess" foreground templates.  Now, while our model is based on empirical data, it does possess uncertainties.  A robust method for foreground subtraction thus needs to be able to account for possible errors in the generalized noise model.  To set the stage for this, we define our best guess model described above to be the mean (amidst uncertainties in the construction of the model) of the $\mathbf{n}^{\textrm{fg}}$ vector in Equation \eqref{noisedecomp}, \emph{i.e.}
\begin{equation}
\label{mean}
 \mathbf{m}^{\textrm{fg}}_{\alpha i} \equiv \langle \mathbf{n}^{\textrm{fg}}_{\alpha i} \rangle ,
\end{equation}
where $\mathbf{m}^{\textrm{fg}}$ is our best guess model.  The covariance of the foregrounds is defined as the error in our foreground model:
\begin{equation}
\label{Nfg}
\mathbf{N}^{\textrm{fg}}_{\alpha i \beta j} \equiv \langle \mathbf{n}^{\textrm{fg}}_{\alpha i} \mathbf{n}^{\textrm{fg}}_{\beta j} \rangle -\langle \mathbf{n}^{\textrm{fg}}_{\alpha i} \rangle \langle \mathbf{n}^{\textrm{fg}}_{\beta i} \rangle= \varepsilon^2 \mathbf{m}^{\textrm{fg}}_{\alpha i} \mathbf{m}^{\textrm{fg}}_{\beta j} \mathbf{R}_{ij} \mathbf{Q}_{\alpha \beta},
\end{equation}
where we have assumed that the error in our foreground model is proportional to the model $\mathbf{m}^{\textrm{fg}}$ itself, with a proportionality constant $\varepsilon$ between $0$ and $1$.  The matrices $\mathbf{R}$ and $\mathbf{Q}$ encode the spatial and spectral correlations in our foreground model errors, respectively.

With this form, we are essentially assuming that there is some constant percentage error in every pixel and frequency of our model.  Note that the model error $\varepsilon$ will in general depend on the angular resolution, for if we start with a high resolution foreground model and spatially downsample it to the resolution of our experiment, some of the errors in the original model will cancel out.  As a simple, crude model, we can suppose that the errors in different pixels average down as the square root of the number of pixels.  Since the number of pixels scales as the inverse of $\theta^2$, where $\theta$ is some generic angular resolution, we have
\begin{equation}
\label{crudeScaling}
\varepsilon = \varepsilon_0 \frac{\theta_\textrm{fg}}{\theta_b},
\end{equation}
where $\theta_b$ is the angular resolution of our current experiment, $\theta_{\textrm{fg}}$ is the ``native" angular resolution of our foreground model, and $\varepsilon_0$ is the fractional error in our model at this resolution.  Equation \eqref{crudeScaling} implicitly assumes a separation of scales, where we deal only with instruments that have coarser angular resolution that the correlation length of modeling errors specified by $\mathbf{R}$ (described below).  We will remain in this regime for the rest of the paper.  Our foreground covariance thus becomes
\begin{equation}
\mathbf{N}^{\textrm{fg}}_{\alpha i \beta j} = \varepsilon_0^2 \left(\frac{\theta_\textrm{fg}}{\theta_b}\right)^2 \mathbf{m}^{\textrm{fg}}_{\alpha i} \mathbf{m}^{\textrm{fg}}_{\beta j} \mathbf{R}_{ij} \mathbf{Q}_{\alpha \beta}.
\end{equation}
Since the angular structure of our foreground model is based on that of the GSM, which conservatively has a $10\%$ accuracy at $5^\circ$ \cite{GSM}, we use $\varepsilon_0=0.1$ and $\theta_{\textrm{fg}} = 5^\circ$ as fiducial values for this paper.

To capture spatial correlations\footnote{We emphasize that the spatial correlations encoded by $\mathbf{R}$ are spatial correlations in the foreground \emph{error}, not correlations in the spatial structure of the foreground emission itself.  The spatial structure of the foregrounds are captured by the $\mathbf{m}^{\textrm{fg}}$ terms of Equations \ref{mean} and \ref{Nfg}, and will typically be correlated over much larger scales than the errors.  The data analysis formalism that we present in Section \ref{measurementformalism} will take into account both types of correlation.} in our foreground modeling errors, we choose the matrix $\mathbf{R}$ to correspond to the continuous kernel
\begin{equation}
R (\mathbf{\widehat{r}},\mathbf{\widehat{r}^\prime}) \equiv \frac{4\pi}{N_{\textrm{pix}}} \frac{\exp(\sigma^{-2} \mathbf{\widehat{r}}\cdot \mathbf{\widehat{r}^\prime})}{4 \pi \sigma^2 \sinh(\sigma^{-2})}.
\end{equation}
Aside from the constant $4 \pi/N_{\textrm{pix}}$ factor (which is needed to make the discrete and continuous descriptions consistent \cite{maxMapping,knox}), this is known as a Fisher function, the analog of a Gaussian on a sphere.  The quantity $\sigma$ measures the spread of the kernel, and since the GSM's spectral fits were performed on pixels of roughly $5^\circ$ resolution, we set $\sigma = 5^\circ$ for this work\footnote{Note that $\sigma$ does not necessarily have to be equal to $\theta_{\textrm{fg}}$.  For instance, if one's foreground model is based on a survey with some instrumental beam size that is oversampled in an attempt to capture all the features in the map, one would be in a situation where $\theta_{\textrm{fg}} < \sigma$.}.

For the spectral correlation matrix $\mathbf{Q}$, suppose we imagine that our foreground model was constructed by spectrally fitting every pixel of a foreground survey to a power law of the form
\begin{equation}
t(\nu) = A (\nu / \nu_*)^{-\alpha}, 
\end{equation}
where $A$ is a normalization constant that will later cancel out, $\alpha$ is a spectral index, and $\nu_*$ is a reference frequency for the fits, which we take to be $150\,\textrm{MHz}$ for the $30$ to $100\,\textrm{MHz}$ observations targeting the first luminous sources, and $50\,\textrm{MHz}$ for the $100$ to $250\,\textrm{MHz}$ observations targeting reionization.  (The reference frequency is somewhat arbitrary, and in practice one would simply adjust it to get the best possible fits when constructing one's foreground model).  The spectral index will have some error associated with it, due in part to uncertainties in the foreground survey and in part to the fact that foreground spectra are not perfect power laws.  We model this error as being Gaussian distributed, so that the probability distribution of spectral indices is given by
\begin{equation}
p(\alpha) = \frac{1}{\sqrt{2\pi \sigma_\alpha^2}} \exp\left[ -\frac{1}{2} \frac{(\alpha-\alpha_0)^2}{\sigma_\alpha^2} \right],
\end{equation}
where $\alpha_0$ is a fiducial spectral index for typical foregrounds, which in this paper we take to be $2.5$.  The parameter $\sigma_\alpha$ controls the spectral coherence of the foregrounds, and we choose the rather large value of $\sigma_\alpha =1$ to be conservative.  

With this, the mean spectral fit to our foreground survey is
\begin{equation}
\langle t (\nu) \rangle = A \int \left( \frac{\nu}{\nu_*} \right)^{-\alpha} p(\alpha) d\alpha.
\end{equation}
Sampling this function at a discrete set of frequencies corresponding to the frequency channels of our global signal experiment, we can form a mean vector $\langle \mathbf{t} \rangle$.  A covariance matrix $\mathbf{C}^{\textrm{survey}}$ of the power law fits is then given by
\begin{equation}
\mathbf{C}^{\textrm{survey}} \equiv \langle \mathbf{t} \mathbf{t}^{t} \rangle - \langle \mathbf{t} \rangle \langle \mathbf{t} \rangle^t,
\end{equation}
where $\mathbf{t}$ is a discretized version of $t(\nu)$, and
\begin{equation}
\langle \mathbf{t} \mathbf{t}^{t} \rangle_{\beta \eta} = A^2 \int \left( \frac{\nu_\beta \nu_\eta}{\nu_*^2} \right)^{-\alpha} p(\alpha) d\alpha.
\end{equation}
Finally, we take the covariance $\mathbf{C}^{\textrm{survey}}$ and insert it into the left hand side of Equation \eqref{whitenedcovar1} (just as we did with $\mathbf{G}_{\textrm{GSM}}$ earlier) to form our spectral correlation matrix $\mathbf{Q}$.  Note that the normalization constant $A$ cancels out in the process, as we claimed.

\begin{figure}
\centering
\includegraphics[width=0.45\textwidth]{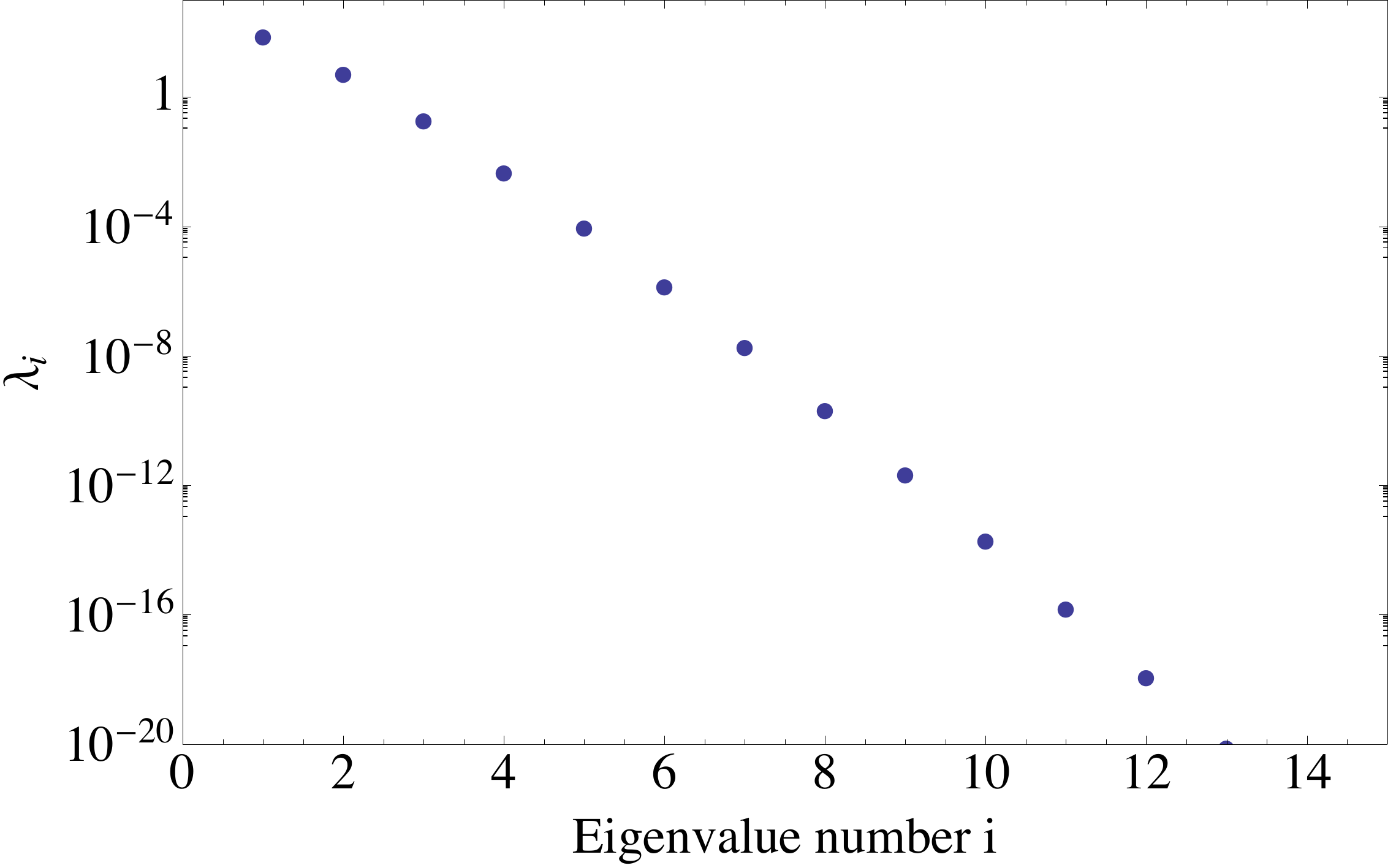}
\caption{First few eigenvalues of the spectral correlation matrix $\mathbf{Q}$.  The eigenvalues decay roughly exponentially, which means that the foreground correlations are dominated by the first few eigenmodes.}
\label{Qeigenvals}
\end{figure}

\begin{figure}
\centering
\includegraphics[width=0.45\textwidth]{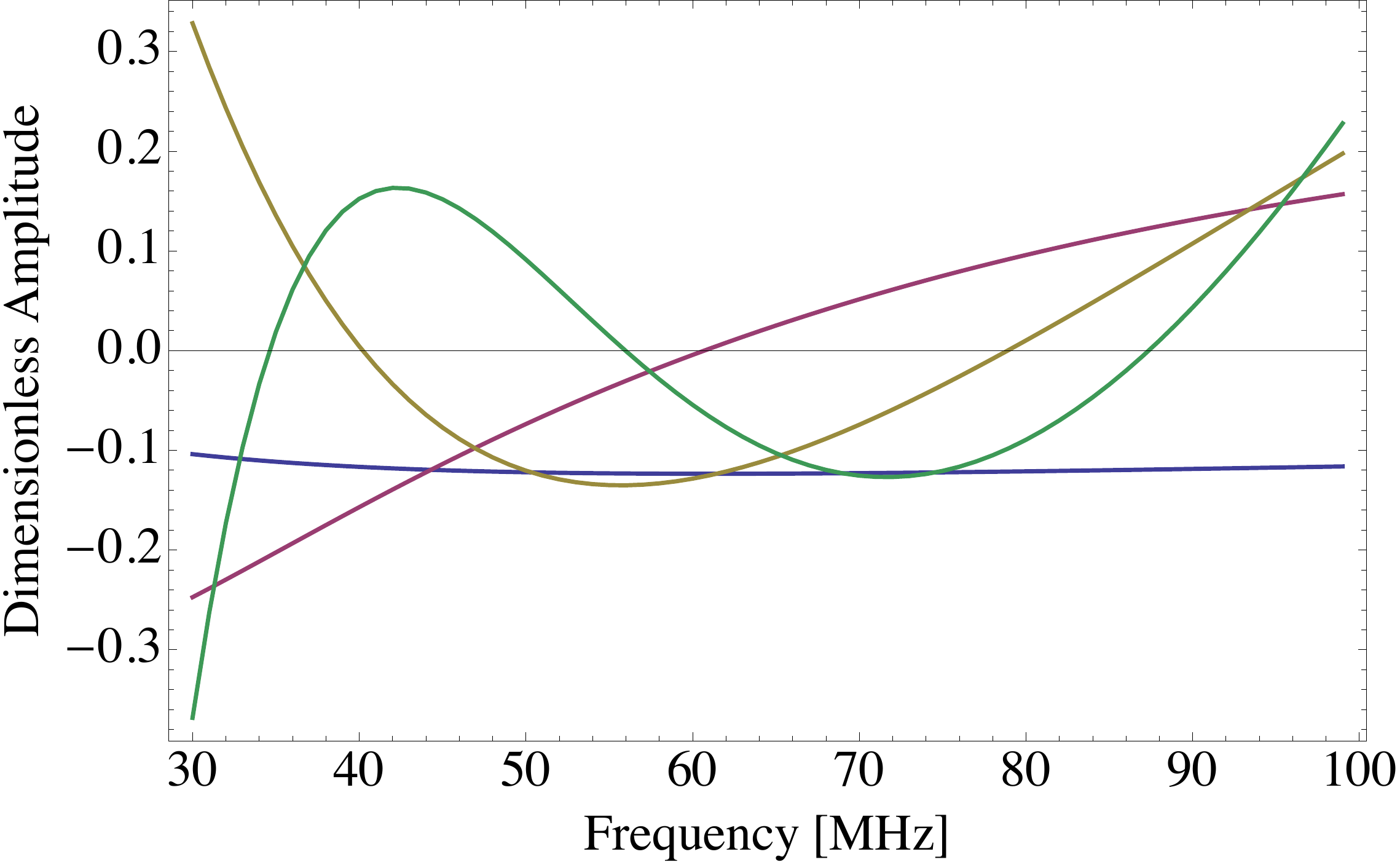}
\caption{First few eigenvectors of the spectral correlation matrix $\mathbf{Q}$.  These spectral eigenmodes are seen to be quite smooth, and have increasing structure as one goes to higher modes.  In conjunction with Figure \ref{Qeigenvals}, this shows that the frequency correlations in the foreground are dominated by smooth spectral eigenmodes.}
\label{Qeigenvects}
\end{figure}

In Figure \ref{Qeigenvals}, we show the eigenvalues of $\mathbf{Q}$, and in Figure \ref{Qeigenvects}, the first few eigenmodes.  Just as with the foregrounds themselves, the roughly exponential decay of the eigenvalues show that the foreground correlations are dominated by the first few eigenmodes, which are smooth functions of frequency.

\subsubsection{Instrumental Noise}
\label{instNoiseDescription}
We model the instrumental noise $\mathbf{n}^{\textrm{inst}}_{i\alpha}$ in every pixel and every frequency channel as uncorrelated.  Additionally, we make the assumption that there are no systematic instrumental effects, so that $\langle \mathbf{n}^{\textrm{inst}}_{i \alpha} \rangle = 0$.  What remains is the random contribution to the noise.  Assuming a sky-noise dominated instrument, the amplitude of this contribution is given by the radiometer equation.  In our notation, this gives rise to a covariance of the form\footnote{In principle, the cosmological signal also contributes to the sky-noise, and so there is in fact some cosmological information in the noise.  For the noise term only, we make the approximation that the cosmological signal contributes negligibly to the sky temperature, since the foregrounds are so much brighter.  For any reasonable noise level, this should be an excellent assumption, and in any case a conservative one, as extra sources of cosmological information will only serve to tighten observational constraints.}
\begin{equation}
\mathbf{N}^{\textrm{inst}}_{\alpha i \beta j} = \langle \mathbf{n}^{\textrm{inst}}_{\alpha i} \mathbf{n}^{\textrm{inst}}_{\beta j} \rangle = \frac{ \mathbf{m}^{\textrm{fg}}_{\alpha i} \mathbf{m}^{\textrm{fg}}_{\beta j} \delta_{ij} \delta_{\alpha \beta}}{\Delta t \Delta \nu},
\end{equation}
where $\Delta t$ is the integration time per pixel and $\Delta \nu$ is the channel width.

The instrumental noise is different from all other signal sources in an experiment (including the cosmological signal and all other forms of generalized noise) in that it is the only contribution to the measurement that is added \emph{after} the sky has been convolved by the beam of an instrument.  All the ``other" contributions should be convolved with the beam.  For instance, if an instrument has a Gaussian beam with a standard deviation $\theta_b$, then these contributions to the covariance are multiplied by $\exp[ -\theta_b^2 \ell (\ell+1)]$ in spherical harmonic space \cite{knox,TE96}.  However, since the instrumental noise is the only part of our measurement that does not receive this correction, it is often convenient to adopt the convention that all measured maps of the sky are already deconvolved prior to the main data analysis steps.  This allows us to retain all of the expressions for the cosmological signal and the foregrounds that we derived above, and to instead modify the instrumental noise to reflect the deconvolution.  In this paper, we will adopt the assumption of already-deconvolved maps.

If the instrumental noise contribution were rotation-invariant on the sky (as is the case for many Cosmic Microwave Background experiments), modifying the noise to reflect deconvolution would be simple.  One would simply multiply the instrumental noise covariance by $\exp[ \theta_b^2 \ell (\ell+1)]$ in spherical harmonic space \cite{knox,TE96}.  Unfortunately, since we are assuming a sky-noise dominated instrument, the assumption of rotation-invariance breaks down, thanks to structures such as the Galactic plane, which clearly contributes to the sky signal in an angularly-dependent way.  The simple prescription of multiplying by $\exp[ \theta_b^2 \ell (\ell+1)]$ thus becomes inadequate, and a better method is required.

Suppose we work in a continuous limit and let $b(\mathbf{\widehat{r}}, \mathbf{\widehat{r}^\prime})$ be an integration kernel that represents a Gaussian instrumental beam, so that
\begin{eqnarray}
T^{\textrm{conv}} (\mathbf{\widehat{r}}) &=& \int b(\mathbf{\widehat{r}}, \mathbf{\widehat{r}^\prime}) T (\mathbf{\widehat{r}}^\prime) d\Omega^\prime \nonumber \\
&=& \sum_{\ell,m} Y_{\ell m} (\mathbf{\widehat{r}}) e^{-\theta_b^2 \ell (\ell+1)/2} \int Y^*_{\ell m} (\mathbf{\widehat{r}}^\prime)T (\mathbf{\widehat{r}}^\prime) \,d\Omega^\prime,\qquad
\end{eqnarray}
where $T^{\textrm{conv}} (\mathbf{\widehat{r}})$ is convolved version of the original sky $T (\mathbf{\widehat{r}}) $ and $Y_{\ell m}$ are the spherical harmonics.  With this, our effective (\emph{i.e.} deconvolved) instrumental noise covariance is given by
\begin{eqnarray}
N^{\textrm{eff}} (\mathbf{\widehat{r}}, \mathbf{\widehat{r}^\prime}) &\propto &\int b^{-1}(\mathbf{\widehat{r}}, \mathbf{\widehat{r}}_1) m^{\textrm{fg}}(\mathbf{\widehat{r}}_1) \delta(\mathbf{\widehat{r}}_1, \mathbf{\widehat{r}}_2) \times\nonumber \\
&& \quad m^{\textrm{fg}} (\mathbf{\widehat{r}}_2) b^{-1}(\mathbf{\widehat{r}}_2, \mathbf{\widehat{r}^\prime}) d\Omega_1 d\Omega_2,
\end{eqnarray}
where $m^{\textrm{fg}}$ is the continuous version of $\mathbf{m}^{\textrm{fg}}$, and $b^{-1}$ the deconvolution kernel (given by the operator inverse of $b$), which in spherical harmonic space simply multiplies by $e^{\theta_b^2 \ell (\ell+1)/2}$.  For notational simplicity, we have omitted constants and suppressed the frequency dependence (since it has nothing to do with the spatial deconvolution).  Now, suppose we make the approximation that the foregrounds are spatially smooth and slowly varying.  In real space, $b^{-1}(\mathbf{\widehat{r}}, \mathbf{\widehat{r}^\prime})$ is a rapidly oscillating function that is peaked around $\mathbf{\widehat{r}} = \mathbf{\widehat{r}}^\prime$.  We may thus move the two copies of $m^{\textrm{fg}}$ outside the integral, setting $\mathbf{\widehat{r}}=\mathbf{\widehat{r}}_1$ and $\mathbf{\widehat{r}}=\mathbf{\widehat{r}}_2$:
\begin{eqnarray}
\label{neff}
N^{\textrm{eff}} (\mathbf{\widehat{r}}, \mathbf{\widehat{r}^\prime}) &\propto& m^{\textrm{fg}}(\mathbf{\widehat{r}})  m^{\textrm{fg}}(\mathbf{\widehat{r}}^\prime)  \int b^{-1}(\mathbf{\widehat{r}}, \mathbf{\widehat{r}}_1) b^{-1}(\mathbf{\widehat{r}}_1, \mathbf{\widehat{r}^\prime}) d\Omega_1 \nonumber \\
&=& m^{\textrm{fg}}(\mathbf{\widehat{r}})  m^{\textrm{fg}}(\mathbf{\widehat{r}}^\prime) \sum_{\ell m} e^{\theta_b^2 \ell (\ell+1)} Y^*_{\ell m}§§ (\mathbf{\widehat{r}}) Y_{\ell m} (\mathbf{\widehat{r}}^\prime).\qquad
\end{eqnarray}
If the $m^{\textrm{fg}}(\mathbf{\widehat{r}})  m^{\textrm{fg}}(\mathbf{\widehat{r}}^\prime)$ terms were not present, this would be equivalent to the ``usual" prescription, where the effective noise kernel involves transforming to spherical harmonic space, multiplying by $e^{\theta_b^2 \ell (\ell+1)/2}$, and transforming back.  Here, the prescription is similar, except the kernel is modulated in amplitude by the sky signal.

In conclusion, we can take instrumental beams into account simply by replacing $\mathbf{N}^{\textrm{inst}}$ with an effective instrumental noise covariance of the form
\begin{equation}
\mathbf{N}^{\textrm{inst,eff}} = \mathbf{D} \mathbf{N}^{\textrm{dec}} \mathbf{D},
\end{equation}
where $\mathbf{D}_{\alpha i \beta j} \equiv \mathbf{m}_{\alpha i } \delta_{\alpha \beta} \delta_{ij}$, and $\mathbf{N}^{\textrm{dec}}$ is a deconvolved white noise covariance (\emph{i.e.} one that is multiplied by $e^{\theta_b^2 \ell (\ell+1)}$ in spherical harmonic space) with proportionality constant $1/ \Delta t \Delta \nu$.

In deriving our deconvolved noise covariance, we made the assumption that the sky emission is spatially smooth and slowly-varying.  This allowed us to treat the deconvolution analytically even in the case of non-white noise, and in Section \ref{both} will allow us to analytically derive an optimal estimator for the cosmological signal.  While our assumption of smooth emission is of course only an approximation, we expect it to be a rather good one for our purposes.  The only component of the sky emission where smoothness may be a bad assumption is the collection of bright point sources.  However, we will see in Section \ref{both} that the optimal estimator will heavily downweight brightest regions of the sky, so extremely bright point sources are effectively excluded from the analysis anyway.

\subsubsection{Cosmological Anisotropy Noise}
\label{cosmonoise}
In measuring the global signal, we are measuring the monopole contribution to the sky.  As mentioned above, any anisotropic contribution to the cosmological power is therefore a noise contribution as far as a global signal experiment is concerned.  By construction, these non-monopole contributions have a zero mean after spatially averaging over the sky, and thus do not result in a systematic bias to a measurement of the global signal.  They do, however, have a non-zero variance, and therefore contribute to the error bars.

Although it is strictly speaking non-zero, we can safely ignore cosmological anisotropy noise because it is negligibly small compared to the foreground noise.  Through a combination of analytic theory \cite{LewisChallinor} and simulation analysis \cite{BittnerLoeb}, the cosmological anisotropies have been shown to be negligible on scales larger than $\sim1^\circ$ to $2^\circ$, which is a regime that we remain in for this paper.
\subsubsection{Generalized noise model summary}
\label{genNoiseModel}
In the subsections above, we have outlined the various contributions to the generalized noise that plagues any measurement of the global signal.  Of these contributions, only foregrounds have a non-zero mean, so the mean of our generalized noise model is just that of the foregrounds:
\begin{equation}
\label{noisemeandef}
 \mathbf{m}_{\alpha i} \equiv \langle \mathbf{n} \rangle = \mathbf{m}^{\textrm{fg}}_{\alpha i}.
\end{equation}
Foregrounds therefore have a special status amongst the different components of our generalized model, for they are the only contribution with the potential to cause a systematic bias in our global signal measurement.  The other contributions appear only in the total noise covariance, taken to be the sum of the foreground covariance and the effective instrumental noise covariance:
\begin{equation}
\label{Ntotal}
\mathbf{N}_{\alpha i \beta j} \equiv \mathbf{N}^{\textrm{fg}}_{\alpha i \beta j} + \mathbf{N}^{\textrm{inst,eff}}_{\alpha i \beta j},
\end{equation}
where as noted above, we are neglecting the cosmological anisotropy noise.

In the foreground subtraction/data analysis scheme that we describe in Section \ref{measurementformalism}, we will think of the mean $\mathbf{m}$ as a foreground template that is used to perform a first subtraction.  However, there will inevitably be errors in our templates, and thus our scheme also takes into account the covariance $\mathbf{N}$ of our model.  In our formalism, the mean term therefore represents our best guess as to what the foreground contamination is, and the covariance quantifies the uncertainty in our guess.  We note that this is quite different from many previous approaches in the literature, where either the foreground modeling error is ignored (\emph{e.g.} when the foreground spectra are assumed to be perfect polynomials), or the mean is taken to be zero and the covariance is formed  by taking the ensemble average of the outer product of the foreground template error.  The former approach is clearly unrealistic, while the latter approach has a number of shortcomings.  For example, it is difficult to compute the necessary ensemble average, since foregrounds are difficult to model from first principles, and empirically the only sample that we have for taking this average is our Galaxy.  As a solution to this, ensemble averages are often replaced by spatial (\emph{i.e.} angular) averages.  But this is unsatisfactory for our purposes, since in Section \ref{both} we will be using the angular structure of foregrounds to aid with foreground subtraction, and this is impossible if the information has already been averaged out.  Even if an ensemble average could somehow be taken (perhaps by running a large suite of radiative transfer foreground simulations), a foreground subtraction scheme that involved minimizing the resulting variance would be non-optimal for two reasons.  First, in such a scheme one would be guarding against foreground power from a ``typical" galaxy, which is irrelevant---all that matters to an experiment are the foregrounds that are seen in our Galaxy, even if they are atypical.  In addition, foregrounds are not Gaussian distributed, and thus a minimization of the variance is not necessarily optimal.

Our approach---taking the mean to be an empirical foreground template and the covariance to be the errors in this template---solves these problems.  Since the covariance arises from measurement errors (which can usually be modeled to an adequate accuracy), taking the ensemble average is no longer a problem.  And with the mean term being a template for the foregrounds as seen by our experiment, our foreground model is tailored to our Galaxy, even if our Galaxy happens to be atypical.  Finally, while the foregrounds themselves are certainly not Gaussian, it is a much better approximation to say that the uncertainties in our model are Gaussian, at least if the uncertainties are relatively small.  Constructing our foreground model in this way thus allows us to take advantage of the optimal data analysis techniques that we introduce in Section \ref{measurementformalism}.
\subsection{Why it's hard}
\begin{figure}
\centering
\includegraphics[width=0.45\textwidth]{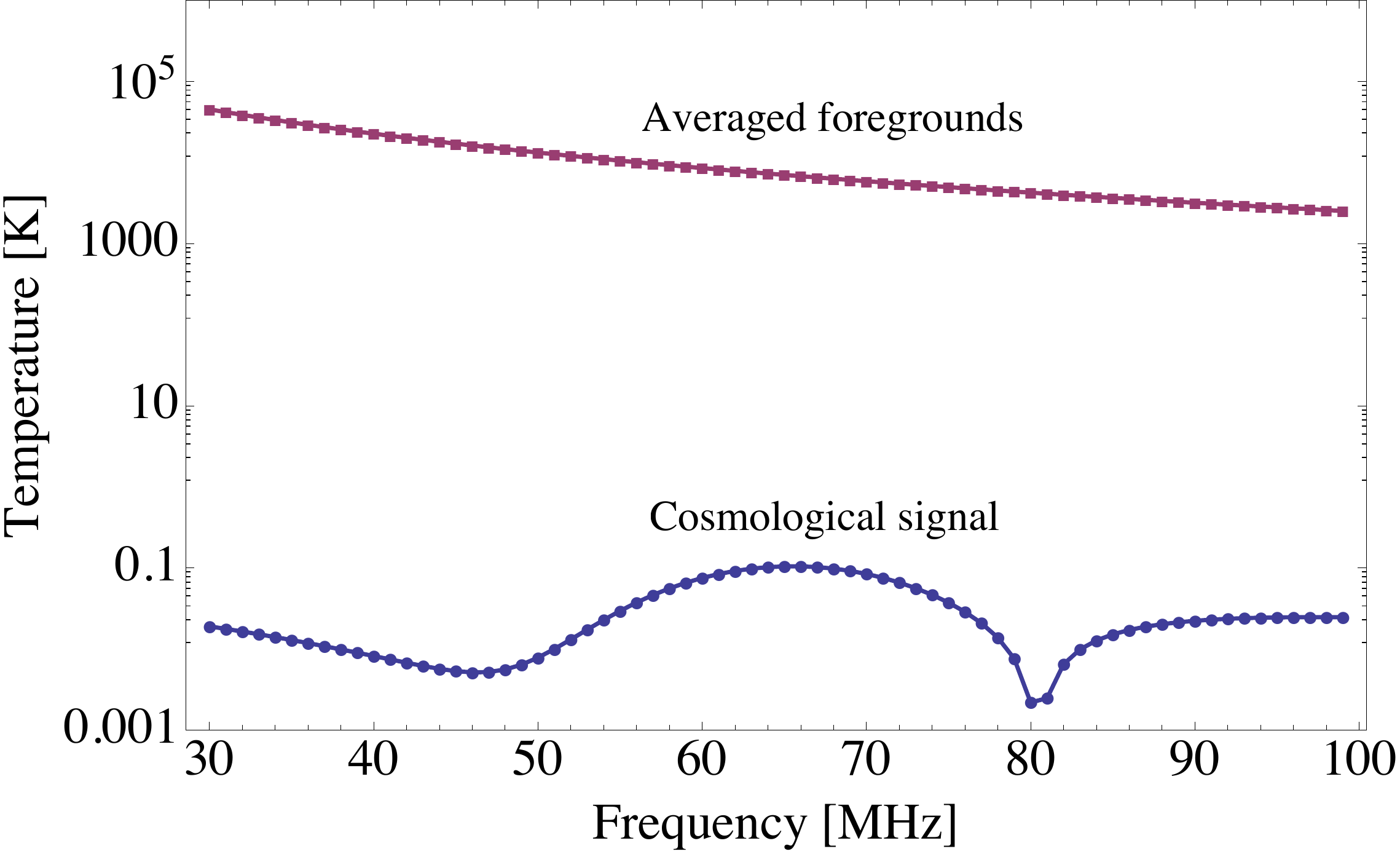}
\caption{A comparison of the absolute value of the expected theoretical signal spectrum (blue circles) and the mean foreground contamination (purple squares).  This highlights the difficulty of global $21\,\textrm{cm}$ signal experiments: the foregrounds are many orders of magnitude brighter than the signal we seek to measure.}
\label{GlobalFGSpec}
\end{figure}

Before we proceed to describe how the global $21\,\textrm{cm}$ signal can be optimally extracted, we pause to  describe the challenges ahead\footnote{Not included in this paper is the fact that an instrument might have a non-trivial frequency response that needs to be calibrated extremely well.   In principle, if one has 1) sufficiently good instrumental calibration and 2) an exquisitely accurate foreground model, then it will always be able to pick out a small cosmological signal from beneath the foreground sources, however bright they might be.  In this paper we concentrate on lessening the second requirement by proposing reliable foreground subtraction algorithms.  Tackling the  problem of calibration is beyond the scope of this paper, but encouraging progress has recently been made in engineering tests \cite{rogersCalib}.}.  As an initial ``straightforward" approach, one can imagine measuring the global $21\,\textrm{cm}$ signal by taking a simple spatial average of a measured sky.  The corresponding foreground contamination would be obtained by spatially averaging our model, which is shown in Figure \ref{GlobalFGSpec} along with our expected theoretical signal from Figure \ref{fig:target_signal}.  A straightforward measurement would thus be completely dominated by the bright foregrounds.  In addition, both the foreground contamination and the theoretical signal are smooth as a function of frequency, making it difficult to use foreground subtraction techniques that have been proposed for tomographic maps, where the cosmological signal is assumed to vary much more rapidly as a function of frequency than the foregrounds.  It is therefore crucial that optimal foreground cleaning methods are employed in the analysis, and in the following section we derive such methods.

\section{A framework for global signal measurements}
\label{measurementformalism}
In this section, we develop the mathematical framework for analyzing data from global signal experiments.  We begin with a measurement equation.  For an experiment with both spectral and angular sensitivity, we have
\begin{equation}
\label{fullMeas}
\mathbf{y} = \mathbf{A} \mathbf{x}_s + \mathbf{n},
\end{equation}
where $\mathbf{y}$ is a vector of length $N_{\textrm{vox}} \equiv N_{\textrm{pix}} N_{\textrm{freq}}$ containing the measurement, $\mathbf{n}$ is the generalized noise contribution of Section \ref{gennoise}, $\mathbf{x}_s$ is a vector of length $N_{\textrm{freq}}$ containing the global signal that we wish to measure, and $\mathbf{A}$ is a vertical stack of $N_{\textrm{pix}}$ identity matrices of size $N_{\textrm{freq}} \times N_{\textrm{freq}}$.  The effect of multiplying the global signal by $\mathbf{A}$ is to copy the theoretical spectrum to every pixel on the sky before the noise and foregrounds are added.  The term $\mathbf{A x}_s$ therefore represents a data ball\footnote{Or perhaps a ``data shell", since there is a lower limit to the redshift of the experiment.} that contains ideal (noiseless and foregroundless) data that depends only on the radial distance from the center, and not on the direction.  To every voxel of this data volume the combined noise and foreground contribution $\mathbf{n}$ is added, giving the set of measured voxel temperatures that comprise $\mathbf{y}$.
Note that by thinking of the measurement vector as a set of voxel temperatures, we have implicitly assumed that a prior mapmaking step has been performed on the raw time-ordered data.  This step should ideally take into account instrumental complications such as instrumental beam sidelobes.

For measurements with no angular sensitivity, we can define $\mathbf{z} \equiv \frac{1}{N_{\textrm{pix}}}  \mathbf{A}^t \mathbf{y}$ to be the globally averaged measurement.  In this notation, our measurement equation becomes
\begin{equation}
\label{reducedmeas}
\mathbf{z} = \mathbf{x}_s + \frac{1}{N_{\textrm{pix}}} \mathbf{A}^t \mathbf{n} = \mathbf{x}_s + \mathbf{c},
\end{equation}
where $\mathbf{c} \equiv \frac{1}{N_{\textrm{pix}}}  \mathbf{A}^t \mathbf{n}$ is the angularly averaged noise and foreground contribution, with mean
\begin{equation}
\langle \mathbf{c} \rangle =\frac{1}{N_{\textrm{pix}}}  \mathbf{A}^t \langle \mathbf{n} \rangle = \frac{1}{N_{\textrm{pix}}}  \mathbf{A}^t \mathbf{m},
\end{equation}
where in the last step we used the definition in Equation \eqref{noisemeandef}, and covariance
\begin{equation}
\label{averagedCovar}
\mathbf{C} \equiv \langle \mathbf{c} \mathbf{c}^t \rangle - \langle \mathbf{c} \rangle \langle \mathbf{c}^t \rangle = \frac{1}{N_{\textrm{pix}}^2}  \mathbf{A}^t \mathbf{N} \mathbf{A}.
\end{equation}

Our goal is to derive a statistically optimal estimator $\mathbf{\widehat{x}}_s$ for the true global signal $\mathbf{x}_s$.  With an eye towards optimizing experimental design, we will construct optimal estimators for both measurement equations, treating the spectral-only measurements in Section \ref{spectralOnly} and the spectral-plus-angular measurements in Section \ref{both}.  We will prove that beyond simply subtracting a best-guess model for the foregrounds, there is formally nothing that can be done to mitigate foreground residuals by using spectral information only.  In contrast, adding angular foreground information partially immunizes one from errors in the best-guess model, and allows the final error bars to be reduced.

\subsection{Methods using only spectral information}
\label{spectralOnly}

For methods that use only spectral information, we write down an arbitrary linear estimator of the form
\begin{equation}
\widehat{\mathbf{x}}_s =  \mathbf{M} \mathbf{z} - \mathbf{d},
\end{equation}
where $\mathbf{M}$ is an $N_{\textrm{freq}} \times N_{\textrm{freq}}$ matrix and $\mathbf{d}$ is a vector of length $N_{\textrm{freq}}$, whose forms we will derive by minimizing the variance of the estimator.  

Taking the ensemble average of our estimator and inserting Equation \eqref{reducedmeas} gives
\begin{equation}
\label{intermediate1}
\langle \mathbf{\widehat{x}}_s \rangle = \mathbf{M} \mathbf{x}_s +  \mathbf{M} \langle \mathbf{c}\rangle  - \mathbf{d},
\end{equation}
which shows that in order for our estimator to avoid having a systematic additive bias, one should select
\begin{equation}
\label{intermediate2}
\mathbf{d} \equiv  \mathbf{M} \langle \mathbf{c}\rangle .
\end{equation}
With this choice, we have $\langle \mathbf{\widehat{x}}_s \rangle = \mathbf{M} \mathbf{x}_s$.  The variance of this estimator can be similarly computed, yielding
\begin{equation}
\label{anothercovar}
\mathbf{\Sigma} = \langle \mathbf{\widehat{x}}_s \mathbf{\widehat{x}}_s^t \rangle - \langle \mathbf{\widehat{x}}_s \rangle \langle \mathbf{\widehat{x}}_s^t  \rangle= \mathbf{M} \mathbf{C} \mathbf{M}^t.
\end{equation}
We can minimize this variance subject to the normalization constraint $\mathbf{M}_{\alpha \alpha} = 1$ by using Lagrange multipliers.  To do so we minimize
\begin{equation}
( \mathbf{M} \mathbf{C} \mathbf{M}^t)_{\alpha \alpha} - \lambda_\alpha \mathbf{M}_{\alpha \alpha}
\end{equation}
with respect to the elements of $\mathbf{M}$.  Taking the necessary derivatives and solving for the Lagrange multiplier $\lambda$ that satisfies the normalization constraint, one obtains
\begin{equation}
\mathbf{M}_{\alpha \beta} = \frac{(\mathbf{C}^{-1})_{\alpha \beta}}{(\mathbf{C}^{-1})_{\alpha \alpha}}.
\end{equation}
Inserting this into our general form for the estimator, we find
\begin{equation}
\label{spectralOnlyMinVar}
\mathbf{\widehat{x}}_s^\alpha =  \frac{1}{(\mathbf{C}^{-1})_{\alpha \alpha}}
\left[ \mathbf{C}^{-1} \left(\mathbf{z}- \langle \mathbf{c} \rangle \right) \right]_\alpha.
 \end{equation}
In words, this prescription states that one should take the data, subtract off the known foreground contamination, and then inverse variance weight the result before re-weighting to form the final estimator.  The inverse variance weighting performs a statistical suppression/subtraction of instrumental noise and foreground contamination.  Loosely speaking, this step corresponds to the subtraction of polynomial modes in the spectra as simulated in \cite{jonathanAvi,barkanaMorandi} and implemented in \cite{bowmanRogersMeasurement}.  The final re-weighting by $\mathbf{C}^{-1}_{\alpha \alpha}$ rescales the modes so that the previous subtraction step does not cause the estimated modes to be biased high or low.  For instance, if a certain mode is highly contaminated by foregrounds, it will be strongly down-weighted by the inverse variance weighting, and thus give an artificially low estimate of the mode unless it is scaled back up.

The corresponding measurement error covariance for this estimator is given by
\begin{equation}
\mathbf{\Sigma}_{\alpha \beta} \equiv \langle \mathbf{\widehat{x}}_s^\alpha \mathbf{\widehat{x}}_s^\beta \rangle - \langle \mathbf{\widehat{x}}_s^\alpha \rangle \langle \mathbf{\widehat{x}}_s^\beta \rangle= \frac{(\mathbf{C}^{-1})_{\alpha \beta}}{(\mathbf{C}^{-1})_{\alpha \alpha}(\mathbf{C}^{-1})_{\beta \beta}}.
\end{equation}
It is instructive to compare this with the errors predicted by the Fisher matrix formalism.  By the Cramer-Rao inequality, an estimator that is unwindowed (\emph{i.e.} one that has $\langle \mathbf{\widehat{x}}_s \rangle = \mathbf{x}_s$) will have a covariance that is at least as large as the inverse of the Fisher matrix.  Computing the Fisher matrix thus allows one to estimate the best possible errors bars that can be obtained from a measurement.  In the approximation that fluctuations about the mean are Gaussian, the Fisher matrix takes the form
\begin{equation}
\label{fisherDef}
\mathbf{F}_{\alpha \beta} = \frac{1}{2} \textrm{Tr} \left[ \mathbf{C}^{-1} \mathbf{C}_{,\alpha} \mathbf{C}^{-1} \mathbf{C}_{,\beta}\right] + \langle \mathbf{z} \rangle_{,\alpha}^\dagger \mathbf{C}^{-1} \langle \mathbf{z}\rangle_{,\beta},
\end{equation}
where commas denote derivatives with respect to the parameters that one wishes to measure.  In our case, the goal is to measure the global $21\,\textrm{cm}$ spectrum, so the parameters are the values of the spectrum at various frequencies.  Put another way, we can write our mean measurement equation as
\begin{equation}
\langle \mathbf{z} \rangle = \sum_\alpha \mathbf{x}^\alpha_s \mathbf{e}_\alpha + \langle \mathbf{c} \rangle,
\end{equation}
where $\mathbf{e}_\alpha$ is a unit vector with 0's everywhere except for a 1 at the $\alpha^{th}$ frequency channel.  The derivative $\langle \mathbf{z} \rangle_{,\alpha}$ with respect to the $\alpha^{th}$ parameter (\emph{i.e.} the derivative with respect to the mean measured spectrum $\mathbf{x}_s^\alpha$ in the $\alpha^{th}$ frequency channel) is therefore simply equal to $\mathbf{e}_\alpha$.  Since the measurement covariance $\mathbf{C}$ does not depend on the cosmological signal $\mathbf{x}_s$, our Fisher matrix reduces to
\begin{equation}
F_{\alpha \beta} = \mathbf{e}_\alpha \mathbf{C}^{-1} \mathbf{e}_\beta,
\end{equation}
which shows that the Fisher matrix is simply the inverse covariance \emph{i.e.} $\mathbf{F} = \mathbf{C}^{-1}$.  This implies that the covariance of the optimal unwindowed method is equal to the original noise and foreground covariance, which means that the error bars on the estimated spectrum are no smaller than if no line-of-sight foreground subtraction were attempted beyond the initial removal of foreground bias [Equations \eqref{intermediate1} and \eqref{intermediate2}].  In our notation, an unwindowed estimator would be one with $\mathbf{M} = \mathbf{I}$ [from Equation \eqref{intermediate1} plus the requirement that the estimator be unbiased, \emph{i.e.} Equation \eqref{intermediate2}].  But this means our optimal unwindowed estimator is
\begin{equation}
\label{sillyEst}
\widehat{\mathbf{x}}_s  = \mathbf{z} -\langle \mathbf{c} \rangle,
\end{equation}
which says to subtract our best-guess foreground spectrum model and to do nothing more!

To perform just a direct subtraction on a spectrum that has already been spatially averaged and to do nothing else [as Equation \eqref{sillyEst} suggests] is undesirable, because the error covariance in our foreground model is simply propagated directly through to our final measurement covariance.  This would be fine if we had perfect knowledge of our foregrounds.  Unfortunately, uncertainties in our foreground models mean that residual foregrounds can result in rather large error bars.  As an example, suppose we plugged our foreground covariance from Section \ref{fgModel} into the spatially averaged covariance $\mathbf{C}$ [Equation \eqref{averagedCovar}].  If we optimistically assume completely uncorrelated foreground model errors (so that they average down), one obtains an error bar of $11\,\textrm{K}$ at $79\,\textrm{MHz}$.  This is far larger than the amplitude of the cosmological signal that one expects.

If one uses the minimum variance estimator [Equation \eqref{spectralOnlyMinVar}] instead, one can reduce the error bars slightly (by giving up on the unwindowed requirement, the estimator can evade the Cramer-Rao bound).  However, this reduction is purely cosmetic, for Equations \ref{spectralOnlyMinVar} and \ref{sillyEst} differ only by the multiplication of an invertible matrix.  There is thus no difference in information content between the two estimators, and the minimum variance estimator will not do any better when one goes beyond the measured spectrum to constrain theoretical parameters.

Intuitively, one can do no better than a ``do nothing" algorithm because the global signal that we seek to measure is itself a spectrum (with unique cosmological information in every frequency channel), so a spectral-only measurement provides no redundancy.  With no redundancy, one can only subtract a best-guess foreground model, and it is impossible to minimize statistical errors in the model itself.

The key, then, is to have multiple, redundant measurements that all contain the same cosmological signal.  This can be achieved by designing experiments with angular information (\emph{i.e.} ones that do not integrate over the entire sky automatically).  Because the global signal is formally a spatial monopole, measurements in different pixels on the sky have identical cosmological contributions, but different foreground contributions, allowing us to further distinguish between foregrounds and cosmological signal.

\begin{widetext}

\subsection{Methods using both spectral and angular information}
\label{both}

For an experiment with both spectral and angular information, the relevant measurement equation is Equation \eqref{fullMeas}, and the minimum variance, unwindowed estimator for the signal can be shown \cite{maxMapmaking} to take the form
\begin{equation}
\label{optsolution}
\mathbf{\widehat{x}}_s = \left[ \mathbf{A}^t \mathbf{N}^{-1} \mathbf{A} \right]^{-1} \mathbf{A}^{t} \mathbf{N}^{-1} (\mathbf{y} -  \mathbf{m}).
\end{equation}
With this estimator, the error covariance can also be shown to be
\begin{equation}
\label{genOptError}
\mathbf{\Sigma} \equiv \langle \mathbf{\widehat{x}}_s \mathbf{\widehat{x}}_s^t \rangle- \langle \mathbf{\widehat{x}}_s \rangle \langle \mathbf{\widehat{x}}_s^t \rangle = \left[ \mathbf{A}^t \mathbf{N}^{-1} \mathbf{A} \right]^{-1}.
\end{equation}

These equations in principle encode all that we need to know for data analysis, and even allow for a generalization of the results that follow in the rest of this paper.  For example, frequency-dependent beams (a complication that we ignore, as noted earlier) can be incorporated into the analysis by making suitable adjustments to $\mathbf{N}$.  Recall from Section \ref{instNoiseDescription} that our convention is to assume that all data sets have already been suitably deconvolved prior to our analysis.  Taking into account a frequency-dependent beam is therefore just a matter of including the effects of a frequency-dependent deconvolution in the generalized noise covariance.  For the Gaussian beams considered in this paper, for instance, we can simply make the $\theta_b$ parameter [in Equation \eqref{neff}] a function of frequency.

Despite their considerable flexibility, we will not be using Equations \ref{optsolution} and \ref{genOptError} in their most general form.  For the rest of the paper, we will make the approximation of frequency-independent beams, which allows us to make some analytical simplifications.
Doing so not only allows us to build intuition for what the matrix algebra is doing, but also makes it computationally feasible to explore a wide region of parameter space, as we do in later sections.  In its current form, Equation \eqref{optsolution} is too computationally expensive to be evaluated over and over again because it involves the inversion of $\mathbf{N}$, which is an $N_{\textrm{vox}} \times N_{\textrm{vox}}$ matrix.  Given that $N_{\textrm{vox}}$ is likely to be quite large for a $21\,\textrm{cm}$ survey with full spectral and angular information, it would be wise to avoid direct matrix inversions.

We thus seek to derive analytic results for $\mathbf{\Sigma} \equiv [\mathbf{A}^t \mathbf{N}^{-1} \mathbf{A}]^{-1}$ and $\mathbf{A}^t \mathbf{N}^{-1}$, where $\mathbf{N}$ is given by Equation \eqref{Ntotal}.  We begin by factoring out the foreground templates from our generalized noise covariance, so that $\mathbf{N} = \mathbf{D} \widetilde{\mathbf{N}} \mathbf{D}$, where\footnote{Recall that throughout this paper, we use Greek indices to signify the spectral dimension (or spectral eigenmodes) and Latin indices to signify angular dimensions.} $\mathbf{D}_{\alpha i \beta j} \equiv \mathbf{m}_{\alpha i } \delta_{\alpha \beta} \delta_{ij}$, just as we defined in Section \ref{instNoiseDescription}.  This is essentially a whitening procedure\footnote{In what follows, $\mathbf{N}$, $\mathbf{\widetilde{N}}$, $\mathbf{\overline{N}}$, and $\mathbf{\widehat{N}}$ all refer to the same matrix, but use different unit conventions and/or are expressed in different bases.  The original generalized noise $\mathbf{N}$ is assumed to be in ``real space" \emph{i.e.} frequency and spatial angles on the sky; $\mathbf{\widetilde{N}}$ is in the same basis, but is in units where the foreground model has been divided out; $\mathbf{\overline{N}}$ is the same as $\mathbf{\widetilde{N}}$, but in a spatial angle and spectral eigenforeground basis; $\mathbf{\widehat{N}}$ is the same as $\mathbf{\widetilde{N}}$, but in a spherical harmonic and spectral eigenforeground basis.}, making the generalized noise independent of frequency or sky pixel.  Since both $\mathbf{\Sigma} \equiv [\mathbf{A}^t \mathbf{N}^{-1} \mathbf{A}]^{-1}$ and $\mathbf{A}^t \mathbf{N}^{-1}$ involve $\mathbf{N}^{-1}$, we proceed by finding $\mathbf{N}^{-1} = \mathbf{D}^{-1} \mathbf{\widetilde{N}}^{-1} \mathbf{D}^{-1}$, and to do so we move into a diagonal basis.  The matrix $\mathbf{D}$ is already diagonal and easily invertible, so our first step is to perform a similarity transformation in the frequency-frequency components of $\mathbf{\widetilde{N}}$ in order to diagonalize $\mathbf{Q}$ (which, recall from Section \ref{gennoise}, quantifies the spectral correlations of the foregrounds).  In other words, we can write $\mathbf{\widetilde{N}}$ as
\begin{equation}
\label{freqwhitenedcovar}
\mathbf{\widetilde{N}}_{\alpha i \beta j} = (\mathbf{V} \overline{\mathbf{N}} \mathbf{V}^t )_{\alpha i \beta j} = \sum_{\eta \lambda k m} \mathbf{V}_{\alpha i \eta k} \left[\varepsilon_0^2 \left( \frac{\theta_{\textrm{fg}}}{\theta_b} \right)^2 \lambda_\eta \mathbf{R}_{km} + \mathbf{N}^{\textrm{dec}}_{km} \right] \delta_{\eta \lambda} \mathbf{V}_{\beta j \lambda m},
\end{equation}
where $\mathbf{V}_{\alpha i \eta k} \equiv (\mathbf{v}_\eta)_\alpha \delta_{ik}$, with $(\mathbf{v_\eta})_\alpha$ signifying the value of the $\alpha^{th}$ component (\emph{i.e.} frequency channel) of the $\eta^{th}$ eigenvector of $\mathbf{Q}$.  The $\eta^{th}$ eigenvalue is given by $\lambda_\eta$.  Note also that $\mathbf{V}^{-1} = \mathbf{V}^t$.

Our next step is to diagonalize $\mathbf{R}$, the spatial correlations of the foregrounds, as well as $\mathbf{N}^{\textrm{dec}}$, which as the whitened instrumental noise covariance is spatially correlated after the data has been deconvolved.  If we assume that the correlations are rotationally invariant\footnote{Recall that $\mathbf{R}$ encodes the spatial correlations in the errors of our foreground model.  It is thus entirely possible to break rotation invariance, for instance by using a foreground model that is constructed from a number of different surveys, each possessing different error characteristics and different sources of error.  For this paper we ignore this possibility in order to gain analytical intuition, but we note that it can be corrected by finding the eigenvectors and eigenvalues of $\mathbf{R}$, just as we did with $\mathbf{Q}$.}, these matrices will be diagonal in a spherical harmonic basis.  For computational convenience, we will now work in the continuous limit.  As discussed above and shown in \cite{maxMapping,knox}, this is entirely consistent with the discrete approach provided one augments all noise covariances with a factor of $4 \pi/N_\textrm{pix}$.  In the continuous limit, the spatial correlations of the foregrounds therefore take the form
\begin{equation}
R (\mathbf{\widehat{r}},\mathbf{\widehat{r}^\prime}) = \frac{4 \pi}{N_\textrm{pix}}\frac{\exp(\sigma^{-2} \mathbf{\widehat{r}}\cdot \mathbf{\widehat{r}^\prime})}{4 \pi \sigma^2 \sinh(\sigma^{-2})},
\end{equation}
which (except for our $4 \pi/N_\textrm{pix}$ factor) is known as a Fisher function, the analog of a Gaussian on a sphere.  For $\sigma \ll 1\,\textrm{rad}$, this reduces to the familiar Gaussian:
\begin{equation}
R (\mathbf{\widehat{r}},\mathbf{\widehat{r}^\prime}) \approx \frac{4 \pi}{N_\textrm{pix}} \frac{1}{2\pi \sigma^2} \exp \left[ -\frac{1}{2} \frac{\theta^2}{\sigma^2}\right],
\end{equation}
where $\theta \equiv \arccos (\mathbf{\widehat{r}}\cdot \mathbf{\widehat{r}^\prime})$ is the angle between the two locations on the sphere.  Switching to a spherical harmonic basis, we have
\begin{equation}
\widehat{R}_{\ell m \ell^\prime m^\prime} \equiv \frac{4 \pi}{N_\textrm{pix}} \int d\Omega d\Omega^\prime Y^*_{\ell m} (\widehat{\mathbf{r}}) R (\mathbf{\widehat{r}},\mathbf{\widehat{r}^\prime}) Y_{\ell^\prime m^\prime} (\mathbf{\widehat{r}^\prime}) \approx  \frac{4 \pi}{N_\textrm{pix}} \exp\left[ -\frac{1}{2} \sigma^2 \ell (\ell +1) \right] \delta_{\ell \ell^\prime} \delta_{m m^\prime},
\end{equation}
where $Y_{\ell m}$ denotes the spherical harmonic with azimuthal quantum number $\ell$ and magnetic quantum number $m$, and the last approximation holds if $\sigma \ll 1\,\textrm{rad}$.

For the instrumental noise term, we saw in Section \ref{instNoiseDescription} that after dividing out the instrument's beam, we have
\begin{equation}
\widehat{N}^{\textrm{dec}}_{\ell m \ell^\prime m^\prime} =  \frac{4 \pi}{N_\textrm{pix}} \frac{e^{\theta_b^2 \ell (\ell+1)}}{\Delta t \Delta \nu} \delta_{\ell \ell^\prime} \delta_{m m^\prime},
\end{equation}
and adding this to the foreground contribution gives us the equivalent of $\overline{\mathbf{N}}$ but in spherical harmonic space:
\begin{equation}
\widehat{\mathbf{N}}_{\ell m \eta \ell^\prime m^\prime \lambda} = \frac{4 \pi}{N_\textrm{pix}} \left[ \varepsilon_0^2 \left( \frac{\theta_{\textrm{fg}}}{\theta_b} \right)^2 \lambda_\eta e^{-\frac{1}{2} \sigma^2 \ell (\ell +1)} +\frac{e^{\theta_b^2 \ell (\ell+1)}}{\Delta t \Delta \nu} \right] \delta_{\ell \ell^\prime} \delta_{m m^\prime} \delta_{\eta \lambda}.
\end{equation}
The diagonal nature of $\widehat{\mathbf{N}}$ in this equation allows a straightforward inversion:
\begin{equation}
(\widehat{\mathbf{N}}^{-1})_{\ell m \eta \ell^\prime m^\prime \lambda} = \frac{N_{\textrm{pix}}}{4 \pi} \left[ \varepsilon_0^2 \left( \frac{\theta_{\textrm{fg}}}{\theta_b} \right)^2 \lambda_\eta e^{-\frac{1}{2} \sigma^2 \ell (\ell +1)} +\frac{e^{\theta_b^2 \ell (\ell+1)}}{\Delta t \Delta \nu} \right]^{-1} \delta_{\ell \ell^\prime} \delta_{m m^\prime} \delta_{\eta \lambda },
\end{equation}
thus allowing us to write the inverse matrix in the original spatial basis as
\begin{equation}
(\overline{\mathbf{N}}^{-1})_{\alpha i \beta j} = (\mathbf{F}^\dagger \widehat{\mathbf{N}}^{-1} \mathbf{F})_{\alpha i \beta j} = \frac{N_{\textrm{pix}}}{4 \pi} \sum_{\eta \lambda \ell \ell^\prime m m^\prime} \mathbf{F}^\dagger_{\alpha i \eta \ell m} \left[ \varepsilon_0^2 \left( \frac{\theta_{\textrm{fg}}}{\theta_b} \right)^2 \lambda_\eta e^{-\frac{1}{2} \sigma^2 \ell (\ell +1)} +\frac{e^{\theta_b^2 \ell (\ell+1)}}{\Delta t \Delta \nu} \right]^{-1} \delta_{\ell \ell^\prime} \delta_{m m^\prime} \delta_{\eta \lambda } \mathbf{F}_{\lambda \ell^\prime m^\prime \beta j},
\end{equation}
where $\mathbf{F}_{\beta \ell m \alpha i } \equiv \delta_{\alpha \beta} \mathbf{Y}_{\ell m i }$ and $\mathbf{Y}$ is a unitary matrix that transforms from a real-space angular basis to a spherical harmonic basis.  Obtaining the inverse of $\mathbf{\widetilde{N}}$ from here is done by evaluating $\mathbf{\widetilde{N}}^{-1} = \mathbf{V} \overline{\mathbf{N}}^{\,-1} \mathbf{V}^t$.

We are now ready to assemble the pieces to form $\mathbf{\Sigma} \equiv [\mathbf{A}^t \mathbf{N}^{-1} \mathbf{A}]^{-1}$ which, in the notation of our various changes of basis can be written as $ [\mathbf{A}^t \mathbf{D}^{-1} \mathbf{F}^\dagger \mathbf{V} \widehat{\mathbf{N}}^{\,-1} \mathbf{V}^t \mathbf{F} \mathbf{D}^{-1} \mathbf{A}]^{-1}$.  We first compute
\begin{equation}
(\mathbf{V}^t \mathbf{F} \mathbf{D}^{-1})_{\alpha \ell m \beta j} = \mathbf{Y}_{\ell m j} (\mathbf{v}_\alpha)_\beta (\mathbf{m}_{\beta j})^{-1} \equiv \mathbf{Y}_{\ell m j} (\mathbf{v}_\alpha)_\beta \mathbf{u}_{\beta j},
\end{equation}
where $\mathbf{u}_{\alpha i} \equiv 1/ \mathbf{m}_{\alpha i}$  is the reciprocal temperature (in units of $\textrm{K}^{-1}$) at the $\alpha^{\textrm{th}}$ frequency and the $i^{\textrm{th}}$ pixel of our foreground templates.  Note that in the above expression, there is no sum over $j$ yet.  This is accomplished by the angular summation matrix $\mathbf{A}_{\alpha i \beta} = \delta_{\alpha \beta}$, giving
\begin{equation}
(\mathbf{V}^t \mathbf{F} \mathbf{D}^{-1} \mathbf{A})_{\alpha \ell m \beta} = (\mathbf{v}_\alpha)_\beta \sum_j \mathbf{Y}_{\ell m j} \mathbf{u}_{\beta j} = (\mathbf{v}_\alpha)_\beta \widehat{\mathbf{u}}_{\beta \ell m},
\end{equation}
where there is similarly no sum over $\beta$, and $\widehat{\mathbf{u}}$ signifies our reciprocal foreground templates in spherical harmonic space.  The inverse covariance $\boldsymbol{\Sigma}^{-1}$ is thus given by
\begin{equation}
(\boldsymbol{\Sigma}^{-1})_{\alpha \beta} = \frac{N_{\textrm{pix}}}{4 \pi} \sum_\eta (\mathbf{v}_\eta)_\alpha (\mathbf{v}_\eta)_\beta \sum_{\ell m} \frac{\widehat{\mathbf{u}}_{\alpha \ell m}\widehat{\mathbf{u}}_{\beta \ell m}}{\varepsilon_0^2 \left( \frac{\theta_{\textrm{fg}}}{\theta_b} \right)^2 \lambda_\eta e^{-\frac{1}{2} \sigma^2 \ell (\ell +1)} +\frac{1}{\Delta t \Delta \nu}e^{\theta_b^2 \ell (\ell+1)}}.
\end{equation}
Defining $t_{\textrm{int}} \equiv N_{\textrm{pix}} \Delta t$ to be the total integration time over the survey
\footnote{That is, $\Delta t$ refers to the amount of integration time spent by a \emph{single} beam on a small patch of the sky with area equal to our pixel size, and $t_\textrm{int}$ refers to the total integration time of a \emph{single} beam scanning across the entire sky.  An experiment capable of forming $N_{\textrm{beams}}$ independent beams simultaneously would require one to replace $t_\textrm{int}$ with $N_{\textrm{beams}} t_\textrm{int}$ in our formalism.}
,  and making the substitution $N_{\textrm{pix}} = 4 \pi / \theta_b^2$ gives
\begin{equation}
(\boldsymbol{\Sigma}^{-1})_{\alpha \beta} = \frac{1}{4 \pi} \sum_\eta (\mathbf{v}_\eta)_\alpha (\mathbf{v}_\eta)_\beta \sum_{\ell m} \frac{\widehat{\mathbf{u}}_{\alpha \ell m}\widehat{\mathbf{u}}_{\beta \ell m}}{\frac{\varepsilon_0^2 \theta_{\textrm{fg}}^2}{4 \pi}  \lambda_\eta e^{-\frac{1}{2} \sigma^2 \ell (\ell +1)} +\frac{1}{t_{\textrm{int}} \Delta \nu}e^{\theta_b^2 \ell (\ell+1)}}.
\end{equation}

At this point, notice that the angular cross-power spectrum $C_\ell^{\alpha \beta}$ between two reciprocal maps $\mathbf{u}_{\alpha i}$ and $\mathbf{u}_{\beta i}$ at frequencies $\alpha$ and $\beta$ respectively is given by
\begin{equation}
C_\ell^{u,\alpha \beta} \equiv \frac{1}{2 \ell +1} \sum_{m = -\ell}^\ell \widehat{\mathbf{u}}_{\alpha \ell m}\widehat{\mathbf{u}}_{\beta \ell m},
\end{equation}
where the ``$u$" superscript serves to remind us that $C_\ell^{u,\alpha \beta}$ is the cross-power spectrum of the reciprocal maps, not the original foreground templates.  With this, our expression for the inverse measurement covariance $\mathbf{\Sigma}^{-1}$ can be written as
\begin{equation}
\label{intuitiveInfo}
(\mathbf{\Sigma}^{-1})_{\alpha \beta} = \frac{1}{4 \pi} \sum_\eta (\mathbf{v}_\eta)_\alpha (\mathbf{v}_\eta)_\beta \sum_\ell \frac{2 \ell + 1}{\frac{\varepsilon_0^2 \theta_{\textrm{fg}}^2}{4 \pi}  \lambda_\eta e^{-\frac{1}{2} \sigma^2 \ell (\ell +1)} +\frac{1}{t_{\textrm{int}} \Delta \nu}e^{\theta_b^2 \ell (\ell+1)}} C_\ell^{u,\alpha \beta}.
\end{equation}
Permuting the sums and recalling that $\lambda_\eta$ and $\mathbf{v}_\eta$ are the $\eta^{\textrm{th}}$ eigenvalue and eigenvector of our foreground spectral correlation matrix $\mathbf{Q}$ respectively, this expression can be further simplified to give
\begin{equation}
\label{grandInfo}
(\boldsymbol \Sigma^{-1})_{\alpha \beta} = \sum_{\ell=0}^\infty \frac{2 \ell +1}{4 \pi} C_{\ell}^{u, \alpha \beta} \left[ \frac{\varepsilon_0^2 \theta^2_{\textrm{fg}}}{ 4 \pi}e^{-\frac{\sigma^2}{2} \ell (\ell+1)} \mathbf{Q} + \frac{\mathbf{I}}{t_{\textrm{int}} \Delta \nu} e^{\theta_b^2 \ell (\ell+1)}\right]^{-1}_{\alpha \beta}.
\end{equation}

This provides a fast prescription for computing $\boldsymbol{\Sigma} \equiv [\mathbf{A}^t \mathbf{N}^{-1} \mathbf{A}]^{-1}$.  One first inverts a series of relatively small matrices (each given by the expression in the square brackets).  These inversions do not constitute a computationally burdensome load, for $\alpha$ and $\beta$ are frequency indices, so the matrices are of dimension $N_{\textrm{freq}} \times N_{\textrm{freq}}$.  One then uses publicly available fast routines for computing the angular cross-power spectrum, multiplies by $(2\ell +1)/4\pi$ and the inverted matrices, and sums over $\ell$.  The resulting matrix is then inverted, a task that is much more computationally feasible than a brute-force evaluation of $\boldsymbol{\Sigma}$, which would involve the inversion of an $N_{\textrm{vox}} \times N_{\textrm{vox}}$ matrix.

Using essentially identical tricks, we can also derive a simplified version of our estimator $\widehat{\mathbf{x}}_s$ [Equation \eqref{optsolution}].  The result is
\begin{equation}
\label{uberEst}
\mathbf{\widehat{x}}^{\alpha}_s =\sum_{\beta} \boldsymbol \Sigma_{\alpha \beta} \int \frac{d\Omega}{m (\widehat{\mathbf{r}}, \nu_\beta)}\sum_{\ell m} \frac{Y_{\ell m} (\widehat{\mathbf{r}})}{4\pi} \sum_{\eta} (\mathbf{v}_\eta)_\beta \mathbf{w}_{\ell \eta} \sum_\gamma (\mathbf{v}_\eta)_\gamma \int d\Omega^\prime \,Y^*_{\ell m} (\widehat{\mathbf{r}}^\prime) \left[\frac{y (\widehat{\mathbf{r}}^\prime, \nu_\gamma)}{m (\widehat{\mathbf{r}}^\prime, \nu_\gamma)}-1\right],
\end{equation}
where $y$ and $m$ are continuous versions of the measured sky signal $\mathbf{y}$ and the foreground model $\mathbf{m}$ respectively, and the weights $\mathbf{w}$ are defined as
\end{widetext}
\begin{equation}
\label{weightyWeights}
\mathbf{w}_{\ell \eta}\equiv\left[{\frac{\varepsilon_0^2 \theta_{\textrm{fg}}^2}{4 \pi}  \lambda_\eta e^{-\frac{1}{2} \sigma^2 \ell (\ell +1)} +\frac{1}{t_{\textrm{int}} \Delta \nu}e^{\theta_b^2 \ell (\ell+1)}}\right]^{-1}.
\end{equation}

In words, this estimator calls for the following data analysis recipe:
\begin{enumerate}
\item Take the measured sky signal $y (\widehat{\mathbf{r}}, \nu)$ and subtract the best-guess foreground model $m(\widehat{\mathbf{r}}, \nu)$.
\item Downweight regions of the sky that are believed to be heavily contaminated by foregrounds by dividing by our best-guess model $m(\widehat{\mathbf{r}}, \nu)$.  [This and the previous step of course simplify to give $y/m-1$, as we wrote in Equation \eqref{uberEst}].
\item Express the result in a spherical harmonic and spectral eigenmode basis.
\item Take each mode (in spherical harmonic and frequency eigenmode space) and multiply by weights $\mathbf{w}_{\ell \eta}$.
\item Transform back to real (angle plus frequency) space. 
\item Divide by $m(\widehat{\mathbf{r}}, \nu)$ to downweight heavily contaminated foreground regions once more.
\item Sum over the entire sky to reduce the data to a single spectrum.
\item Normalize the spectrum by applying $\boldsymbol{\Sigma} \equiv [\mathbf{A}^t \mathbf{N}^{-1} \mathbf{A}]^{-1}$ [the inverse of Equation \eqref{grandInfo}] to ensure that modes that were heavily suppressed prior to our averaging over the sky are rescaled back to their correct amplitudes.  (Note that the error bars will also be correspondingly rescaled so that heavily contaminated modes are correctly labeled as lower signal-to-noise measurements).
\end{enumerate}
\begin{figure*}
\centering
\includegraphics[width=1.0\textwidth]{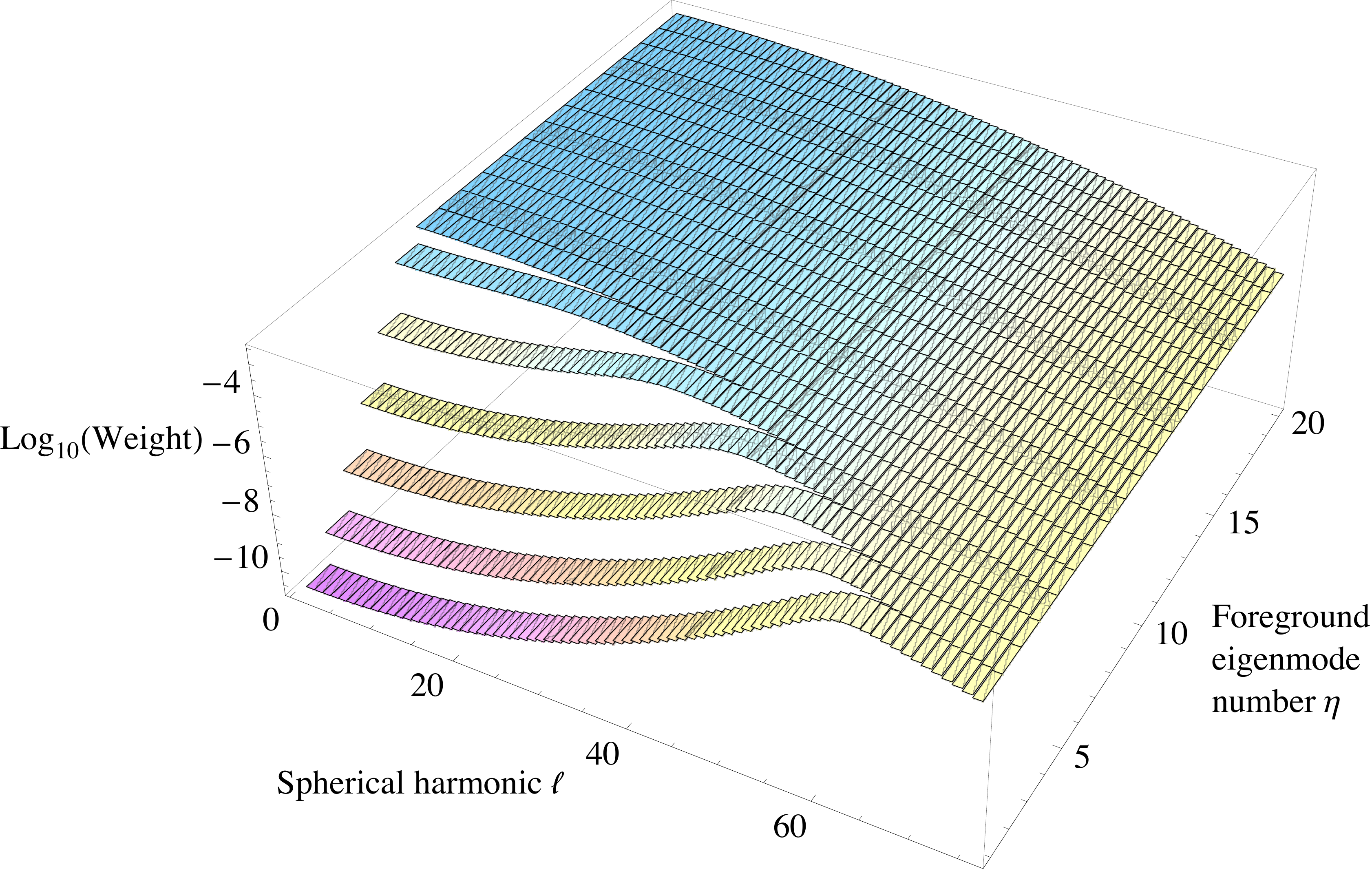}
\caption{Weights [Equation \eqref{weightyWeights}] in spherical harmonic $\ell$ and eigenmode $\eta$ space for combining data in our optimal estimator of the global signal [Equation \eqref{uberEst}].  Low $\eta$ and high $\ell$ modes are downweighted because of contamination by foregrounds and instrumental noise, respectively.}
\label{optimalWeights}
\end{figure*}

The recipe outlined here takes full advantage of spectral and angular information to mitigate foregrounds and produce the smallest possible error bars under the constraint that there be no multiplicative or additive bias \emph{i.e.} under the constraint that $\langle \mathbf{\widehat{x}}_s \rangle = \mathbf{x}_s$.  To see how this recipe works intuitively, we can plot the weights $\mathbf{w}_{\ell \eta}$ to see how our optimal prescription uses various spherical harmonic $\ell$ and eigenforeground $\eta$ modes.  This is shown in Figure \ref{optimalWeights}.  At low spherical harmonic $\ell$, the first few (\emph{i.e.} smoothest) eigenmodes are dominated by foregrounds, so they are severely downweighted.  At high $\ell$, the limited angular resolution of our instrument means that the instrumental noise dominates regardless of spectral eigenmode, so all eigenmodes are weighted equally in an attempt to average down the instrumental noise.  At high $\eta$ (spectrally unsmooth eigenmodes), the foreground contamination is negligible, so the weightings are dictated entirely by the instrumental noise, with the only trend being a downweighting of the noisier high $\ell$ modes.

It is important to emphasize that these weights are applied to whitened versions of the data, \emph{i.e.} data where the best-guess foreground model was divided out in Step 2 above.  If this weren't the case, the notion of weights as a function of $\ell$ would be meaningless, for our goal is to estimate the monopole (in other words, the global signal), so all that would be required would be to pick the $\ell =0$ component.  The monopole in our original data is a linear combination of different $\ell$ modes in the whitened units, and the weights tell us what the optimal linear combination is, taking into account the interplay between foreground and instrumental noise contamination.  In contrast, the spectral-only methods are a ``one size fits all" approach with no whitening and a simple weighting that consists solely of the $\ell=0$ row of Figure \ref{optimalWeights}, ignoring the fact that at some $(\ell, \eta)$ the foregrounds dominate, whereas at others the instrumental noise dominates.

The rest of this paper will be dedicated to examining the error properties of our optimal estimator.  The goal is to  gain an intuitive understanding for the various trade-offs that go into designing an experiment to measure the global $21\,\textrm{cm}$ spectrum.

\subsection{Making the cosmological signal more apparent}
\label{Wiener}

In Sections \ref{spectral} and \ref{reionizationDesigner}, we will find that despite our best efforts at foreground mitigation, the resulting error bars on our measured signal can sometimes still be large, particularly for instruments that lack angular resolution.  These errors will be dominated by residual foregrounds at every frequency, and are unavoidable without an exquisitely accurate foreground model.  The resulting measurements will thus look very much like foreground spectra.

Often, we will find that this happens even when the detection significance is high.  This apparent paradox can be resolved by realizing that often the large errors are due to a small handful of foreground modes that dominate at every single frequency channel.  Put another way, the contaminants are fundamentally  sparse in the right basis, since there are a small number of modes that are responsible for most of the contamination.  Plotting our results as a function of frequency is therefore simply a bad choice of basis.  By moving to a better basis (such as one where our measurement covariance $\mathbf{\Sigma}$ is diagonal, as we will discuss in Section \ref{spectral}), it becomes apparent that the cosmological signal can be detected to high significance.

Still, it is somewhat unfortunate that the frequency basis is a bad one, as the cosmological signal is most easily interpreted as a function of frequency.  It would thus be useful to be able to plot, as a function of frequency, a measured spectrum that is not dominated by the largest amplitude foreground modes, even if we are unable to remove them.  Essentially, one can arrive at a measurement that is closer to the ``true" cosmological signal by simply giving up on certain modes.  This means resigning oneself to the fact that some modes will be forever lost to foregrounds, and that one will never be able to measure those components of the cosmological signal.  As discussed for an analogous problem in \cite{Paper5}, this can be accomplished by subjecting our recovered signal to a Wiener filter $\mathbf{W}$:
\begin{equation}
\mathbf{\widehat{x}_{s}^{\textrm{Wiener}}} = \mathbf{S} [ \mathbf{S} + \mathbf{\Sigma}]^{-1} \mathbf{\widehat{x}_{s}} \equiv \mathbf{W} \mathbf{\widehat{x}_{s}} ,
\end{equation}
where $\mathbf{S}$ is the signal covariance matrix and $\mathbf{\widehat{x}_{s}}$ is our estimator from Equation \eqref{uberEst}.  Roughly speaking, this amounts to weighting the data by ``signal over signal plus noise", which means well-measured (low noise/foreground) modes are given a weight of unity, whereas poorly measured modes are downweighted and effectively excluded from our estimate.

The Wiener-filtered result has the desirable property of minimizing the quantity $\langle |\boldsymbol{\varepsilon}_i|^2 \rangle$, where $\boldsymbol{\varepsilon} \equiv \mathbf{\widehat{x}}_s - \mathbf{x}_s$ is the error vector.  It thus represents our ``best guess" as to what the cosmological signal looks like, at the expense of losing the information in highly contaminated modes.  Since these modes are irretrievably lost (without better foreground modeling), one must also Wiener-filter the theoretical models in order to make a fair comparison.  We show such Wiener-filtered theoretical spectra in Section \ref{darkDesigner}.

In contrast to the Wiener filter, our previous estimator minimized $\langle |\boldsymbol{\varepsilon}_i|^2 \rangle$ only under the constraint that $\langle \mathbf{\widehat{x}}_s \rangle = \mathbf{x}_s$.  It also had the property that it minimized $\chi^2 \equiv (\mathbf{y} - \mathbf{A} \mathbf{\widehat{x}}_s)^t \mathbf{N}^{-1} (\mathbf{y} - \mathbf{A} \mathbf{\widehat{x}}_s)$ \cite{maxMapmaking}, and so the estimator constructed an unbiased model that best matched the observations.  The model will thus include modes that are so contaminated that there is no hope of measuring the cosmological signal in them, since these modes, however contaminated and error-prone they might be, are in fact part of the observation.  The result is a foreground contaminated spectrum, which the Wiener-filtered result avoids.

It must be emphasized, however, that because the Wiener filter $\mathbf{W}$ is an \emph{invertible} matrix, there is no change in information content.  The information about the cosmological signal was always there, and moving to a different basis or Wiener filtering simply made it more apparent.  Wiener filtering is simply a convenient post-processing visualization tool for building intuition, and will not change our ability to constrain the physics in our theoretical models.


\section{A designer's guide to experiments that probe the Dark Ages}
\label{darkDesigner}

Having established a general framework for analyzing data from global signal experiments, we now step back and tackle the problem experimental design.  In this section, we will consider experiments that are designed to target the trough in brightness temperature between $30\,\textrm{MHz}$ and $100\,\textrm{MHz}$ (\emph{i.e.} probing the dark ages).  In Section \ref{reionizationDesigner} we will discuss experiments that target the reionization window from $100$ to $250\,\textrm{MHz}$.

Our guide to experimental design will be the quantity $\boldsymbol \Sigma^{-1}$, given by Equation \eqref{intuitiveInfo} [or equivalently, Equation \eqref{grandInfo}].  As an inverse covariance, this quantity represents an information matrix.  Our goal in what follows will be to design an experiment that maximizes the information.  One approach to maximizing the information would be a brute-force exploration of parameter space.  However, this is a computationally intensive, ``black box" method that does not yield intuition for the various trade-offs that go into experimental design.  Instead, we take the following approach.  We first rewrite our information matrix by separating out the monopole ($\ell = 0$) term in our spherical harmonic expansion:
\begin{widetext}
\begin{equation}
\label{decomposedInfo}
\mathbf{\Sigma}^{-1}_{\alpha \beta} = \Bigg \langle \frac{1}{T(\nu_\alpha)} \Bigg \rangle \Bigg \langle \frac{1}{T(\nu_\beta)} \Bigg \rangle \left[\frac{\varepsilon_0^2  \theta_{\textrm{fg}}^2}{4 \pi} \mathbf{Q} + \frac{\mathbf{I}}{t_{\textrm{int}} \Delta \nu}\right]^{-1}_{\alpha \beta} + \frac{1}{4\pi} \sum_\eta \sum_{\ell=1}^\infty \frac{ (2 \ell + 1) C_\ell^{u,\alpha \beta} }{\frac{\varepsilon_0^2  \theta_{\textrm{fg}}^2}{4 \pi} \lambda_\eta e^{-\frac{\sigma^2}{2} \ell (\ell +1)} + \frac{e^{\theta_b^2 \ell (\ell+1)}}{t_{\textrm{int}} \Delta \nu}} (\mathbf{v}_\eta)_\alpha (\mathbf{v}_\eta)_\beta,
\end{equation}
\end{widetext}
where we have taken advantage of the fact that when $\ell=0$, the power spectrum takes the value $C_\ell^u = 4 \pi \langle u(\mathbf{\widehat{r}}) \rangle^2 = 4 \pi \langle 1/T(\mathbf{\widehat{r}}) \rangle^2$, with expectation values $\langle \cdots \rangle$ denoting a spatial average in this case.  In the next few subsections we will interpret each piece of Equation \eqref{decomposedInfo} individually, using quantitative arguments to arrive at qualitative rules of thumb for global signal experiment design.

Unless otherwise stated, in what follows we will perform calculations with $t_{\textrm{int}} = 100\,\textrm{hrs}$ and assume an instrument with a channel width of $\Delta \nu = 1 \,\textrm{MHz}$.  As stated in Section \ref{fgModel}, we will imagine that our foreground model has roughly the same accuracy as that of the global sky model of \cite{GSM}, and thus set $\varepsilon_0 = 0.1$ at a resolution of $\theta_{\textrm{fg}} = 5^\circ$.  Since this is also the ``native" resolution of the GSM, we assume that the errors are correlated over $\sigma=5^\circ$.  Because we do not possess angular foreground models that have much finer resolution than this (the GSM can at most be pushed to a $1^\circ$ resolution by locking to the Haslam map \cite{Haslam} at $408\,\textrm{GHz}$), we will not be analyzing experiments with beams that are smaller than $5^\circ$.  The $5^\circ$ figure that will be referenced repeatedly in the following sections thus does not represent some optimized beam size, but should instead be thought of as a fiducial value chosen to highlight the advantages of having a small beam.  We do note, however, that thanks to the cosmological anisotropy noise described in Section \ref{cosmonoise}, an instrument with finer beams than a few degrees will likely be suboptimal.

\subsection{How much does angular information help?}
We begin by providing intuition for the role of angular information in a global signal experiment where such information is available.  In general, angular information can potentially help one's measurement by:
\begin{itemize}
\item Allowing regions of the sky that are severely contaminated by foregrounds to be downweighted or discarded in our estimate of the global signal.
\item Providing information on the angular structure of the foregrounds, which can be leveraged to perform spatial foreground subtraction.
\end{itemize}
In what follows we will assess the effectiveness each of these strategies.

\subsubsection{Downweighting heavily contaminated regions}
\label{downweighting}
\begin{figure}
\centering
\includegraphics[width=0.5\textwidth]{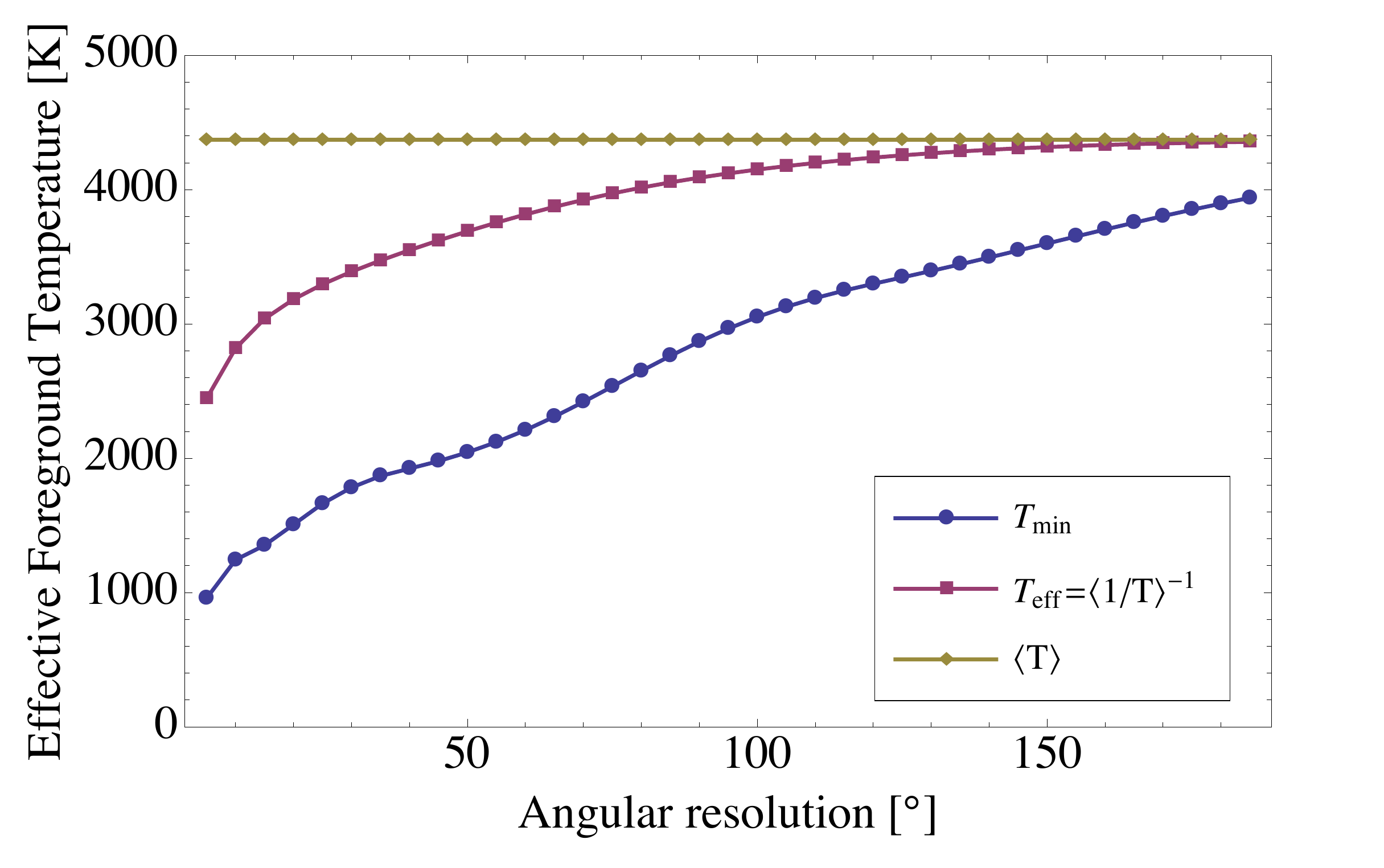}
\caption{Effective foreground temperature $T_{\textrm{eff}} = \langle 1/T \rangle^{-1}$ (purple squares), average foreground temperature $\langle T \rangle$ (gold diamonds), and minimum foreground temperature $T_\textrm{min}$ (blue circles) in the sky as a function of the angular resolution at $79\,\textrm{MHz}$.  In each case, a high resolution foreground map was convolved with a Gaussian beam with a full-width-half-max given by values on the horizontal axis.  When interpreted with the information matrix [Equation \eqref{decomposedInfo}], this shows that having angular sensitivity allows an optimal downweighting of regions heavily contaminated by foregrounds, reducing the effective foreground contamination in the global signal.  The effect is particularly pronounced with fine angular resolution.}
\label{recipStats}
\end{figure}

We begin by addressing the first point.  Consider the first term of Equation \eqref{decomposedInfo}, the information matrix, which contains two factors of $\langle 1/T \rangle $.  Since $1/T$ is a convex function for $T>0$ (almost always the case for our foregrounds\footnote{Foregrounds that can be negative, such as those from the Sunyaev-Zeldovich Effect, are completely negligible compared to the sources considered in this paper.}), we know from Jensen's inequality that
\begin{equation}
\Big \langle \frac{1}{T} \Big \rangle \ge \frac{1}{\langle T \rangle},
\end{equation}
where equality holds only in the limiting case of uniform foregrounds over the entire sky.  Put another way, equality holds only when an experiment has no angular sensitivity, which is mathematically equivalent to an experiment that measures the same value in all pixels of the sky.  With angular sensitivity, $\langle 1/T \rangle$ rises above $1/\langle T \rangle$, increasing the information content via the monopole term of Equation \eqref{decomposedInfo}.  More intuitively, we can define the quantity:
\begin{equation}
\label{Teffdef}
T_{\textrm{eff}} \equiv \Big \langle \frac{1}{T} \Big \rangle^{-1},
\end{equation}
which can be thought of as the \emph{effective foreground temperature} of the sky after we have taken advantage of our angular sensitivity to downweight heavily contaminated regions.  In Figure \ref{recipStats} we show this quantity as a function of angular resolution at $79\,\textrm{MHz}$ (the behavior at different frequencies is qualitatively similar).  This is shown along with the spatially averaged foreground temperature, which is the relevant temperature scale for experiments with no angular sensitivity, and the foreground temperature in the coolest pixel of the sky, which will be useful later in Section \ref{excludePixels}.  As one expects from the preceding discussion, the effective foreground temperature is strictly less than the averaged temperature, demonstrating that residual foreground errors can be reduced by downweighting heavily contaminated regions of the sky in our analysis, which is only possible if one has sufficiently fine angular resolution.  The benefits become increasingly pronounced as one goes to finer and finer angular resolution.  Intuitively, this occurs because the finer the angular resolution, the more ``surgically" small regions of heavy contamination (\emph{e.g.} bright point sources) can be downweighted.

\subsubsection{Using spatial structure to perform angular foreground subtraction}

Now consider the second potential use of angular information, namely, to perform spatial foreground subtraction by taking advantage of the angular structure or correlations of foregrounds.  The extra information content that would be provided by such a procedure is represented by the second term of Equation \eqref{decomposedInfo}, where the detailed spatial properties of the foregrounds enter in the quantity $C_\ell^u$.

\begin{figure*}
\centering
\includegraphics[width=1.0\textwidth]{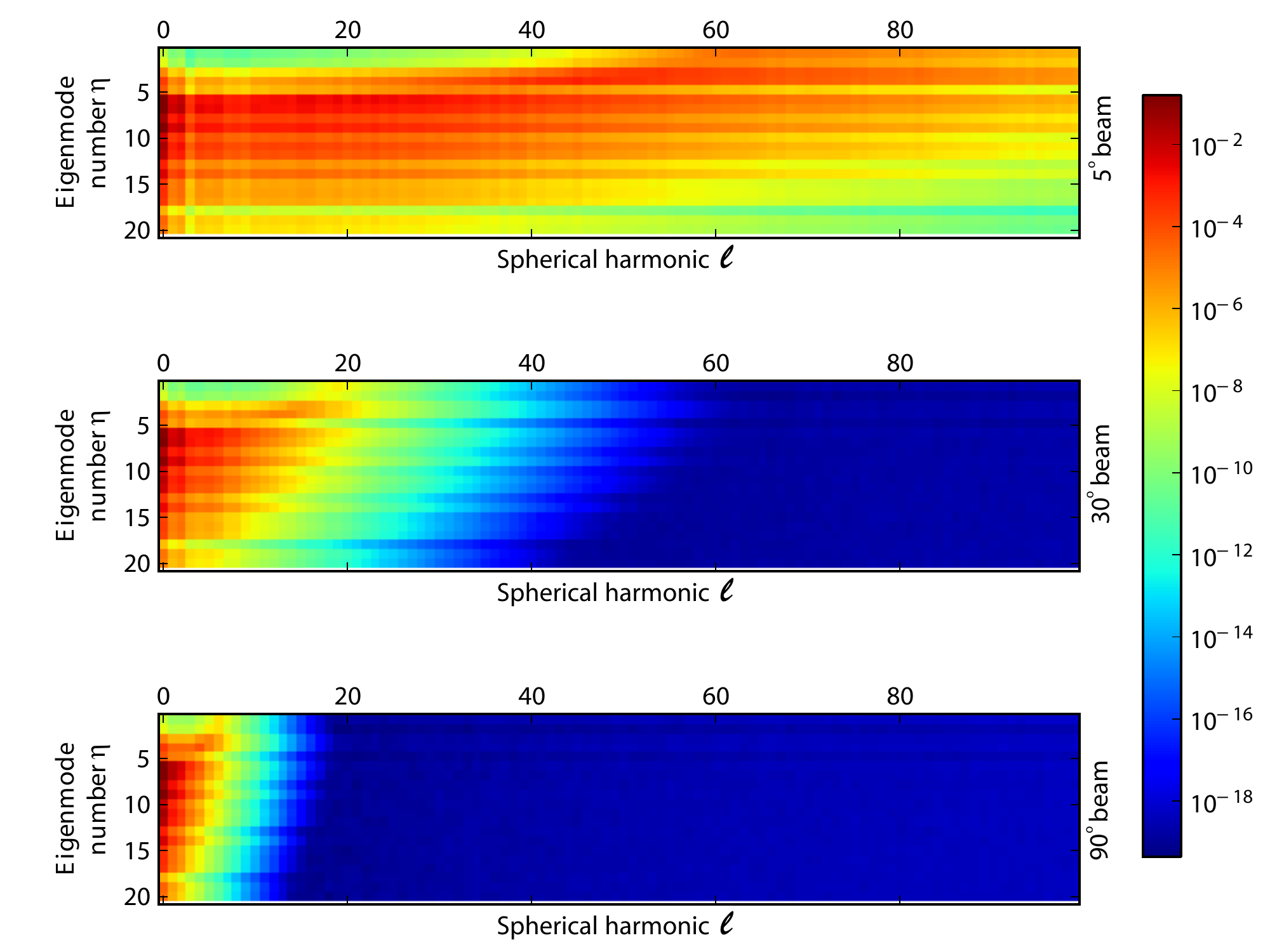}
\caption{Contributions to the detection significance statistic $\gamma^2$, as a function of the foreground spectral eigenmode number $\eta$ (vertical axis, increasing downwards) and spherical harmonic $\ell$ (horizontal axis, increasing to the right).  Top panel: instrumental beam with FWHM of $5^\circ$; middle panel: $30^\circ$; bottom: $90^\circ$.  The plots are normalized to give unity when summed over all $\eta$ and $\ell$.  In all three cases, almost all of the detection significance comes from the $\ell=0$ modes beyond the first few $\eta$ modes (note the logarithmic color scale).  As expected, the dominance of low $\ell$ modes is more pronounced for experiments with low angular resolution.}
\label{SigmasEtaEllDarkAges}
\end{figure*}

To quantify the added benefit of including angular correlations, we compute the quantity
\begin{equation}
\label{gammaDef}
\gamma \equiv \left( \mathbf{x}^t_s \boldsymbol \Sigma^{-1} \mathbf{x}_s \right)^{\frac{1}{2}},
\end{equation}
which can be thought of as the ``number of sigmas" with which the cosmological signal $\mathbf{x}_s$ can be detected.  Since Equation \eqref{decomposedInfo} gives us $\boldsymbol \Sigma^{-1}$ in a form that is decomposed  both in angular scale $\ell$ and foreground mode number $\eta$, we can express $\gamma^2$ in the same decomposition.  This is shown in Figure \ref{SigmasEtaEllDarkAges} for instrumental beams with FWHMs of $5^\circ$ (top panel), $30^\circ$ (middle panel), and $90^\circ$ (bottom panel).  Several trends are immediately apparent.  First, in all cases the detection significance is concentrated in the low $\ell$ modes, in particular the $\ell = 0$ mode (note the logarithmic color scale).  As expected, the use of higher $\ell$ information to boost detection significance is more prevalent when the instrumental beam is narrow.  It is also more prevalent for the first few (\emph{i.e.} spectrally smoother) foreground eigenmodes.

To gain intuition for this we can examine the second term of Equation \eqref{decomposedInfo}.  At high $\ell$, this term is exponentially suppressed by the $\exp \left[\theta_b^2 \ell (\ell+1)\right]$ factor in the denominator, a result of our instrument having finite angular resolution.  At low $\ell$, this behavior is counteracted by the other term in the denominator, which has the opposite behavior with $\ell$ and is proportional to the strength of the $\eta^{\textrm{th}}$ foreground mode as quantified by the eigenvalue $\lambda_\eta$.  A conservative estimate for $\ell_{\textrm{max}}$, the largest $\ell$ mode for which there may be significant information content, can be obtained by equating these two terms, since at higher $\ell$ the noise term suppresses the information content.  This estimate is a conservative one, for the foreground correlations themselves (encapsulated by $C_\ell^u$) are almost always a decreasing function of $\ell$, pushing the peak of useful information content to lower $\ell$.  Ignoring this complication to obtain our rough estimate, we have
\begin{equation}
\label{lmax}
\ell_{max} \sim  \sqrt{ \frac{1}{\theta_b^2+\frac{1}{2}\sigma^2} \ln \left( \frac{\varepsilon_0^2 \theta_{\textrm{fg}}^2 t_{\textrm{int}} \Delta \nu}{4 \pi}\lambda_\eta\right)+\frac{1}{4}}\,-\,\frac{1}{2}.
\end{equation}
Several conclusions can be drawn from this.  First, we see that quantities such as $\varepsilon_0$ appear under the square root of the logarithm.  Thus, the dependence of $\ell_{max}$ on most model parameters is extremely weak, and so our qualitative conclusion that most of the detection significance comes from low $\ell$ should be quite robust.  The exception to this is the dependence on $\lambda_\eta$, which we saw from Figure \ref{Qeigenvals} decays exponentially.  We therefore expect a non-negligible decrease in $\ell_{max}$ as one goes to higher and higher foreground eigenmodes, a trend that is visually evident in Figure \ref{SigmasEtaEllDarkAges}.  Intuitively, the higher foreground eigenmodes are an intrinsically small contribution to the contamination, so they are difficult to measure with high enough signal-to-noise to be deemed sufficiently ``trustworthy" for foreground mitigation purposes.  In contrast, the first few foreground eigenmodes are easy to measure, but this is of course only because they were a large contaminating influence to begin with.  From Equation \eqref{lmax}, we can see that extra integration time does allow the weak foreground modes to be measured sufficiently well for their fine angular features to be used in foreground removal, but as we suggested above, the logarithmic dependence on $t_{\textrm{int}}$ makes this an expensive strategy for experiments with coarse (high $\theta_b$) beams.

\begin{figure}
\centering
\includegraphics[width=0.5\textwidth]{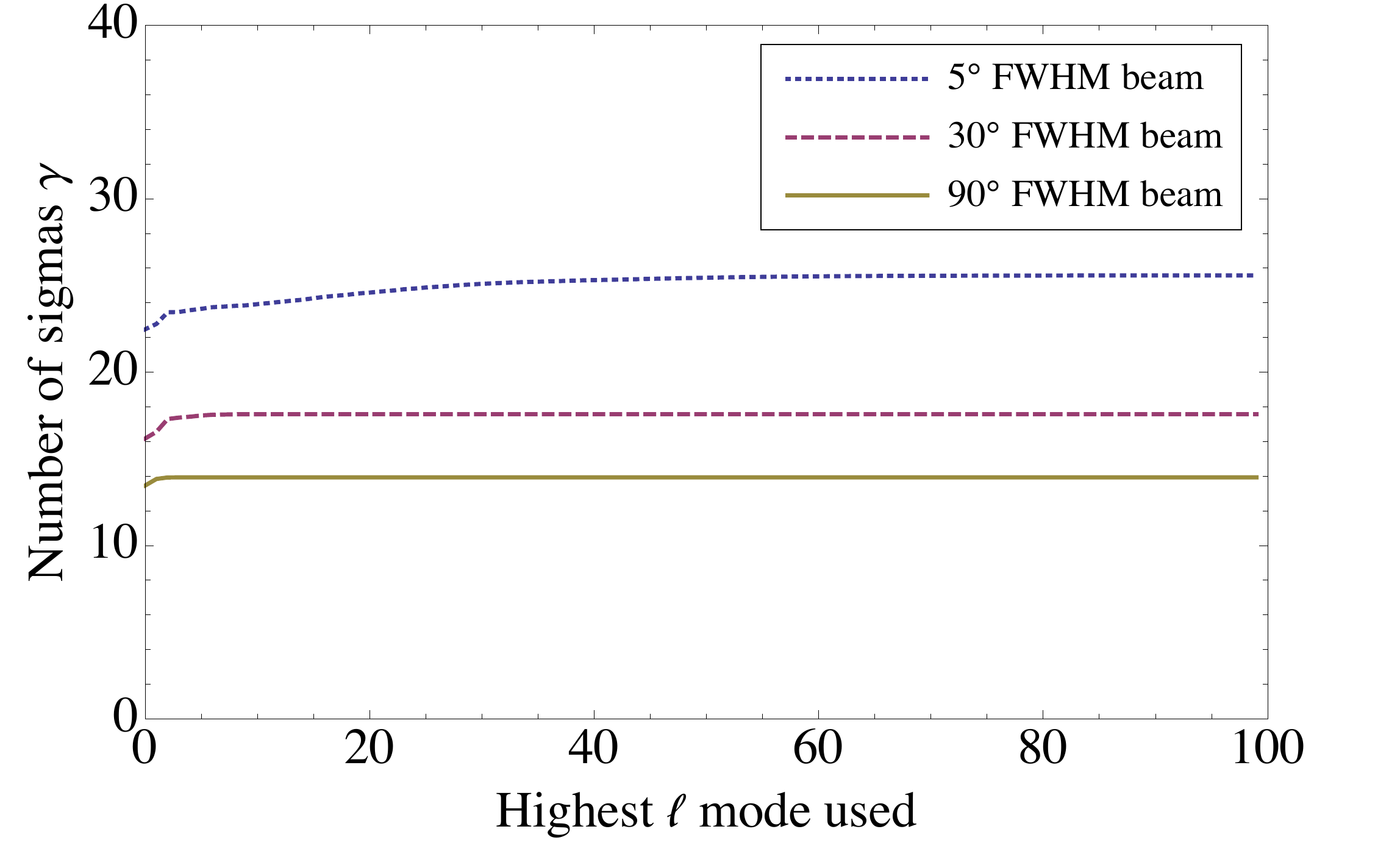}
\caption{Detection significance $\gamma$ for experiments of different levels of angular resolution, plotted as a function of various $\ell$ cutoffs, beyond which the angular correlation information is not used for foreground mitigation.  In all cases, the detection significance is high, which we shall see in Section \ref{spectral} is a result of spectral signatures.  Since the curves in this figure are essentially flat, one may initially conclude that there is little added benefit to using angular information when dealing with foregrounds.  However, as Figure \ref{errorsVsEllCutoff} shows, this conclusion is incorrect.}
\label{sigmasEllCutoff}
\end{figure}

Based on our discussion so far, one might be tempted to conclude that it is unnecessary for $21\,\textrm{cm}$ global signal experiments to take advantage of angular resolution at all.  Another argument for this is presented in Figure \ref{sigmasEllCutoff}, where we once again compute our significance statistic $\gamma$, but this time vary the number of $\ell$ modes used in our angular foreground subtraction.  In other words, to produce Figure \ref{sigmasEllCutoff} we took Equation \eqref{decomposedInfo} and performed the sum of $\ell$ only to some cutoff to produce a truncated information matrix.  One sees that taking advantage of angular correlations in our foreground model does very little to boost the detection significance.  As far as this statistic is concerned, most of the gain in having angular information comes from the effect discussed in Section \ref{downweighting}, namely the downweighting of heavily contaminated regions to suppress foregrounds before we average.  This is responsible for the way the finer angular resolution experiments have higher $\gamma$ in Figure \ref{sigmasEllCutoff} even if the higher $\ell$ modes are not used for angular foreground mitigation at all.

The simple downweighting of contaminated regions, however, is a strategy that is of limited power.  Figure \ref{recipStats} tells us that even with $5^\circ$ beams, we can expect at most a factor of $2$ mitigation in foreground contamination (compared to an experiment with no angular resolution at all).  Thus, one may argue that angular resolution is simply not worth the financial cost or the instrumental challenges, since Figure \ref{sigmasEllCutoff} shows that a statistically comfortable detection can be made even without angular resolution.  As we shall see in Section \ref{spectral}, the high statistical significance seen in Figure \ref{sigmasEllCutoff} is the result of \emph{spectral} methods, which may lead one to abandon angular methods entirely.

\begin{figure}
\centering
\includegraphics[width=0.48\textwidth]{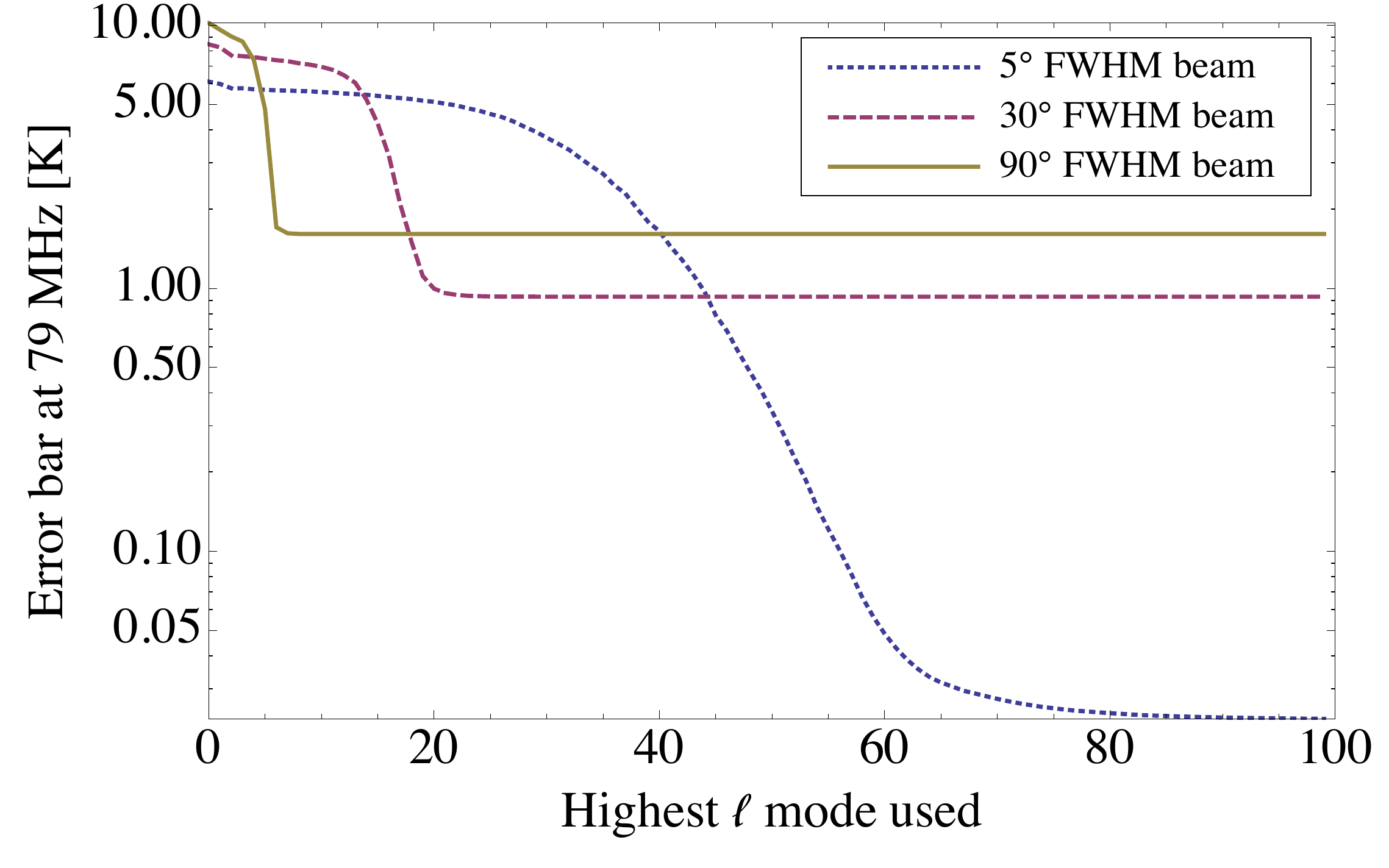}
\caption{Error bars on the measured spectrum at $79\,\textrm{MHz}$, shown as a function of various $\ell$ cutoffs, beyond which the angular correlation information is not used for foreground mitigation.  Excluding any information of angular correlations in the analysis, the errors are highest for the experiments with coarse angular resolution, because such experiments are less able to downweight heavily contaminated regions of the sky, as discussed in Section \ref{downweighting}.  The situation reverses at moderate cutoffs in $\ell$, and the coarser experiments have an advantage because higher $\ell$ modes of the foregrounds were never measured in the first place.  However, information in these unmeasured modes are lost forever, and cannot be used for angular subtraction, unlike in experiments with fine angular resolution.  At high cutoffs in $\ell$, where one is using all available information, fine angular resolution once again wins, giving the smallest error bars.}
\label{errorsVsEllCutoff}
\end{figure}

Such a conclusion, however, would be a misguided one.  For while $\gamma$ (the ``number of sigmas" of our detection) is a useful and commonly quoted statistic, it is somewhat crude in that it only tells us whether we are seeing a significant departure from our residual noise and foreground models.  In other words, it tells us that if the emission from the sky did in fact take the form of foregrounds plus the fiducial global signal assumed in Section \ref{signalmodel}, we would be able to detect \emph{something} above pure foregrounds and noise in our measurements.  It does not tell us \emph{what} global signal has been detected, or whether our final result is indeed correct.

More precisely, suppose that in addition to Figure \ref{sigmasEllCutoff}, we also consider the error bars $(\boldsymbol \Sigma_{\alpha \alpha})^{1/2}$ on our measured frequency spectrum.  This is shown in Figure \ref{errorsVsEllCutoff}, where we show the error bars at $79\,\textrm{MHz}$, again as a function of the maximum $\ell$ mode used for angular foreground mitigation\footnote{For Figures \ref{errorsVsEllCutoff} and \ref{errorsVsEllCutoffReion} \emph{only}, we make the approximation that the angular cross-power spectrum between frequency channels $\alpha$ and $\beta$, $C_{\ell}^{u,\alpha \beta}$, is a separable function of the two frequencies.  There is no \emph{a priori} reason to expect this to be the case, but we find that it is an excellent approximation.  We make use of this approximation for purposes of numerical stability.  In any case, it is a conservative assumption, for a non-separable cross-correlation (which might occur, for instance, if different foreground sources dominate at different frequencies) only provides \emph{more} information with which to subtract foregrounds.}.  In this case, we see large changes as we vary our $\ell$ cutoff.  In all cases, the errors go down as more and more angular correlation information is used.  At the lowest $\ell$, the finer angular resolution experiments do better because they are able to downweight foreground contaminated regions in a more spatially precise way (as we discussed in Section \ref{downweighting}).  Going to moderate $\ell$ cutoffs, the coarser experiments do better because they sampled a blurred version of the sky to begin with, and thus do not need to go to as high $\ell$ in their data analysis to make the most of their angular correlation information.  However, the blurry measurements do not possess the fine details needed perform a high-precision angular foreground subtraction, so at higher $\ell$ cutoffs, the fine angular resolution experiments give the lowest errors.

Figures \ref{sigmasEllCutoff} and \ref{errorsVsEllCutoff} together tell us that even though we do not need angular correlation information to tell that our measurement contains something more than just pure noise and foregrounds, the error bars are large enough (being in the $5$ to $10\,\textrm{K}$ range at $\ell =0$) for our results to be consistent with a wide variety of theoretical spectra.  Since we expect the cosmological signal to have a maximum amplitude of $\sim 0.1 \,\textrm{K}$, such error bars make it impossible to use our measurements to distinguish between different models.

In short, we conclude that the use of angular correlations is crucially necessary, not for the purposes of a mere detection of \emph{something} above the noise and foreground model, but to further reduce error bars to the point where interesting constraints can be placed on theoretical models.

\subsubsection{Excluding heavily contaminated regions from observations}
\label{excludePixels}

While we argued in the previous section that making use of angular correlations is crucial for bringing down residual foregrounds in one's error bars, in this section we briefly consider an alternative: selectively discarding  heavily contaminated foregrounds from one's analysis entirely (or equivalently, simply not observing the dirtiest portions of the sky, such as the Galactic plane).  By excluding the dirtiest foreground pixels one can reduce the effective foreground temperature below $\langle 1/T \rangle^{-1}$, bringing one closer to the \emph{minimum} foreground temperature in the sky, plotted as the $T_{\textrm{min}}$ curve in Figure \ref{recipStats}.  In this section, we examine whether the exclusion of dirty sky regions does indeed produce smaller error bars in one's final measurement.

We begin by considering just the $\ell=0$ term in our expression for the information matrix [Equation \eqref{decomposedInfo}].  Again, this is equivalent to omitting angular correlation information from our analysis\footnote{In this section we are considering a scenario where angular correlations are not used.  One could imagine alternatively a situation where we not only discard (or do not observe) the most heavily contaminated regions, but in addition also take advantage of angular correlations between the remaining pixels.  For the purposes of this paper, we do not analyze this scenario, for the necessary correlation models that this would entail involves additional experiment-specific details such as an experiment's scanning strategy.}.  Discarding all but the first term in the sum over $\ell$, we can simplify our expression for the information matrix using the spectral coherence matrix $\mathbf{Q}$:
\begin{equation}
\label{simpleInfo}
\mathbf{\Sigma}^{-1}_{\textrm{mono}} = \mathbf{T}^{-1} \left(\frac{\varepsilon_0^2  \theta_{\textrm{fg}}^2}{f_{\textrm{sky}} \Omega_{\textrm{sky}}} \mathbf{Q} +  \frac{\mathbf{I}}{t_{int} \Delta \nu}\right)^{-1} \mathbf{T}^{-1},
\end{equation}
where we have used the fact that $\mathbf{Q}_{\alpha \beta} = \sum_\eta (\mathbf{v}_\eta)_\alpha (\mathbf{v}_\eta)_\beta \lambda_\eta$, and have made the substitution $4\pi \!\!\rightarrow\!\! f_{\textrm{sky}} \Omega_{\textrm{sky}}$ (that this is an appropriate step can be verified by a more detailed derivation of $\boldsymbol \Sigma^{-1}$ that assumes incomplete sky coverage from the beginning).  We have also defined $\mathbf{T}^{-1}$ as a diagonal matrix containing the $\langle 1/T \rangle$ values, \emph{i.e.}
\begin{equation}
\mathbf{T}^{-1}_{\alpha \beta} \equiv \delta_{\alpha \beta} \Bigg \langle \frac{1}{T(\nu_\alpha)} \Bigg \rangle.
\end{equation}
From all this, we can easily compute the error bars at a specific frequency, which are given by $\mathbf{\Sigma}_{\alpha \alpha}^{1/2}$.  Expressing this in terms of our definition of the effective foreground temperature [Equation \eqref{Teffdef}], we have
\begin{equation}
\label{simpleError}
\Big{(} \mathbf{\Sigma}_{\alpha \alpha}^{\textrm{mono}} \Big{)}^{1/2} \approx T_{\textrm{eff}} (\nu_\alpha) \sqrt{\frac{\varepsilon_0^2 \theta_{\textrm{fg}}^2}{f_{\textrm{sky}} \Omega_{\textrm{sky}}}  +  \frac{1}{t_{int} \Delta \nu}},
\end{equation}
where we have taken advantage of the fact that $\mathbf{Q}_{\alpha \alpha} = 1$.  This equation reveals that our measurement error consists of two terms added in quadrature.  The first is from residual foregrounds, and scales as $1/\sqrt{f_{\textrm{sky}}}$, since using more pixels allows us to average down the error in our foreground model.  The second is from instrumental noise.  It depends on the bandwidth $\Delta \nu$ and the total integration time $t_{int}$, but not on the number of pixels $f_{\textrm{sky}}$.  This is because our goal is to measure a cosmological monopole, which means the instrumental noise is averaged down at the same rate regardless of whether all the time is spent integrating on a small patch of the sky or the observation time is spread over the entire sky.  The entire expression is modulated by the effective foreground temperature.

Armed with Equation \eqref{simpleError}, we can answer the question of whether it is wise to spend our integration time observing only the cleanest regions in the sky.  Suppose we took the pixels in our sky and sorted them from least foreground contaminated to most foreground contaminated.  One can then imagine (excluding practical logistics such as scanning strategy) observing in only the $f_{\textrm{sky}}$ cleanest portions of the sky.  The effective foreground temperature $T_{\textrm{eff}}$ is an increasing function of $f_{\textrm{sky}}$, since it increases from $T_{\textrm{min}}$ to $\langle 1/T \rangle^{-1}_{\textrm{full sky}}$.  Thus, if the measurement errors were determined by $T_{\textrm{eff}}$ alone, it would be best to spend all our integration time on the single cleanest pixel of the sky.  However, we can see in Equation \eqref{simpleError} that there is a competing influence: the $\sqrt{\frac{\varepsilon_0^2 \theta_{\textrm{fg}}^2}{f_{\textrm{sky}} \Omega_{\textrm{sky}}} +  \frac{1}{t_{int} \Delta \nu}}$ is a decreasing function of $f_{\textrm{sky}}$, because sampling more of the foreground sky allows one to average down the errors in our foreground model.

\begin{figure}
\centering
\includegraphics[width=0.5\textwidth]{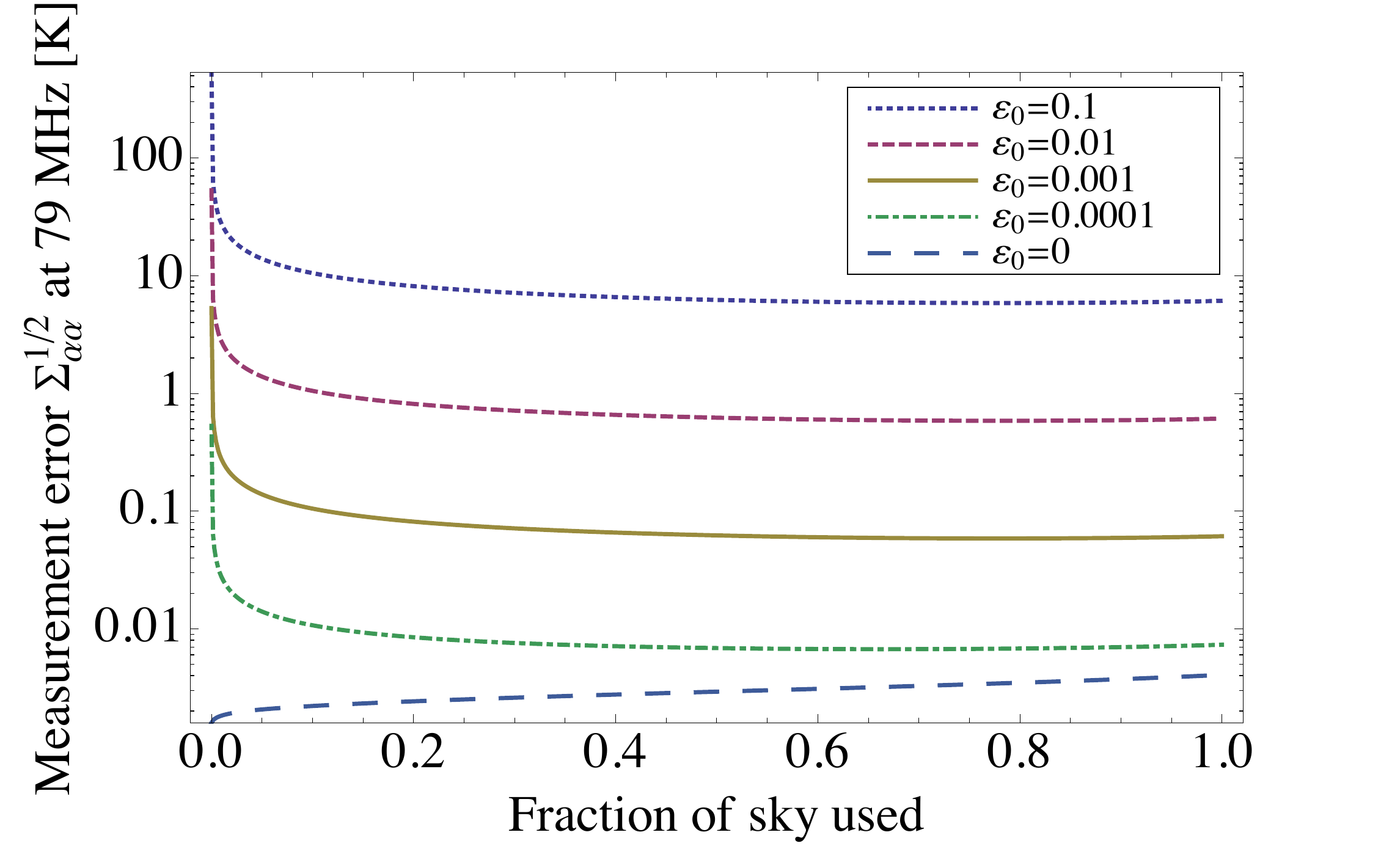}
\caption{Expected error bars on a global signal measurement at $79\,\textrm{MHz}$, as a function of the fraction of the sky observed.  The observations are assumed to be of the cleanest parts of the sky (\emph{e.g.} a fraction of $0.3$ means observing the cleanest $30\%$ of pixels, even if they are not contiguous) and are simulated at a resolution of $5\,\textrm{degrees}$.  Different curves denote different values of the fiducial foreground model error parameter $\varepsilon$.  At any realistic level of foreground modeling error, it is advantageous to cover as much of the sky as possible to reduce the modeling error, even if it means dealing with a higher effective foreground temperature.}
\label{discardingPixels}
\end{figure}

In Figure \ref{discardingPixels} we examine this trade-off, where we show the expected error bars on a global signal measurement at $79\,\textrm{MHz}$, as a function of the fraction of the sky observed.  The observations are assumed to include only the cleanest parts of the sky.  For instance, a coverage fraction of $0.3$ means the observations include only the cleanest $30\%$ of pixels.  From the plot, one can see that for any reasonable level of fiducial foreground modeling error $\varepsilon_0$, the errors are minimized by including as much of the sky as possible, even if that means sampling some highly contaminated foreground regions.  In words, this is a statement that the foreground modeling errors average down more quickly than the foreground amplitudes rise as we go to dirtier parts of the galaxy.  Mathematically, the instrumental noise term turns out to be negligible compared to the residual foreground error term.  The second piece of Equation \eqref{simpleError} therefore ends up scaling as $1/\sqrt{f_{\textrm{sky}}}$, which happens to decay more quickly than $T_{\textrm{eff}}$ grows with $f_{\textrm{sky}}$.  The net result is that overall measurement error is reduced by covering as much of the sky as possible, although in practice the gains appear to be minimal beyond a $30\%$ coverage of the sky.  The trend of decreasing error with increasing sky coverage changes only if the foreground modeling error is unphysically low (\emph{e.g.} the $\varepsilon_0=0$ curve in Figure \ref{discardingPixels}).  Only then is it advantageous to avoid observing the dirtier parts of the sky\footnote{While Figure \ref{discardingPixels} was produced using simulations at a $5^\circ$ resolution, we expect our conclusion---that one should cover as much of the sky as possible, even if it means observing regions that are highly contaminated by foregrounds---to hold at other relevant resolutions.  Finer resolutions are unlikely to be achievable at such low frequencies, and at coarser resolutions the act of discarding pixels does not reduce the effective foreground temperature by very much (as one can see from Figure \ref{recipStats}).  The benefits from going to higher sky coverage fraction are thus even more pronounced.}.
%

\subsection{The role of spectral information}
\label{spectral}

In the previous section, we examined the extent to which angular information can reduce the error bars on a global signal measurement.  In this section, we highlight the crucial role that spectral information plays.

Recall from Figure \ref{sigmasEllCutoff} that even an experiment with a FWHM instrumental beam that is as wide as $90^\circ$ can make a statistically significant ($\sim 14\sigma$) detection of a signal above residual noise and foregrounds.  Since a $90^\circ$ beam is essentially equivalent to having no angular information at all, it follows that the statistical significance of such a detection must come from spectral information.  To see this clearly, let us express our measurement covariance and the cosmological signal in a basis where the covariance is diagonal.  Intuitively, this can be thought of as a ``residual noise and foregrounds eigenbasis".  Such a basis is convenient because without off-diagonal elements, the error bars (given by the square root of the diagonal elements of the covariance) can be directly compared to the signal to arrive at a signal-to-noise ratio.

\begin{figure}
\centering
\includegraphics[width=0.45\textwidth]{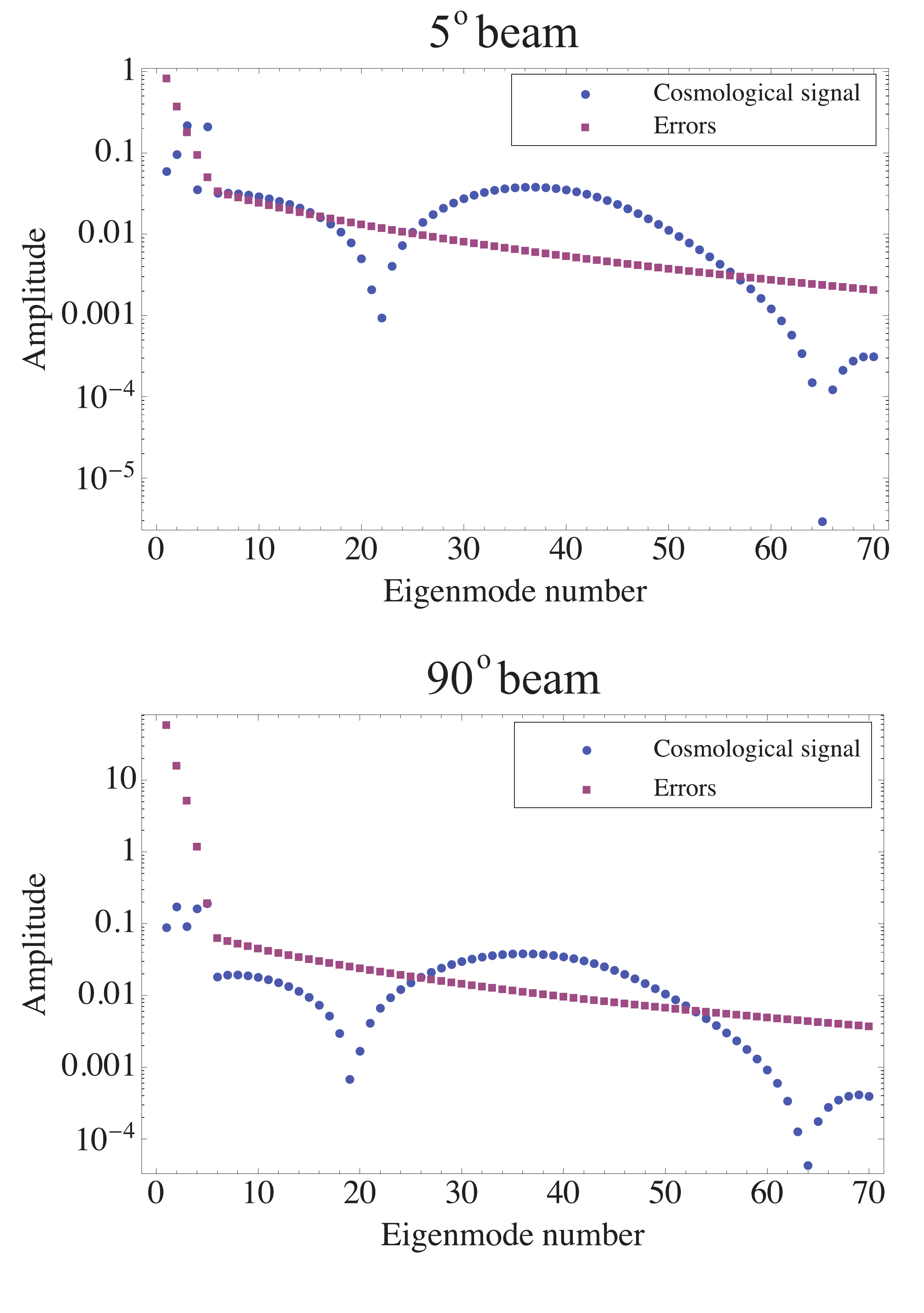}
\caption{Absolute value of the theoretical cosmological signal (blue circles) and predicted error bars (purple squares), shown in a residual foreground and noise eigenmode basis for an experiment with a FWHM instrumental beam of $5^\circ$ (top panel) and a FWHM of $90^\circ$ (bottom panel).  In the $90^\circ$ case, high signal-to-noise detections of the cosmological signal are possible in a small handful of modes, and it is these modes that provide all the detection significance.  However, the inability to constrain any other modes results in large error bars in the final measured global signal spectrum.  With a $5^\circ$ beam, our angular foreground subtraction methods allow more spectral modes to be measured at high signal-to-noise, contributing to a more faithful reconstruction of the shape of the global spectrum, with smaller error bars.  Note that the cosmological signal in the two panels of this figure look slightly different because the measurement covariances change when one changes the beam size, which in turn changes our eigenbasis.}
\label{combinedEigenSNRplot}
\end{figure}

In the bottom panel of Figure \ref{combinedEigenSNRplot}, we show the absolute amplitude of the theoretical signal in our residual noise and foreground basis for an instrument with a $90^\circ$ beam.  Also on the plot is the square root of the (now diagonal) covariance, \emph{i.e.} the error bars on our measurement in this basis.  Still focusing on just the bottom panel for now, we note several features.  As expected, the error bars show an exponential decay over the first few modes, corresponding to an error budget that is dominated by residual foregrounds.  In such a regime, the cosmological signal is completely sub-dominant.  As one goes to higher modes, the errors are determined by a more balanced mixture and noise and foregrounds, and the decay is less rapid.  From this figure, we have a simple explanation for why it is possible to have a high detection significance in spite of large error bars in our frequency spectrum---plotting our measurement as a function of frequency represents a bad choice of basis, with foregrounds dominating every frequency channel.  With a better basis choice, it is clear that there are a small number of modes (eigenmodes $\sim 25$ to $\sim 50$) that can be measured with signal-to-noise ratio greater than unity, and it is these modes that are providing all the detection significance.  However, our inability to measure any of the other modes limits our ability to accurately constrain the shape of the global signal spectrum.

\begin{figure}
\centering
\includegraphics[width=0.45\textwidth]{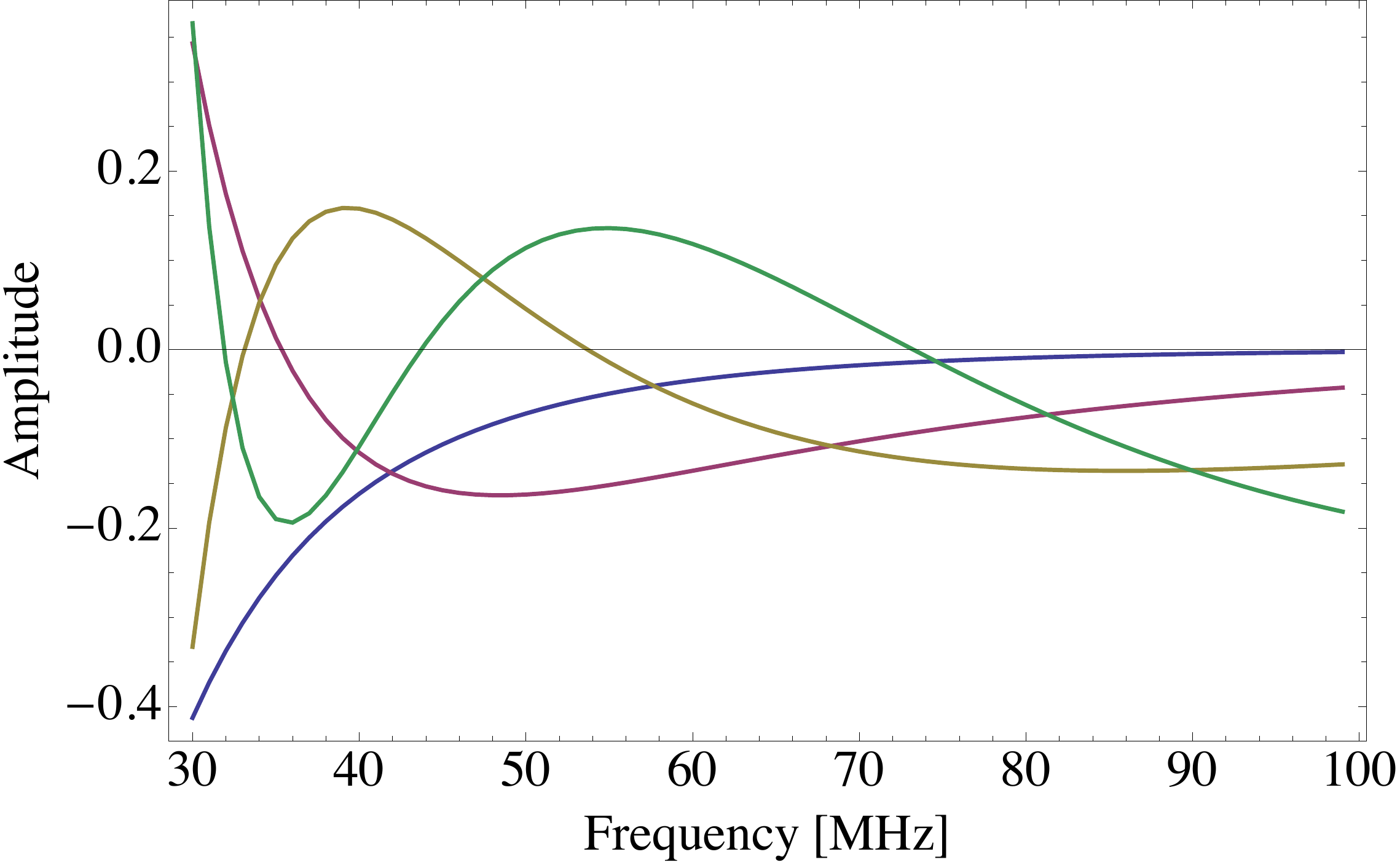}
\caption{First few (foreground residual dominated) eigenmodes of the measurement covariance $\boldsymbol \Sigma$ for an experiment with an instrumental beam of FWHM $90^\circ$.}
\label{badResidFG}
\end{figure}

The difficulty in constraining the shape of the spectrum can be further understood by examining the eigenvectors corresponding to our residual noise and foreground eigenmodes.  In Figure \ref{badResidFG} we show the first few eigenmodes, which from Figure \ref{combinedEigenSNRplot} we know are essentially unmeasurable.  These eigenmodes are all spectrally smooth, and thus any such patterns in the measured spectrum will be lost in the measurement.  In the language of our residual noise and foreground basis, these modes are very poorly constrained, and so our experiment will be unable to tell the difference between two cosmological scenarios that differ only in these unmeasurable modes.  Given that the cosmological signal is expected to be reasonably smooth spectrally, being unable to measure these smooth modes is problematic, for it means that there will be a wide class of reasonable cosmological models that our experiment will not be able to tell apart.  Mathematically, this is why we found that the error bars on the final measured spectrum were large for experiments with no angular resolution.  Such experiments rely too heavily on spectral information for their foreground subtraction, and since both the foregrounds and global cosmological signal are rather smooth, a sharp constraint on the signal is very difficult.

\begin{figure}
\centering
\includegraphics[width=0.45\textwidth]{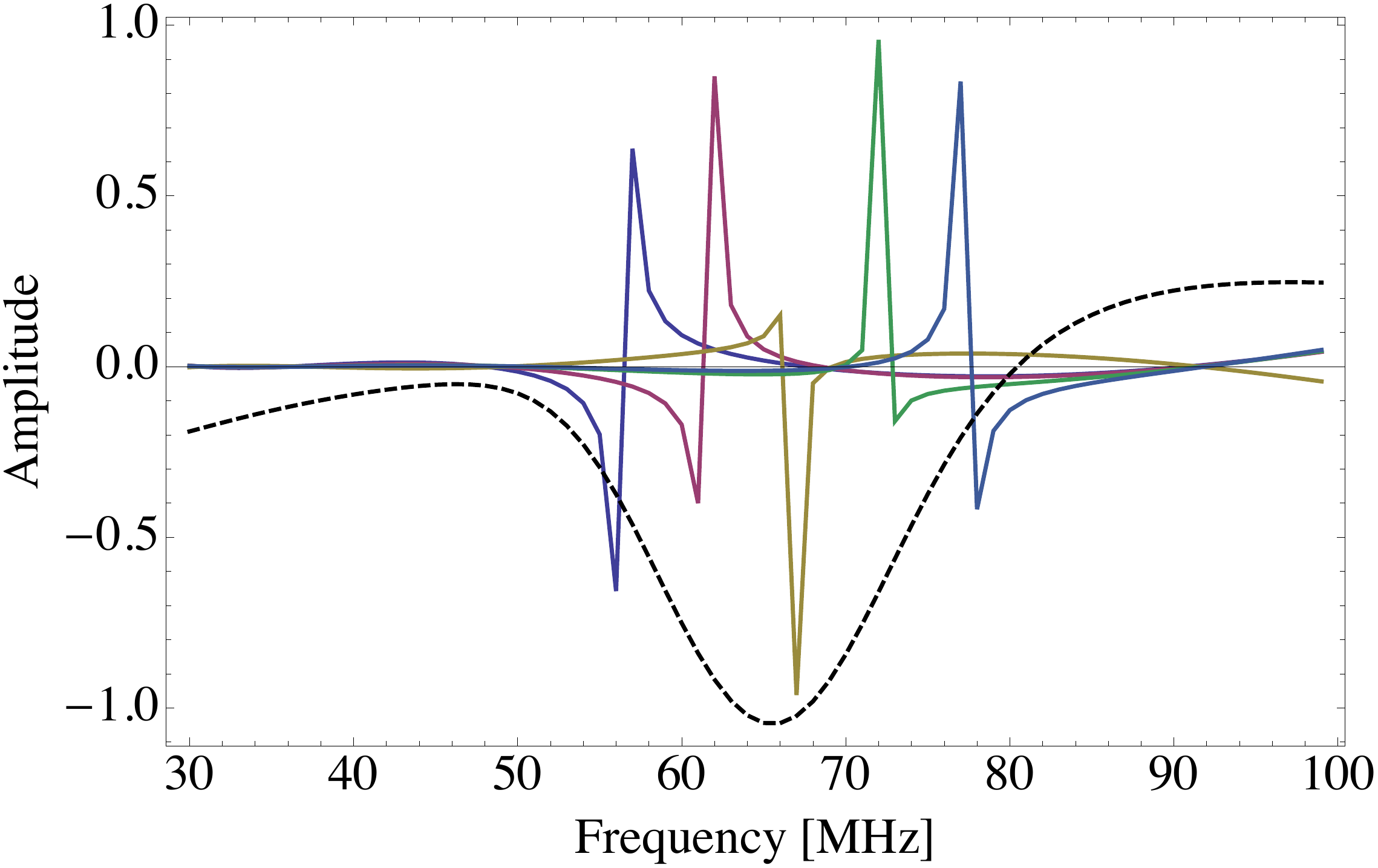}
\caption{The $30^{\textrm{th}}$, $35^{\textrm{th}}$, $40^{\textrm{th}}$, $45^{\textrm{th}}$, and $50^{\textrm{th}}$ eigenmodes of the measurement covariance $\Sigma$, all of which are modes that can be measured with signal-to-noise ratio greater than one.  The computations were performed for an experiment with a FWHM instrumental beam of $90^\circ$.  Shown in dotted black is our fiducial cosmological signal, arbitrarily rescaled in amplitude.}
\label{90degGoodResidFG}
\end{figure}

In Figure \ref{90degGoodResidFG}, we show several of the eigenmodes that in contrast \emph{can} be reasonably well-measured.  Immediately striking is the fact that these eigenmodes are all localized in the trough in the cosmological signal between $50$ and $80\,\textrm{MHz}$.  It is therefore information from the trough that is providing most of the detection significance.  Next, we notice that many of the eigenmodes have a double-peaked structure, with one peak positive and one negative.  With such a structure, each eigenmode is probing \emph{differences} between neighboring frequencies.  Looking at differences rather than raw values has the advantage of being more immune to foreground contamination, since the foregrounds are monotonically decreasing functions of frequency, so differencing neighboring frequency channels will cancel out some of the foregrounds.  An alternative, but equivalent way to understand this heuristic argument is to think of these ``difference modes" as finite-difference versions of derivatives.  Measuring these modes is thus equivalent to measuring the derivative of the spectrum.  The algorithm is simply taking advantage of the fact that even though the foreground spectrum may be larger than the cosmological signal in amplitude, the reverse may be true of the derivatives\footnote{The fact that measurements seem to be quite sensitive to the derivative of the global signal is intriguing, for it has been suggested \cite{globalSigDM} that the derivative could be a way to distinguish between the X-ray heating from dark matter annihilations and that from more conventional astrophysical sources.}.  Of course, here the modes are only approximately derivative-like, since it is clear from Figure \ref{90degGoodResidFG} that there is an asymmetry in height and width between the positive and negative peaks.  This is an indication that the algorithm is doing a little more than just taking the derivative, but in any case, all its behaviors can be captured by calculating the eigenvectors as we have done.

\begin{figure}
\centering
\includegraphics[width=0.45\textwidth]{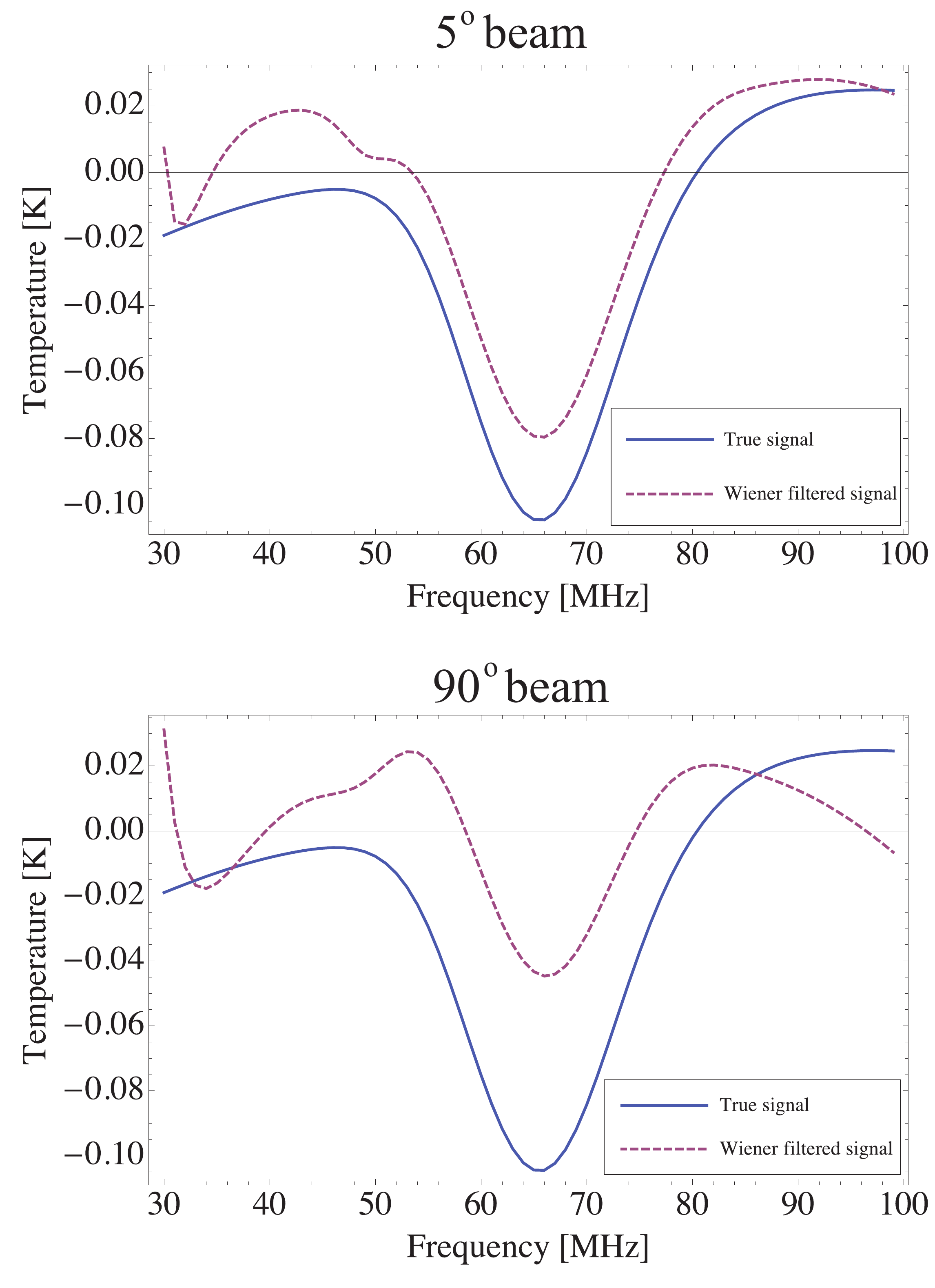}
\caption{Comparisons between the true global signal and ones that have been Wiener-filtered.  The top panel is for an experiment with an instrumental beam of of $5^\circ$ width, while the bottom panel is for one with a $90^\circ$ beam.  The Wiener filter eliminates modes that cannot be measured without an extremely accurate foreground model.  The filtered curves in this figure thus encapsulate the best spectra that one can expect to measure, given some of the unavoidable similarities between the signal and foregrounds, as viewed by an experiment with different beam widths.  With a $90^\circ$ resolution, many of the measured spectral modes are heavily contaminated, and in eliminating them from our final result, we have also washed out a non-negligible fraction of the cosmological signal.  In comparison, foregrounds can be cleaned more aggressively using angular information from a $5^\circ$ resolution experiment, and fewer features in the cosmological signal are washed out.}
\label{combinedWienerDarkAges}
\end{figure}

Figure \ref{90degGoodResidFG} may seem to suggest a strategy for extracting useful cosmological information out of an experiment that has insufficient angular resolution to produce small error bars in a measured spectrum.  Since the (relatively) high signal-to-noise eigenmodes are all concentrated in the trough of the cosmological signal, perhaps one can simply give up on information from the low signal-to-noise modes, and limit oneself to constraining properties of the trough.  Unfortunately, the trough itself is quite smooth, and thus much of its amplitude comes from the unmeasurable low signal-to-noise modes.  To see this quantitatively, we can use the Wiener filtering technique described in Section \ref{Wiener}, which (as we discussed above) is designed to automatically exclude poorly measured modes by appropriately weighting the data.  In the bottom panel of Figure \ref{combinedWienerDarkAges}, we show the expected cosmological signal, as well as the same signal after it has been Wiener-filtered appropriately for an instrument with a $90^\circ$ FWHM beam.  The filtered result represents the best we can possibly do after downweighting the heavily contaminated modes.  We see from this that in our attempt to eliminate heavily contaminated modes, we have also inadvertently destroyed a good portion of the cosmological signal.

By now, it should be clear that in the presence of foregrounds, angular information is necessary for suppressing error bars to an acceptable level.  We now turn to analyzing an experiment with angular information.  In the top panel of Figure \ref{combinedEigenSNRplot} we show the signal and the errors in a residual noise and foreground eigenbasis for an instrument with a FWHM beam of $5^\circ$.  Comparing this the bottom panel, we see that many more modes can be measured at high signal-to-noise, including some of the smoothest (lowest eigenmode number) modes.  We thus expect to be able to be able to mitigate foreground contamination without destroying as many features in the cosmological power spectrum, and indeed we can see in the top panel of Figure \ref{combinedWienerDarkAges} that the Wiener-filtered cosmological signal more accurately reflects the shape of the fiducial spectrum.



\subsubsection{Summary: the role of spectral information}

It is important to stress that in discussing the role of spectral information in this section, the result from Section \ref{spectralOnly} remains true: aside from a direct subtraction of a foreground model spectrum, purely spectral methods are formally unable to reduce errors from residual foregrounds.  Without angular information to help with foreground subtraction, one must simply hope that the foreground spectra look sufficiently different from the cosmological signal that interesting constraints on theoretical models can be placed without using modes that are known to be heavily contaminated.  Unfortunately, these modes also contain a significant fraction of the cosmological signal, and in bottom panel of Figure \ref{combinedWienerDarkAges} we saw that this meant that many of the interesting features of the cosmological signal would be washed out along with the foregrounds.  This also manifested itself in Figure \ref{errorsVsEllCutoff}, where we saw that with no angular resolution, the error bars on the final measurement ended up being unacceptably high.  In order to get small error bars (and subsequently put constraints of theoretical scenarios), one must use angular information to aid with foreground subtraction.

\subsection{Instrumental noise and integration time}

In this section, we consider the effects of varying the integration time.  Increasing the integration time decreases the instrumental noise in our measurement, and we now examine exactly how this affects our detection significance and error bars.

\begin{figure}
\centering
\includegraphics[width=0.45\textwidth]{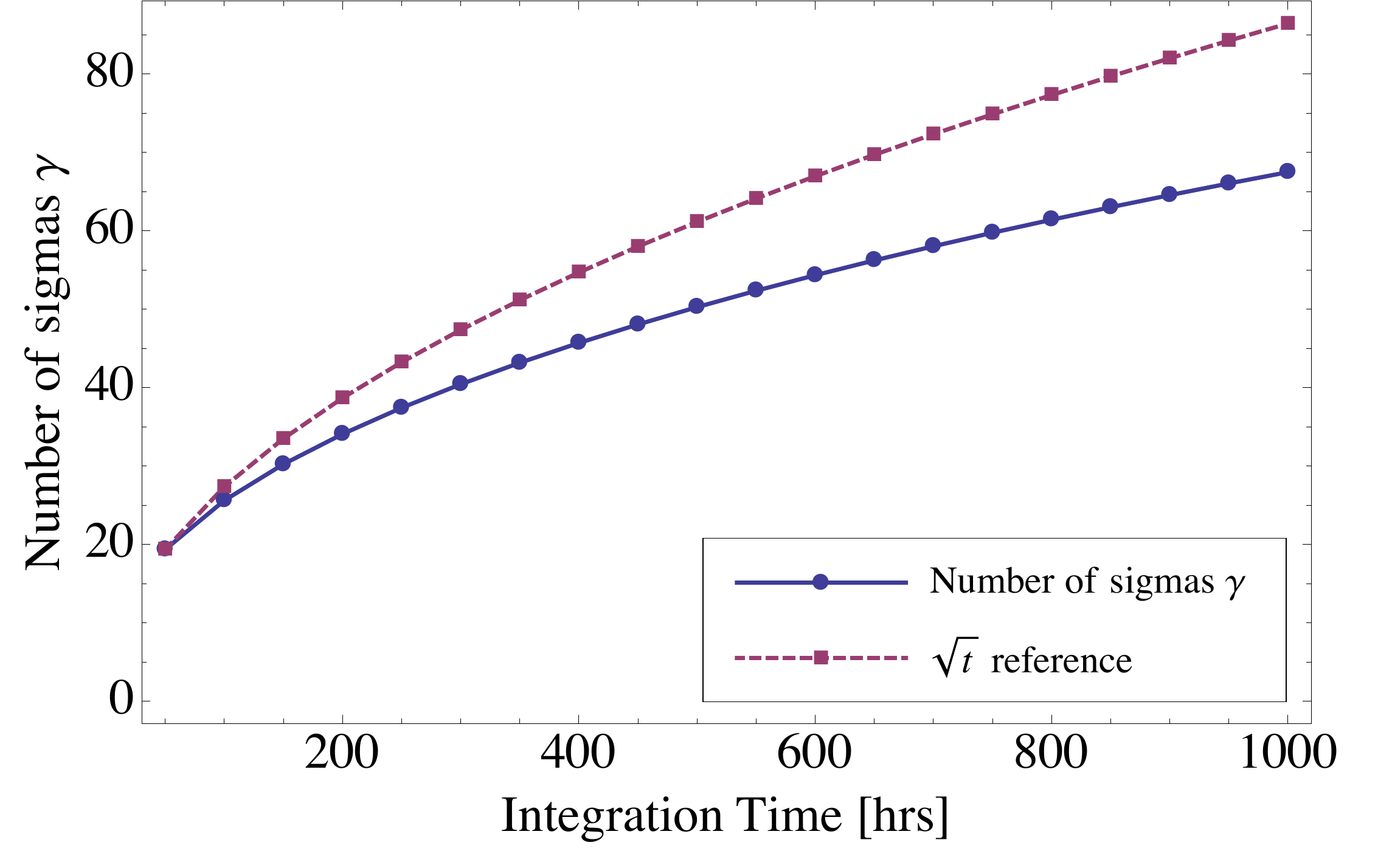}
\caption{A plot of $\gamma \equiv \sqrt{\mathbf{x}_s^t \boldsymbol \Sigma^{-1} \mathbf{x}_s}$, the ``number of sigmas" with which our fiducial cosmological signal can be detected, as a function of integration time.  The detection significance is seen to rise with greater integration, but because the errors involve residual foregrounds in addition to instrumental noise, the rise is less rapid than what would be expected from a simple $\sqrt{t}$ scaling.  A fiducial foreground model error of $\varepsilon_0=0.1$ and an instrumental beam with FWHM of $5^\circ$ was assumed to make this plot.}
\label{sigmasIntTime}
\end{figure}

In Figure \ref{sigmasIntTime}, we show the ``number of sigmas" $\gamma$ with which the cosmological signal can be detected, as a function of time.  The plot is for an experiment with a $5^\circ$ instrumental beam, but apart from a simple rescaling of the amplitude, the plot is essentially identical for wider beams.  As expected, the detection significance increases with extra integration time, although not as $\sqrt{t}$, as would be the case if our experiment were limited only by instrumental noise.  A power law fit to the curve reveals that $\gamma$ scales roughly as $t^{0.4}$, suggesting that residual foregrounds cannot be perfectly sequestered into a few eigenmodes, and affect even those high signal-to-noise spectral modes that are responsible for giving us most of our detection significance.

\begin{figure}
\centering
\includegraphics[width=0.45\textwidth]{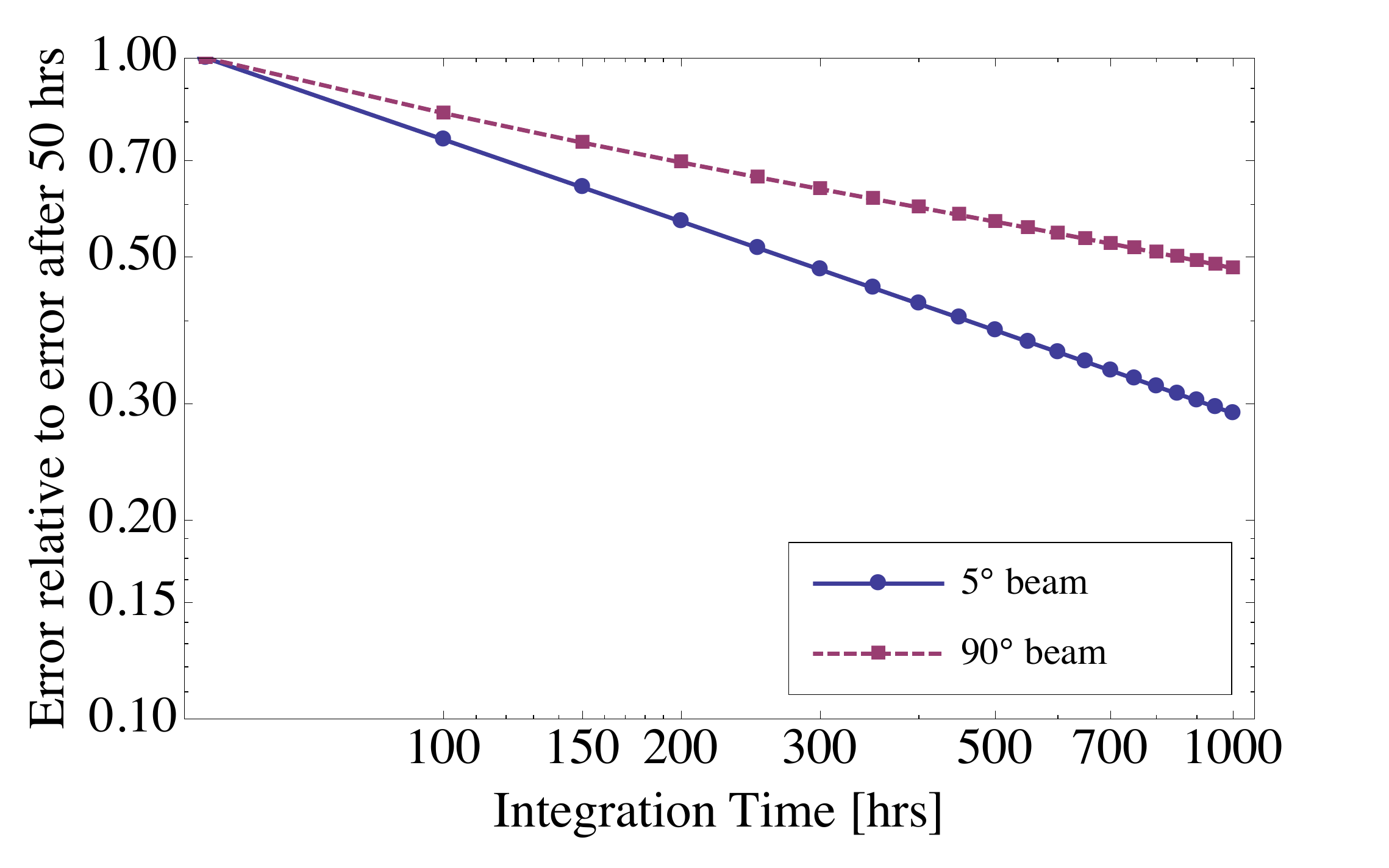}
\caption{Errors at $79\,\textrm{MHz}$ as a function of integration time, normalized to the error after $50\,\textrm{hrs}$ of integration.  The error bars are integrated down more rapidly for the experiment with the $5^\circ$ beam than the one with the $90^\circ$ beam, but both scalings are slower than $t^{-1/2}$, which would be the case if the measurements were instrumental noise dominated.  A fiducial foreground model error of $\varepsilon_0 = 0.1$ was assumed to make this plot.}
\label{errorsVsIntTime}
\end{figure}

Turning to the error bars, we find that the behavior is different for experiments with angular sensitivity compared to those without.  In Figure \ref{errorsVsIntTime} we show the measurement error at $79\,\textrm{MHz}$ as a function of time, normalized to the error of each experiment after $50\,\textrm{hours}$ of integration.  Performing fits to the curves reveals a $t^{-0.4}$ scaling for the $5^\circ$ case and a $t^{-0.24}$ scaling for the $90^\circ$ case.  Thus, in neither case do we have the $t^{-1/2}$ scaling that one would expect from pure instrumental noise, but the errors are integrated down more rapidly when there is angular information.  Intuitively, integrating for longer without angular information allows us to reduce our errors in the high signal-to-noise eigenmodes that we can already access, but does not provide us with the ability to measure new spectral modes that were limiting our ability to constrain the shape of the global signal.  The situation is different when there is angular information, because in that case the foreground residuals are sufficiently well controlled for instrumental noise reduction to produce an appreciable improvement in the error budget.  

In summary, then, we find that extra integration helps to increase the detection significance whether there is angular information or not.  However, the decrease in the error bars is more pronounced when the angular resolution is fine.  This is because in such an experiment, the spectral eigenmodes that are the most useful for constraining the shape of the global signal are not completely residual foreground dominated, and so a reduction in instrumental noise has a more significant effect on the errors.

\subsection{Reference foreground survey uncertainty}

\begin{figure}
\centering
\includegraphics[width=0.45\textwidth]{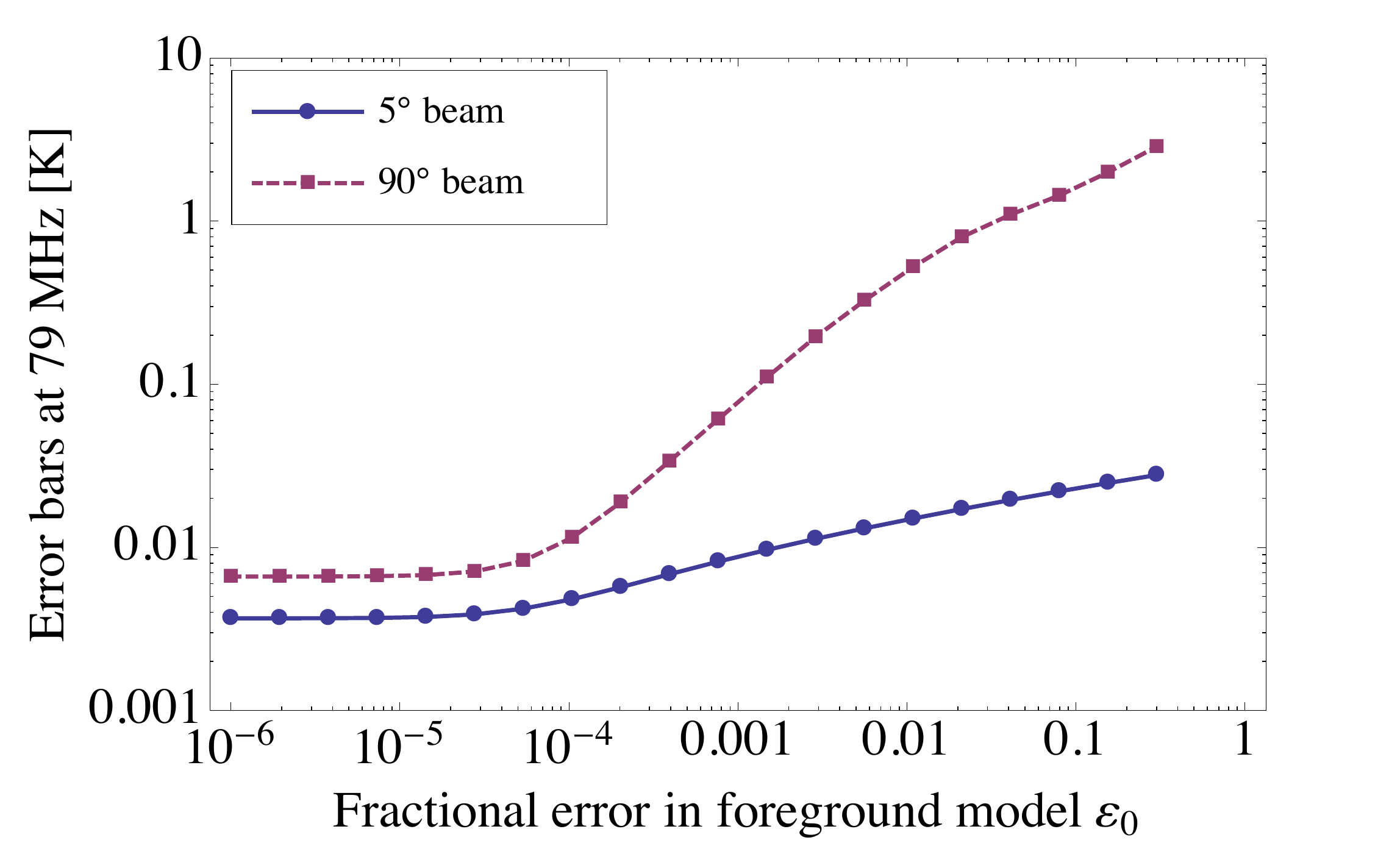}
\caption{A plot of the error at $79\,\textrm{MHz}$ in the extracted global $21\,\textrm{cm}$ signal, shown as a function of the fractional error in our foreground model $\varepsilon_0$.  At unrealistically low $\varepsilon_0$, instrumental noise is the dominant source of error, and the final errors do not vary with $\varepsilon_0$.  As we increase the foreground model errors, residual foregrounds dominate and the errors increase with $\varepsilon_0$.  An integration time of $100\,\textrm{hrs}$ was assumed to make this plot.}
\label{errorsVsEpsilon}
\end{figure}

To further our understanding of how the residual foregrounds affect our measurement, we now consider the effects of varying $\varepsilon_0$, the fractional error in our foreground model.  In Figure \ref{errorsVsEpsilon}, we plot, as a function of $\varepsilon_0$, the error bars on the measured global signal at $79\,\textrm{MHz}$.  At unrealistically low foreground modeling error, the observations are dominated not by residual foregrounds, but instead by instrumental noise.  The errors are therefore roughly independent of $\varepsilon_0$.  Once $\varepsilon_0 \theta_{\textrm{fg}} / \sqrt{\Omega_{\textrm{sky}}} \sim 1/t_{\textrm{int}} \Delta \nu$, the errors become increasingly dominated by residual foregrounds, and therefore rise with $\varepsilon_0$.  Note that the rise occurs much more rapidly when the angular information is poor.  This can be understood through the lens of our result from Section \ref{spectralOnly}, which stated that with only spectral information, nothing can be formally done to mitigate errors in the foreground model.  As $\varepsilon_0$ rises, larger errors must simply be accepted by those experiments without angular resolution.

\begin{figure}
\centering
\includegraphics[width=0.45\textwidth]{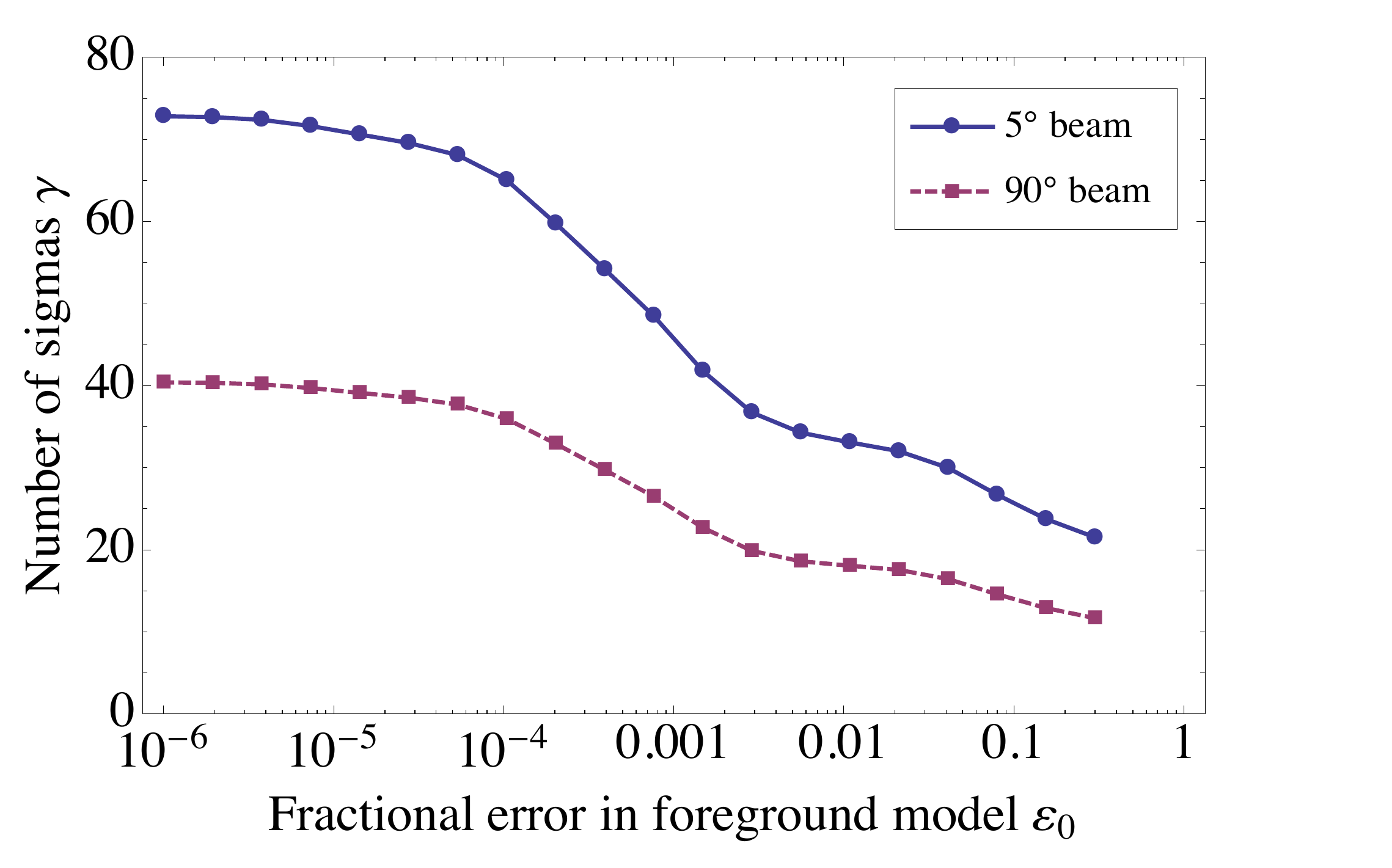}
\caption{The detection significance $\gamma$ of the fiducial global $21\,\textrm{cm}$ signal, plotted against the fractional error in our foreground model $\varepsilon_0$.  At extremely low $\varepsilon_0$, foreground residuals are a negligible source of error and the detection significance is constant.  At higher $\varepsilon_0$, foregrounds become more important, and $\gamma$ drops until the detection is driven by the trough feature between $30$ and $80\,\textrm{MHz}$.  At the highest $\varepsilon_0$, $\gamma$ drops further as foreground modeling errors become so high that even the trough is difficult to detect.  An integration time of $100\,\textrm{hrs}$ was assumed to make this plot.}
\label{sigmasVsEpsilon}
\end{figure}

The transition from a noise-dominated experiment to a foreground-residual dominated experiment can also be seen in Figure \ref{sigmasVsEpsilon}, where we show the detection significance $\gamma$ as a function of $\varepsilon_0$.  There are several regimes of interest.  At very low $\varepsilon_0$, the foreground contribution to the covariance is negligible, and the information matrix [Equation \eqref{grandInfo}] is diagonal to a good approximation.  Its eigenvectors (our residual noise and foreground basis vectors) are thus delta functions in frequency, and every frequency band can be measured easily.  As $\varepsilon_0$ increases beyond $\sim 10^{-4}$ (\emph{i.e.} beyond the noise dominated regime), the detection significance drops, and an examination of the eigenmodes reveals the double-peaked structure seen in Figure \ref{90degGoodResidFG}.  Intuitively, the residual foregrounds are becoming more of a contaminating influence, and precautions (such as differencing neighboring frequency channels, which is what the double-peaks do) need to be taken.  As one approaches $\varepsilon_0 \sim 0.05$ and beyond, the detection significance drops more sharply because one ceases to be able to make statistically significant measurements beyond the $50$ to $80\,\textrm{MHz}$ trough.

As with any experiment that is plagued by foreground contamination, the better one can model the foregrounds, the better the final results.  However, Figure \ref{sigmasVsEpsilon} reveals an interesting result---unless fractional errors in foreground models can be suppressed beyond the $\sim 10^{-2}$ level to the $\sim 10^{-3}$ level, the improvements in detection significance are not dramatic.

\section{A designer's guide to experiments that probe reionization}
\label{reionizationDesigner}

We now turn to examining experiments that target the reionization regime, from $100$ to $250\,\textrm{MHz}$.  The analysis formalism remains essentially the same as that in the previous section, and most results remain unchanged.  We therefore focus on highlighting the differences.  We will find that whereas a significant detection (but not a well-constrained measurement) could be achieved without angular resolution for the Dark Ages, for reionization there are scenarios where a positive detection simply cannot be made without angular information.

\begin{figure}
\centering
\includegraphics[width=0.45\textwidth]{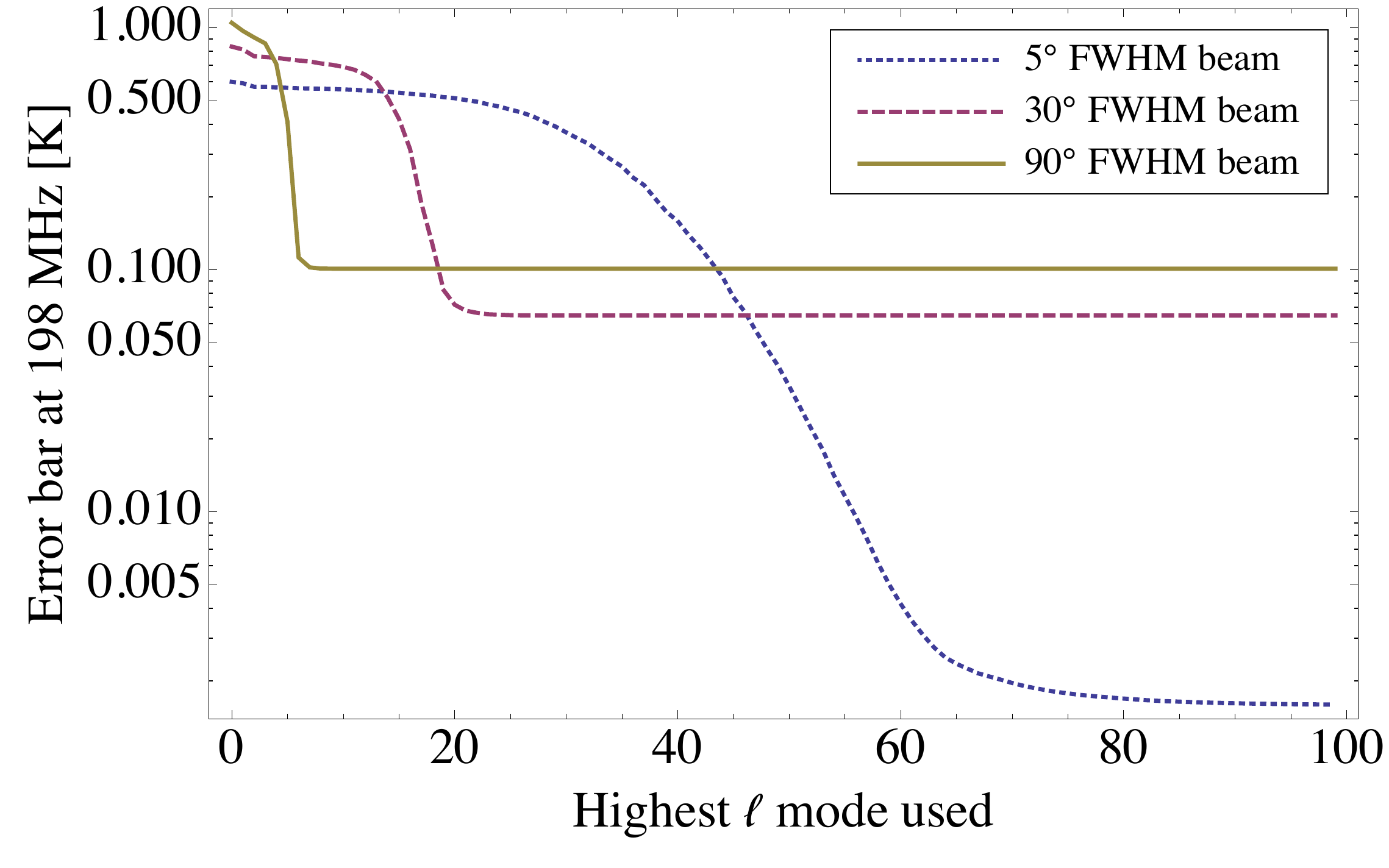}
\caption{Same as Figure \ref{errorsVsEllCutoff}, showing error bars as a function of the highest $\ell$ mode used, but for $\nu=198\,\textrm{MHz}$ (more relevant for reionzation).}
\label{errorsVsEllCutoffReion}
\end{figure}

Consider first the predicted error bars.  In Figure \ref{errorsVsEllCutoffReion} we show the projected errors during reionization as a function of the maximum $\ell$ mode used in the analysis, just as we did for the Dark Ages in Figure \ref{errorsVsEllCutoff}.  Like before, the errors are seen to be relatively large if angular correlations are not used in the analysis (for comparison, recall that the signal is now $\sim 30\,\textrm{mK}$ and gradually decaying to zero).  With angular correlations, the error bars can be suppressed to a level that allows different theoretical models to be distinguished from each other, exactly as we saw for the Dark Ages.

\begin{figure}
\centering
\includegraphics[width=0.45\textwidth]{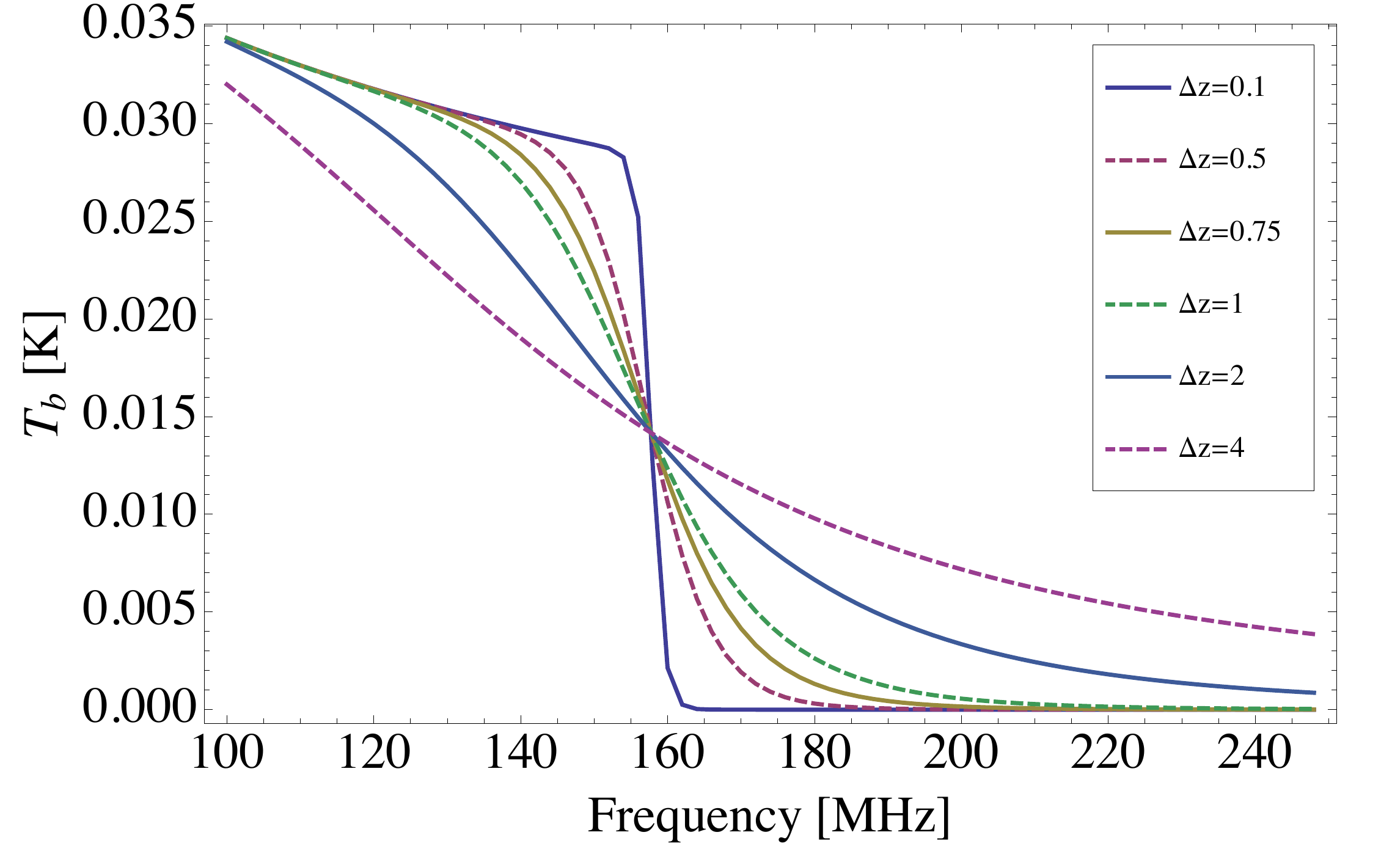}
\caption{Some sample histories of the brightness temperature $T_b$ during reionization.  Our samples here all have a midpoint of reionization $z_r$ of $10$.  Different curves show different reionization durations.}
\label{sampleReionHistories}
\end{figure}

However, when we consider the statistical significance of our detection, the story departs from what we found previously.  In particular, we will find here that the ability to make a statistically significant detection of a signal above noise and foregrounds is dependent on the duration of reionization $\Delta z$.  This parameter enters into a general form of the reionization global signal that we adopt:
\begin{equation}
\label{reionCosmoSignal}
T_b(z) = \frac{T_{21}}{2} \left( \frac{1+z}{10} \right)^{1/2} \left[ \tanh \left( \frac{z-z_r}{\Delta z} \right) +1 \right],
\end{equation}
where $T_{21} = 27\,\textrm{mK}$ and $z_r$ is the redshift at the midpoint of reionization \cite{jonathanAvi}.  Comparing this to Equation \eqref{modelTb}, we see that this arises from assuming that the mean ionized fraction takes the form
\begin{equation}
\label{tanhxi}
\overline{x}_i (z) = \frac{1}{2} \left[1 + \tanh \left( \frac{z_r-z}{\Delta z} \right)\right].
\end{equation}
This parameterization is identical to the one adopted by \cite{jonathanAvi} and \cite{bowmanRogersMeasurement}, and differs from that employed by WMAP \cite{WMAP7} and the CAMB software package \cite{CAMB} only in that their $\tanh$ parameterization is in conformal time rather than redshift.  Some sample brightness temperature signals with differing $\Delta z$ but fixed $z_r$ (equal to $10$) are shown in Figure \ref{sampleReionHistories}.

\begin{figure}
\centering
\includegraphics[width=0.45\textwidth]{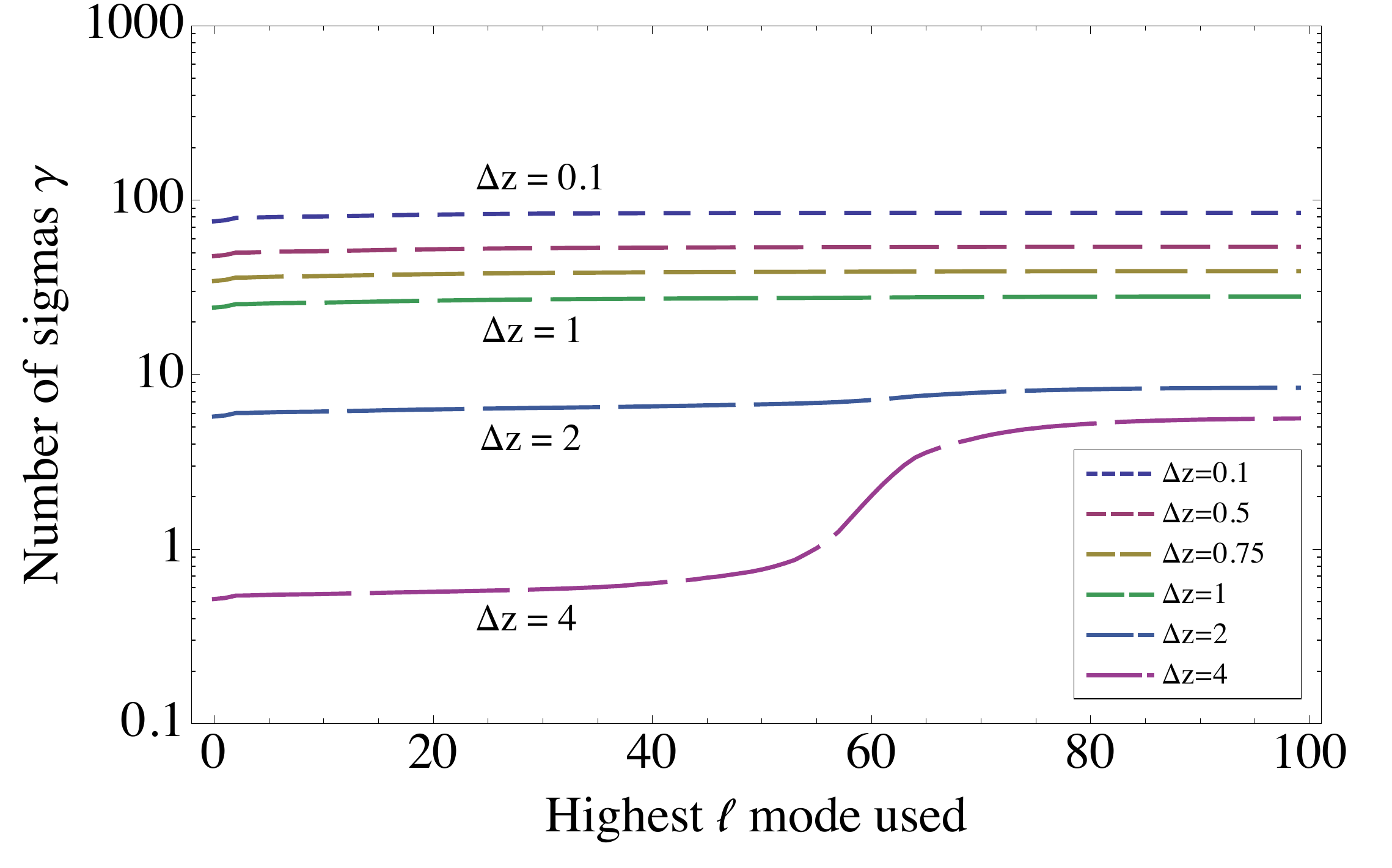}
\caption{Detection significance $\gamma$ for various reionization durations $\Delta z$ (longer dashes correspond to greater $\Delta z$), plotted as a function of various $\ell$ cutoffs, beyond which the angular correlation information is not used for foreground mitigation.  Whereas abrupt reionization scenarios can be easily distinguished from foregrounds and to yield high-significance detections, for gradual reionization scenarios it is necessary to utilize angular correlation information to make a detection.}
\label{sigmasEllCutoffReion}
\end{figure}

Intuitively, one would expect reionization histories with larger $\Delta z$ to be more difficult to detect, since they give rise to cosmological signals that monotonically decrease in smooth, gradual ways with frequency that are very similar to typical foreground spectra such as that shown in Figure \ref{GlobalFGSpec}.  This is exactly what one sees in Figure \ref{sigmasEllCutoffReion}.  As reionization becomes more and more abrupt, its signature in the global signal becomes increasingly distinct from that of the smooth foregrounds, and becomes easier to detect.  For abrupt reionization, using higher $\ell$ modes in the analysis improves the error bars, but does not increase the detection significance, just as we found for the Dark Ages.  As we move to more gradual reionization scenarios, however, we see that the foregrounds become so difficult to distinguish from the cosmological signal that it becomes necessary to use angular correlations simply to obtain a detection.

\begin{figure*}
\centering
\includegraphics[width=1.0\textwidth,trim=3cm 0cm 5cm 0cm,clip]{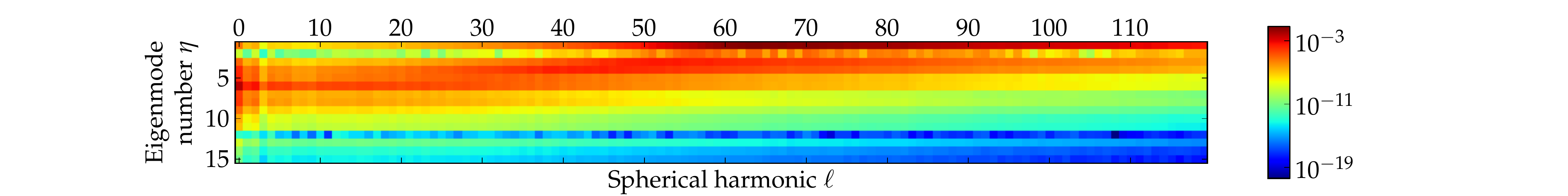}
\caption{Same as Figure \ref{SigmasEtaEllDarkAges}, except for an experiment probing an extended reionization scenario with $z_r=10$ and $\Delta z =4$.  Whereas most of the detection significance comes from the $\ell=0$ information in experiments that target the Dark Ages, an extended reionization scenario has a spectrum that is so similar to a typical foreground spectrum that most of the detection significance comes from using high $\ell$ modes to aid foreground mitigation.}
\label{5degSigmasEtaEllReion1}
\end{figure*}

The necessity of angular information for probing extended reionization histories is further demonstrated in Figure \ref{5degSigmasEtaEllReion1}, where we show the contribution to $\gamma^2$ as a function of $\ell$ and foreground eigenmode $\eta$, for an experiment with a FWHM beam of $5^\circ$.  The assumed reionization scenario has a midpoint $z_r =10$ and duration $\Delta z = 4$.  Unlike for the Dark Ages, where fine angular correlations certainly helped---but did not dominate---the statistical significance of our detection, we see that for extended reionization most of our ability to detect a signal above noise and foregrounds comes from high $\ell$ information.  Angular information is thus crucial.

\section{Expected performance of fiducial experiments}
\label{fidExperiments}

In Sections \ref{darkDesigner} and \ref{reionizationDesigner}, we considered the various trade-offs in the experimental design of global $21\,\textrm{cm}$ signal experiments.  Having gained an intuition for the best types of experiments to build, we now consider some fiducial experiments and analyze their expected performance.

\subsection{A fiducial dark ages experiment}

As our fiducial dark ages experiment, we imagine an instrument with a FWHM beam of $5^\circ$ and a channel width of $\Delta \nu = 1\,\textrm{MHz}$.  We assume a total integration time of $100\,\textrm{hrs}$ over the entire sky.  For our foreground model, we assume that current constraints are accurate to the $10\%$ level at an angular resolution of $5^\circ$.

\begin{figure}
\centering
\includegraphics[width=0.45\textwidth]{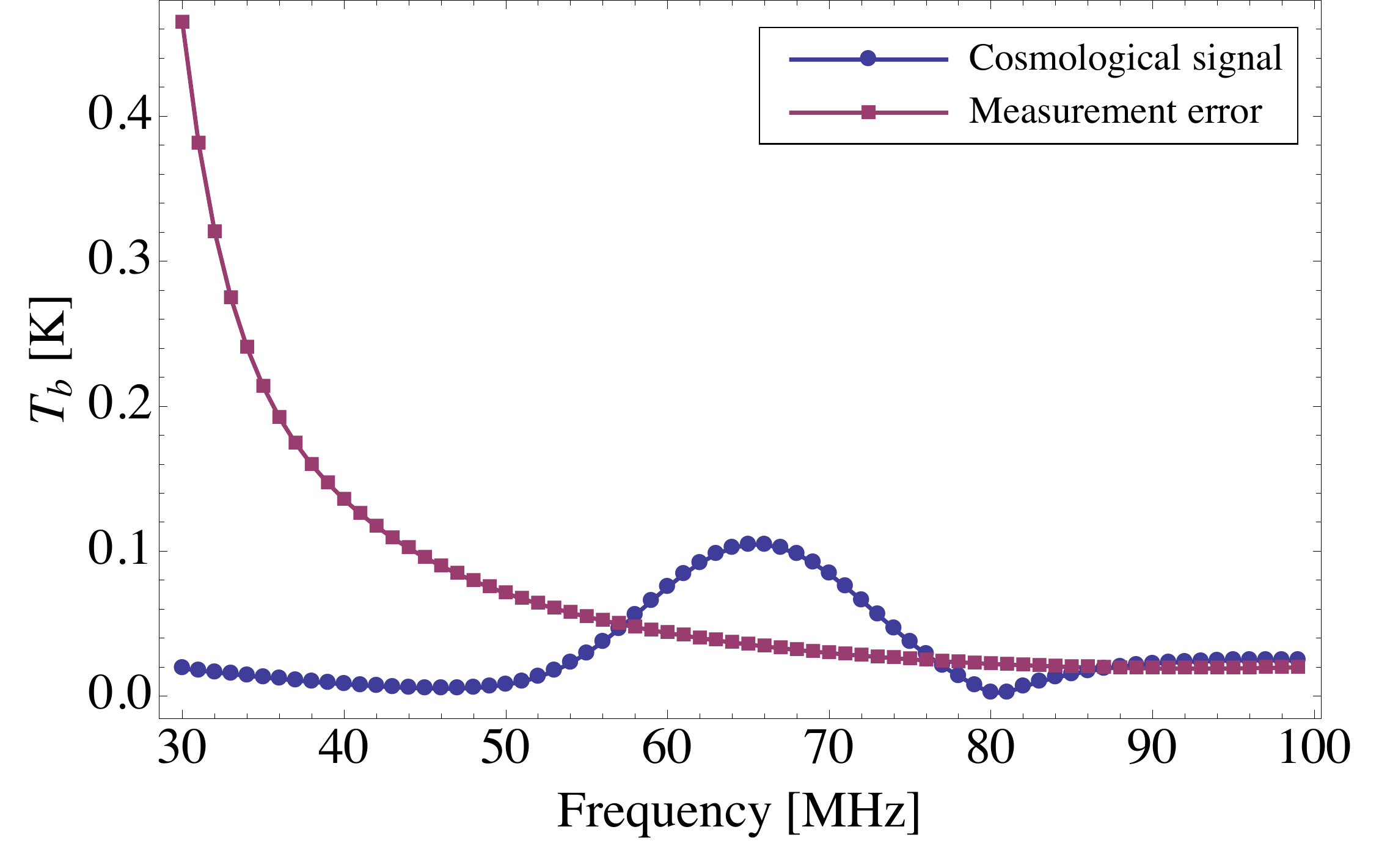}
\caption{Measurement errors and the absolute value of the fiducial cosmological model.  The fiducial experiment has a channel width of $1\,\textrm{MHz}$, a FWHM beam of $5^\circ$, and covers the full sky over a $100\,\textrm{hr}$ integration.  The trough between $50$ and $80\,\textrm{MHz}$ can clearly be measured accurately, allowing different cosmological scenarios to be distinguished from each other.  These results have \emph{not} been Wiener filtered, since the errors are small thanks to the use of angular correlations.}
\label{darkAgeFiducial}
\end{figure}

In Figure \ref{darkAgeFiducial} we show the expected measurement errors, as well as our fiducial cosmological model.  With the error suppressed below the cosmological signal in the $50$ and $80\,\textrm{MHz}$ trough region, it is likely that detailed constraints can be placed.  (The errors can be further suppressed by binning the data points shown in the figure, although there do exist non-negligible correlations between the errors, so the improvement does not quite scale as $1/\sqrt{N}$, where $N$ is the number of data points).  The results shown in Figure \ref{darkAgeFiducial} represent a $25\sigma$ detection of our fiducial cosmological signal.

\subsection{A fiducial reionization experiment}
\label{subsec:fidreion}
As our fiducial reionization experiment, we also imagine an instrument with FWHM beam of $5^\circ$.  We assume a channel width of $\Delta \nu = 2 \,\textrm{MHz}$ and an integration time of $500\,\textrm{hrs}$.  Unlike the Dark Ages where the trough signature aided one's detection, for reionization we find that it is necessary to integrate for longer to be able to detect extended reionization scenarios.

\begin{figure}
\centering
\includegraphics[width=0.45\textwidth]{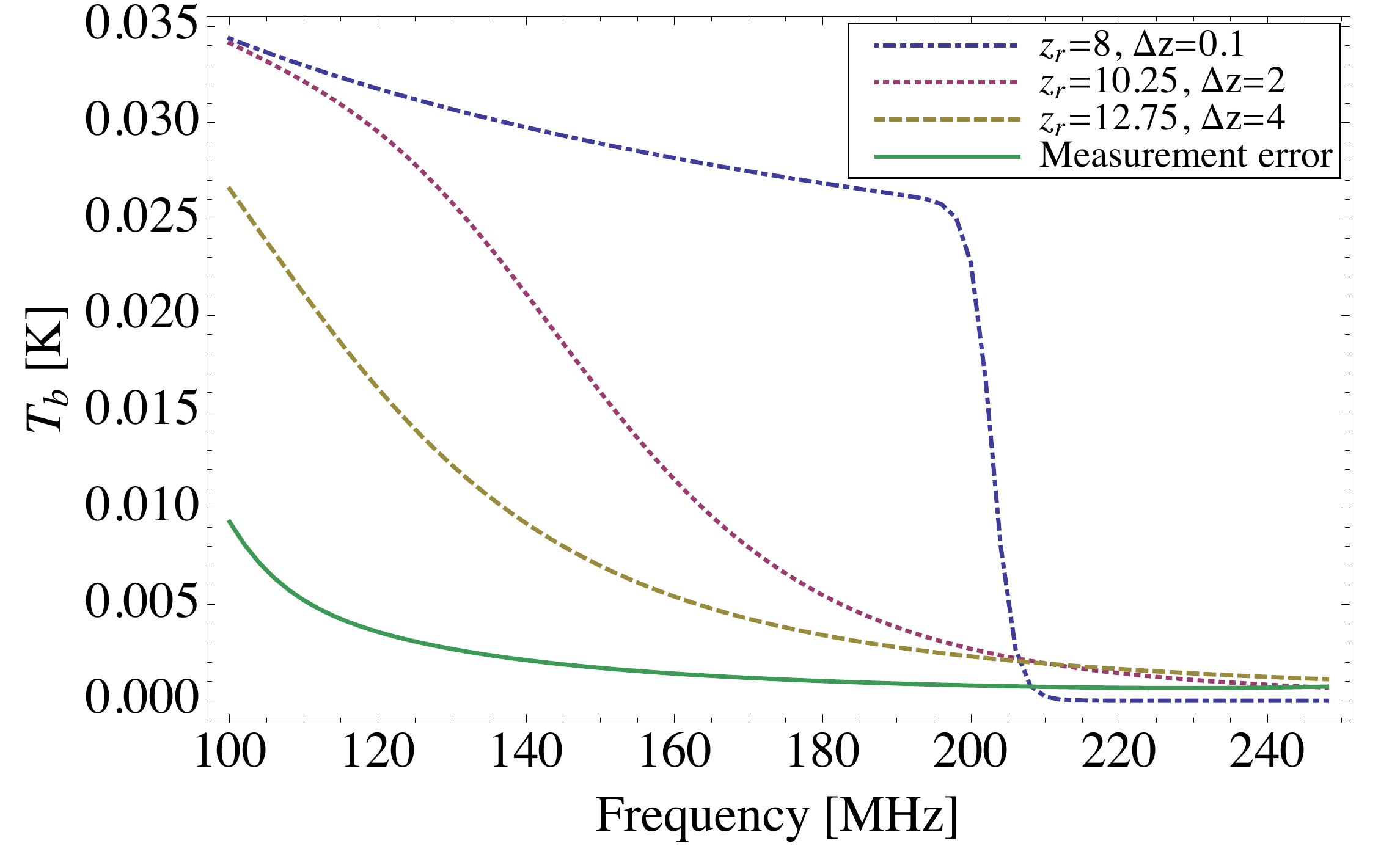}
\caption{Measurement errors and several fiducial reionization scenarios.  The fiducial experiment has a channel width of $2\,\textrm{MHz}$, a FWHM beam of $5^\circ$, and covers the full sky over a $500\,\textrm{hr}$ integration.  In all cases the measurement error is far below the signal, ensuring a positive detection.  The results have \emph{not} been Wiener filtered, since this is unnecessary with small errors.}
\label{fidModelsAndErrors}
\end{figure}

\begin{figure}
\centering
\includegraphics[width=0.45\textwidth]{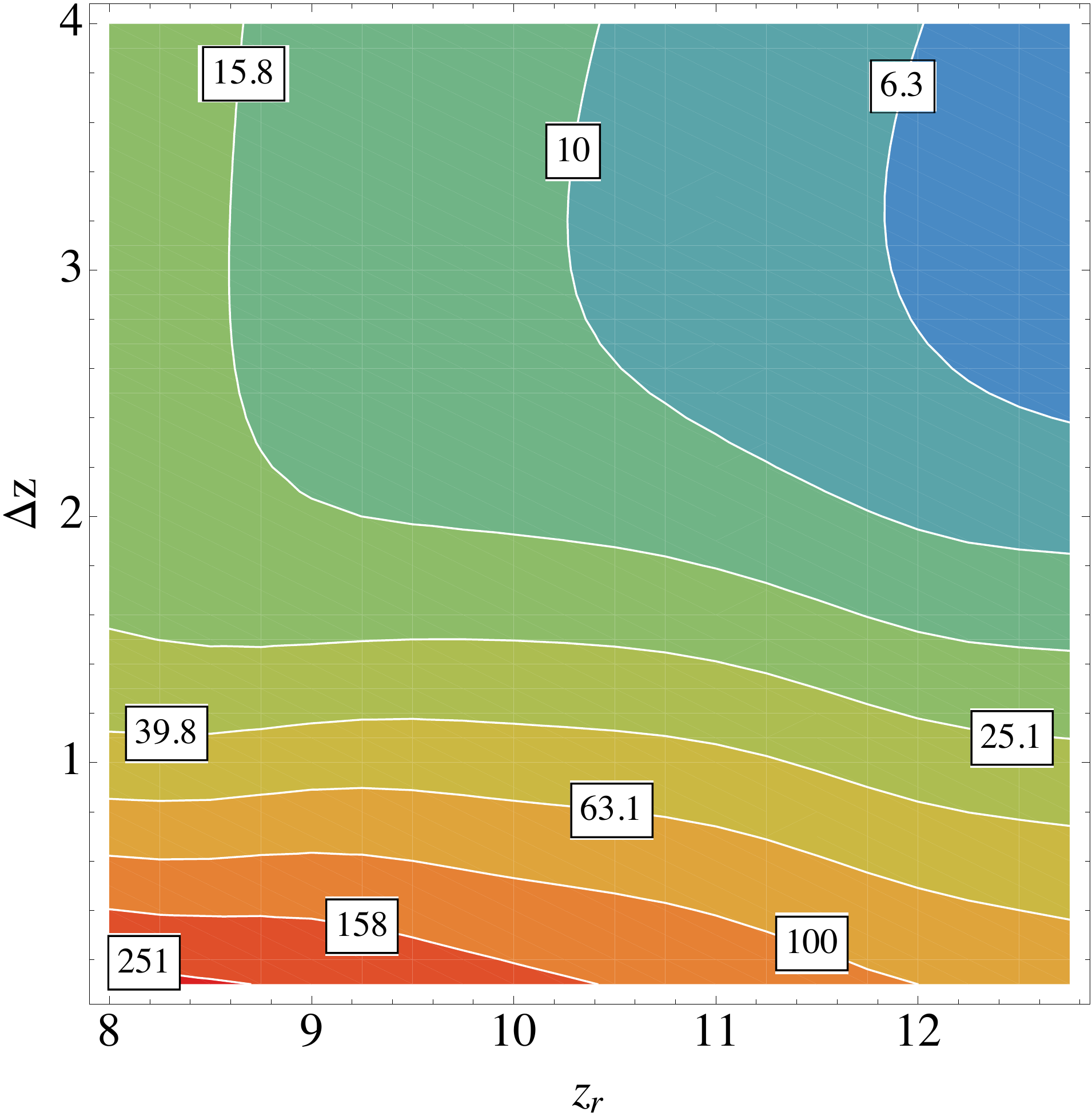}
\caption{Detection significance $\gamma$ [Equation \eqref{gammaDef}] as a function of the midpoint of reionization $z_r$ and the duration of reionization $\Delta z$, plotted with a logarithmic color scale for an instrument with a $5^\circ$ beam.  The boxed numbers show $\gamma$ of each contour.  A positive detection is possible over a wide range of reionization scenarios.}
\label{zrDeltazSigmaPlane}
\end{figure}

In Figure \ref{fidModelsAndErrors} we show the measurement errors along with several reionization scenarios.  The errors are seen to be well below the signal in all cases, suggesting that it should be possible detect a wide variety of reionization scenarios.  This is further illustrated in Figure \ref{zrDeltazSigmaPlane}, where we show $ \gamma$ [recall from Equation \eqref{gammaDef} that this is the ``number of sigmas" of the detection] as a function of the reionization midpoint $z_r$ and duration $\Delta z$ of our fiducial model from Equation \eqref{reionCosmoSignal}.  As expected, sharper reionization histories (small $\Delta z$) are more easily distinguishable from smooth foregrounds, and thus are more easily detectable.  Above $\Delta z \sim 2$, however, we see that this no longer applies, and one must simply hope that reionization took place at lower redshifts, where the overall amplitude of foregrounds is lower.  In any case, we see from the figure that a positive ($3\sigma$ or above) detection can be made over a wide region of parameter space, including rather extended reionization scenarios at high redshift. 

\begin{figure}
\centering
\includegraphics[width=0.45\textwidth]{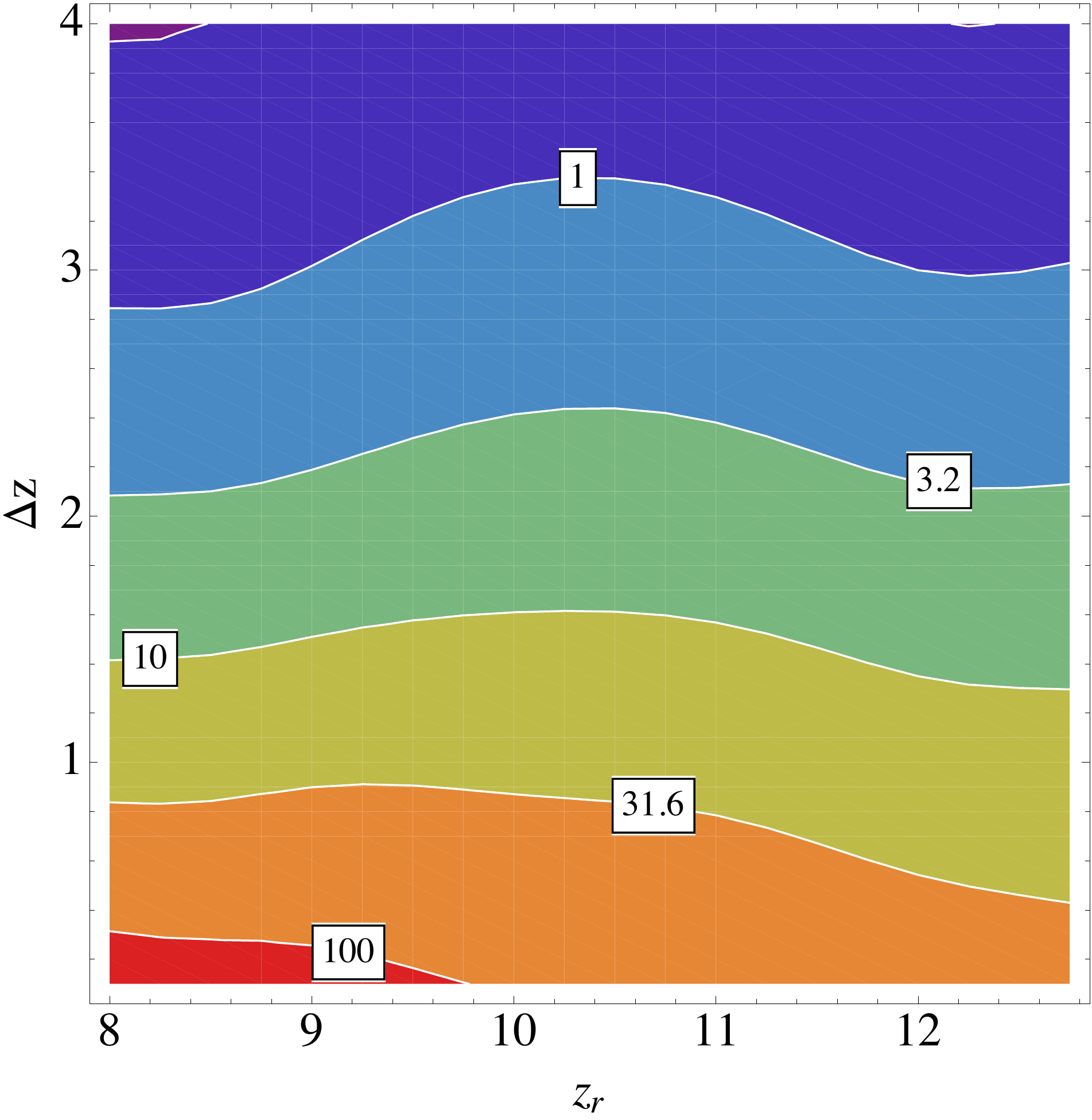}
\caption{Same as Figure \ref{zrDeltazSigmaPlane}, but for an instrument with a $90^\circ$ beam (\emph{i.e.} one with very little angular information).  The detection significance clearly decreases compared to Figure \ref{zrDeltazSigmaPlane}, where angular information was used to improve foreground subtraction.   The horizontal orientation of the contours in this figure indicate that foreground separation in spectral-only experiments rely primarily on reionization happening in an abrupt fashion.}
\label{zrDeltazSigmaPlane90deg}
\end{figure}

We emphasize that angular information is the key to making positive detections of the global $21\,\textrm{cm}$ signal even when reionization is extended.  To see this, consider Figure \ref{zrDeltazSigmaPlane90deg}, where we show the detection significance contours for an instrument with a $90^\circ$ beam.  With very little angular information available, the detection significance $\gamma$ goes down throughout parameter space.  In addition, the rather horizontal orientation of the contours suggests that without angular information, $\gamma$ is driven primarily by the duration of reionization, with abrupt reionization scenarios more easily distinguishable over foreground spectra.  Only with angular information is it possible to detect extended reionization scenarios.

\begin{figure*}
\centering
\includegraphics[width=1.0\textwidth]{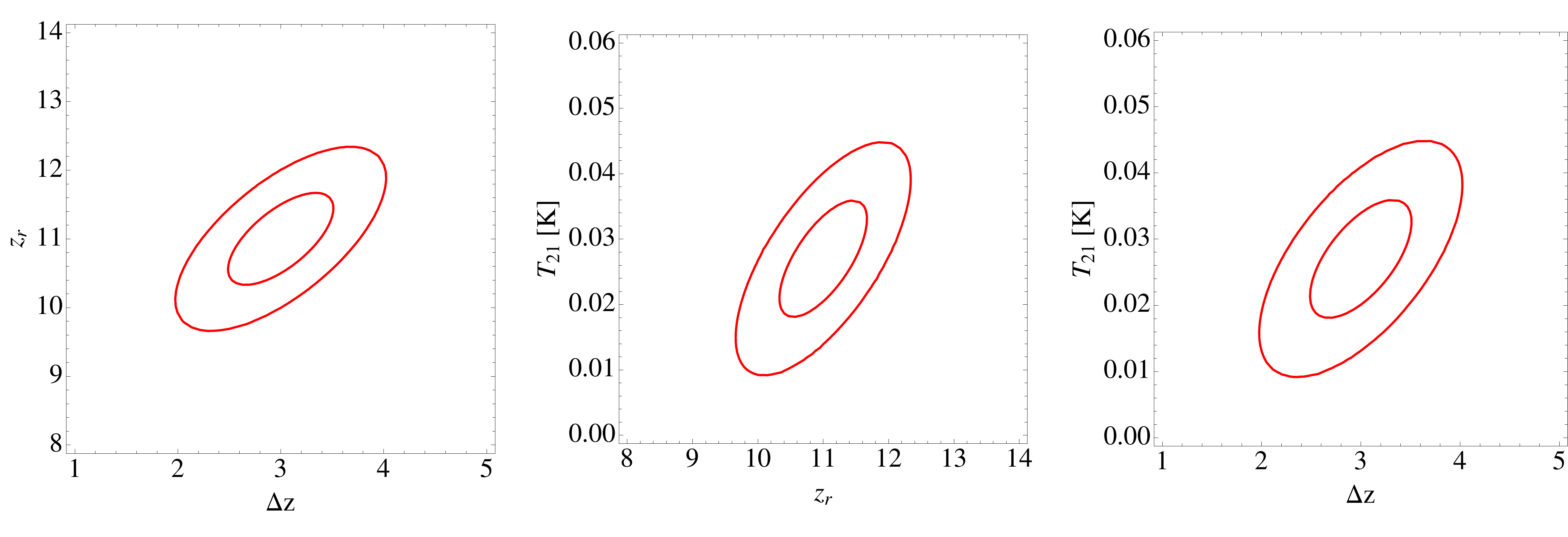}
\caption{Pairwise projections of the likelihood contours ($1\sigma$ and $2\sigma$) for $\Delta z$, $z_r$, and $T_{21}$ from Equation \eqref{reionCosmoSignal}, assuming fiducial values of $\Delta z =3$, $z_r=11$, and $T_{21} = 27\,\textrm{mK}$.  In each case, the projections were obtained by marginalizing over the unplotted third variable.  Experimentally, an angular resolution of $5^\circ$ and an integration time of $500\,\textrm{hrs}$ is assumed.}
\label{contourRow}
\end{figure*}

As a further demonstration of this, suppose one were to fit the data from an experiment to Equation \eqref{reionCosmoSignal}, with $T_{21}$, $z_r$, and $\Delta z$ as free parameters.  To estimate the precision with which the parameters can be constrained, we use the Fisher matrix formalism.  Starting with our equation for the Fisher matrix [Equation \eqref{fisherDef}] we simply select $T_{21}$, $z_r$, and $\Delta z$ as our parameters and use our measurement covariance $\boldsymbol \Sigma$ in place of the noise covariance $\mathbf{C}$.  The inverse of the Fisher matrix then provides us with the best possible covariance on our three parameters.  Performing the calculation for our fiducial experiment assuming a reionization model with $T_b=27\,\textrm{mK}$, $z_r=11$ and $\Delta z =3$ yields error bars of $8.9\,\textrm{mK}$, $0.67$, and $0.51$ for those parameters respectively.  Pairwise projections of the error ellipsoid (obtained by marginalizing over the unplotted third variable) are shown in Figure \ref{contourRow}.  While in this case of rather extended reionization the detection is perhaps not as statistically significant as one might hope, it is at least possible, which is not the case if only spectral methods are employed.  In addition, extra integration time results in tighter constraints, since the angular foreground subtraction techniques allow one to access some of the smooth spectral modes.

%

\subsection{Comparisons to cosmic microwave background constraints}

Constraints on reionization can also be derived from cosmic microwave background (CMB) experiments, and in this section we compare current and upcoming CMB constraints to those that can be expected from global signal measurements.  As CMB photons travel from the surface of last scattering to us, they are Thomson scattered by free electrons from reionization, giving rise to an optical depth $\tau$.  In particular, the optical depth is given by
\begin{equation}
\label{optDepth}
\tau = \sigma_T \int_0^{z_{\textrm{rec}}} \frac{\overline{n}_e (z)}{1+z} \frac{ds}{dz}dz,
\end{equation}
where $\sigma_T$ is the Thomson cross-section, $z_{\textrm{rec}}$ is the recombination redshift, $\overline{n}_e (z)$ is the mean free electron number density, and $ds$ is the comoving line element.  The reionization history is encoded in the functional form of $\overline{n}_e (z)$, which is proportional to the mean ionization fraction $\overline{x}_i(z)$:
\begin{equation}
\label{freeElectronDensity}
\overline{n}_e = \frac{\overline{x}_i (z) \overline{\rho}_b(z)}{\mu_e m_p},
\end{equation}
where $\rho_b$ is the mean baryon density, $\mu_e = 1.22$ is the mean mass per free electron in atomic units, and $m_p$ is the proton mass.  Equations \eqref{optDepth} and \eqref{freeElectronDensity} relate the reionization model [in our case, Equation \eqref{tanhxi}] to the optical depth.  A measurement of the optical depth from the temperature and polarization power spectra of the CMB will thus constrain reionization.

Unfortunately, optical depth measurements have a major shortcoming: as integrated line-of-sight measurements, their dependence on the duration of reionization $\Delta z$ is extremely weak, since all that matters is the column density of free electrons to the last scattering surface.  Optical depth constraints are therefore primarily constraints on $z_r$.  This is certainly true for WMAP, and \cite{MukherjeeLiddle} finds the same to be true for Planck.  The latest WMAP fits find $z_r = 10.5 \pm 1.2$ \cite{WMAP7}.

More recently, the South Pole Telescope (SPT) has used limits on the kinetic Sunyaev-Zel'dovich (kSZ) effect to place constraints on the duration of reionization.  The kSZ effect refers to the Doppler shifting of CMB photons as they scatter off coherently streaming free electrons, and contributes to the CMB anisotropy because reionization is not spatially homogeneous (\emph{i.e.} it is ``patchy").  By combining reionization simulations with SPT measurements of the high $\ell$ portions of the CMB power spectrum, a tentative $2\sigma$ constraint of $(\Delta z)^{\textrm{SPT}} < 7.9$ was obtained \cite{OliverSPT}, where the SPT team defined $(\Delta z)^{\textrm{SPT}}$ to mean the difference in redshift between $x_i = 0.2$ and $0.99$.  \emph{If one assumes the reionization model of Equation \eqref{tanhxi}}, this translates to an upper limit of $\Delta z < 2.64$ with our parameterization\footnote{This is a smaller number than the SPT figure because $z=z_r - \Delta z /2$ and $z=z_r + \Delta z /2$ correspond to ionization fractions of $x_i = 0.269$ and $0.731$ respectively, giving a narrower time period than the one the SPT team used in their parameterization.}.  We use this SPT constraint for illustration purposes, but note that it is non-trivial to convert a measurement of the kSZ contribution to the CMB into a reionization constraint. Allowing for more general reionization models may weaken the constraint on $\Delta z$ \cite{MesingerkSZInterp}.
\begin{figure}
\centering
\includegraphics[width=0.48\textwidth,trim=0cm 0cm 0cm 0cm,clip]{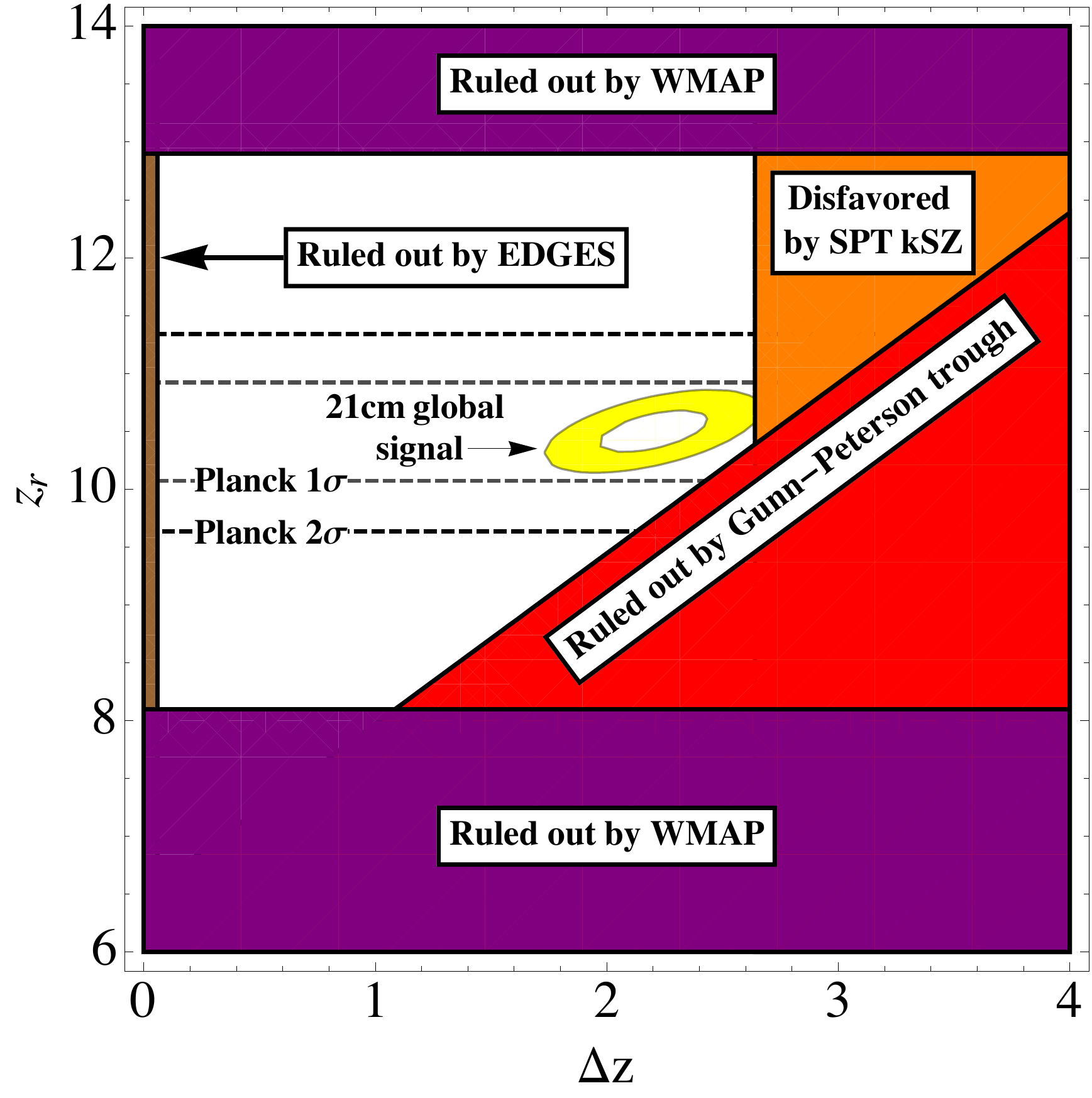}
\caption{A comparison between CMB constraints on reionization and the projected performance of our fiducial global signal experiment, which assumes an angular resolution of $5^\circ$ and an integration time of $500\,\textrm{hrs}$.  Projected Planck constraints are given by the dashed lines.  Existing WMAP constraints \cite{WMAP7} are given by the dark purple bands.  SPT measurements of the kSZ effect disfavor extremely long duration reionization scenarios, whereas EDGES rules out the most rapid scenarios.  Also included is the constraint (from quasar measurements of the Gunn-Peterson trough) that reionization is completed by $z=6.5$.  Assuming a reasonably pessimistic reionization duration of $\Delta z = 2.2$, our fiducial reionization experiment gives tight $1\sigma$ (white) and $2\sigma$ (yellow) error ellipses.  Again, we caution that the SPT constraint here is both aggressive and model dependent, so may prove weaker when more general models of inhomogeneous reionization are considered.}
\label{CMBcomp}
\end{figure}

In Figure \ref{CMBcomp}, we show the constraints set by the WMAP optical depth measurement (in purple), along with the SPT kSZ constraint (in orange), both at $2\sigma$.  From observations of the Gunn-Peterson trough, we also know that reionization is complete by $z=6.5$ \cite{BeckerGP}, so we include a prior that $x_i (z =6.5) > 0.95$ (in red).  The Planck satellite is expected to improve on the WMAP measurements of $\tau$, potentially capable of achieving a precision of $\Delta \tau = 0.005$ \cite{ColomboPierpaoliPritchard}.  This translates to a tighter set of limits on $z_r$, which we also show in Figure \ref{CMBcomp} as a set of dashed lines.

Finally, the EDGES global signal experiment has ruled out very rapid ($\Delta z < 0.06$) reionization \cite{bowmanRogersMeasurement}.  Being a single-dipole experiment, EDGES does not have sufficient angular resolution to provide constraints on more extended reionization scenarios, as we discussed in Sections \ref{reionizationDesigner} and \ref{subsec:fidreion}.  On the other hand, an experiment with a $5^\circ$ instrumental beam and an integration time of $500\,\textrm{hrs}$ (\emph{i.e.} our fiducial reionization experiment from Section \ref{subsec:fidreion}) can mitigate foregrounds much more effectively, leading to the white ($1\sigma$) and yellow ($2\sigma$) likelihood contours in Figure \ref{CMBcomp}.  In computing these contours, we assumed $\Delta z = 2.2$ in order to examine a reasonably extended---and therefore quite  observationally pessimistic---reionization scenario that has not yet been ruled out by the current SPT observations.  Encouragingly, we see that even under these conservative assumptions, global signal experiments can improve on the projected Planck $z_r$ constraints, and in addition provide tight limits on $\Delta z$.  This holds true even if the SPT limits become less stringent with future analyses, as we can see in Figures \ref{zrDeltazSigmaPlane} and \ref{contourRow}.  In any case, a global signal measurement should provide an interesting independent cross-check on all the aforementioned probes of reionization.

It must be noted that the quantitative details in this subsection hinge on the specific parametric form of the ionization fraction that we assumed [Equation \eqref{tanhxi}].  A different parameterization would affect all of the limits in Figure \ref{CMBcomp} except for the WMAP and Planck constraints (since optical depth measurements are insensitive to ionization history).  However, we expect the overall message---that a global signal experiment with good angular resolution can significantly improve our constraints---to be a robust one.

\section{Conclusions: Lessons Learned}
\label{conc}

In this paper, we have considered the general problem of extracting the global $21\,\textrm{cm}$ signal from measurements that are buried in foreground and instrumental noise contaminants.  We developed a mathematical formalism that extends the spectral-only signal extraction methods existing in the literature to incorporate angular methods for separating the cosmological signal from foregrounds.  Crucially, our proposed data analysis algorithms do not require \emph{a priori} information about the form of the cosmological signal.  This makes our methods immune to possible mismatches between theoretical expectations and observational reality, complementing other approaches in the literature that assume a parametric form for the signal.  One might imagine using our methods to make a first measurement of cosmological signal, which would allow a theoretical parameterization to be confirmed or revised; then, one could make use of the parameterization to precisely determine reionization parameters.

We also used our formalism in conjunction with a foreground model that was constructed from a combination of empirical data and analytical extensions to explore the various tradeoffs that one encounters when designing a global $21\,\textrm{cm}$ signal experiment.  The following is list of ``lessons learned" from our exploration:
\begin{itemize}
\item The simplest use of angular information is to downweight heavily contaminated region in the sky.  This can reduce the effective foreground contamination by as much as a factor of $2$ (Figure \ref{recipStats}).
\item Focusing on just the cleanest parts of the sky and simply excluding the dirtiest pixels from observations/analysis has two effects.  On one hand, the average amplitude of foreground contamination is lower.  On the other hand, having fewer pixels makes it harder to average down the \emph{errors} in one's foreground model, as well as to take advantage of angular correlations in the foregrounds to mitigate their influence.  Numerically, we find that the second effect dominates, which suggests that one should cover as much of the sky as possible.
\item Most of the statistical significance in one's detection comes from the $\ell=0$ angular mode, as opposed to higher $\ell$ modes in our whitened (\emph{i.e.} foreground model pre-divided) data.  However, this significance comes mostly from high signal-to-noise measurements of certain spectral modes that alone are insufficient to accurately constrain the shape of the cosmological spectrum.  To get small enough error bars to faithfully reconstruct the shape of the spectrum, it is necessary to take advantage of angular information to clean foreground-contaminated modes that would be inaccessible using spectral-only methods.
\item The most easily measurable feature during the dark ages is (as expected) the trough between $50$ and $80\,\textrm{MHz}$.
\item Errors integrate down more quickly with time for experiments possessing fine angular sensitivity than those that do not.  In the latter case, the extra integration time allows one to achieve even higher significance measurements of spectral modes that were already measured, without having a substantial effect on the detectability of the spectral modes that were not measurable due to large foreground contamination.  As mentioned above, these were the modes that were previously limiting the accuracy of the extracted global signal, and since their measurability was limited by foregrounds and not instrumental noise, extra integration time does little to help.  For experiments with angular resolution, angular foreground mitigation methods allow the residual foregrounds in those modes to be sufficiently suppressed for instrumental noise to be an important contribution to the errors, and so extra integration time has a larger effect.
\item Reducing the error in foreground modeling reduces the final measurement error.  However, unless the fractional foreground modeling error $\varepsilon_0$ can be reduced significantly beyond $0.01$, dramatic gains are unlikely.
\end{itemize}

Taking advantage of these lessons, we examined some fiducial experiments in Section \ref{fidExperiments}.  Incorporating the various recommendations listed above, we showed that small errors could be achieved for both the Dark Ages and reionization.  With reionization in particular, our angular methods allow extended reionization scenarios to be detected, which is encouraging since there are reasonable expectations that the duration of reionization $\Delta z > 2$ \cite{jonathanBayes}.
Crucially, we point out that the angular resolution and integration time requirements that are needed to utilize the methods of this paper are modest compared to those of large interferometric arrays such as LOFAR, MWA, PAPER, or GMRT.  Thus, while it is necessary to go beyond a traditional single-dipole experiment if one is to make full use of angular information, it should be possible to do so using a mid-scale experiment without incurring the expense and technical difficulties of a large experiment \cite{BittnerLoeb}.  If time (and autocorrelation information) is available on the large interferometers, they can of course be used to attempt a global signal measurement (such as that being currently pursued by the LOFAR team \cite{LOFARPrivComm}), but this may be an unnecessarily expensive option unless the relevant data has already been taken for other purposes.

Aside from instrumental noise and foreground contamination, global $21\,\textrm{cm}$ signal experiments have many other issues to contend with before becoming an experimental reality.  For example, instrumental calibration remains a challenge \cite{rogersCalib}, although the accuracy requirements may be reduced by the results of this paper, since one no longer needs to rely solely on an exquisite spectral measurement for foreground subtraction.  In any case, it is encouraging to see that by suitable experimental design and optimal data analysis, the potential show-stopper of instrumental noise and foreground contamination appears to be a controllable problem, paving the way for global $21\,\textrm{cm}$ signal experiments to make some of the first direct measurements of a mostly unconstrained era of cosmic evolution.


\section*{Acknowledgments}
The authors would like to thank Judd Bowman, Josh Dillon, Matt Dobbs, Geraint Harker, Jacqueline Hewitt, Scott Hughes, Matt McQuinn, Aaron Parsons, Ue-Li Pen, Jonathan Pober, Katelin Schutz, Josiah Schwab, Leo Stein, Eric Switzer, Dan Werthimer, Chris Williams, and Oliver Zahn for useful discussions.  A. Liu acknowledges the support of the Berkeley Center for Cosmological Physics.  This work was supported in part (for A. Loeb) by NSF grant AST-0907890, and NASA grants NNX08AL43G and NNA09DB30A, as well as by NSF grant AST-1105835 (for MT).

\bibliography{globalsignal}

\end{document}